\documentclass[english,twocolumn,tighten,times,twocolappendix
]{aastex63}
\pdfoutput=1 	
\usepackage{longtable}
\usepackage{amsmath}
\usepackage{amssymb}
\usepackage{graphicx}
\usepackage{hyperref}

\makeatletter

\def\aap{{\it Astronomy and Astrophys.}}

\def\apj{{\it Astrophys.~J.}}
\def\apjs{{\apj\ \it Suppl.}}		
\def\apjl{{\apj\ \it Lett.}}

\def\ga{\mathrel{\mathchoice {\vcenter{\offinterlineskip\halign{\hfil
\(\displaystyle##\)\hfil\cr>\cr\sim\cr}}}
{\vcenter{\offinterlineskip\halign{\hfil\(\textstyle##\)\hfil\cr
>\cr\sim\cr}}}
{\vcenter{\offinterlineskip\halign{\hfil\(\scriptstyle##\)\hfil\cr
>\cr\sim\cr}}}
{\vcenter{\offinterlineskip\halign{\hfil\(\scriptscriptstyle##\)\hfil\cr
>\cr\sim\cr}}}}}

\makeatother

\begin{document}

\title{
The TeV Cosmic Ray Bump: a Message from Epsilon Indi or Epsilon Eridani Star?}
\author{Mikhail A. Malkov }
\affiliation{Department of Physics and CASS, University of California
San Diego, La Jolla, CA, 92093, USA} 
\author{Igor V. Moskalenko}
\affiliation{Hansen Experimental Physics Laboratory and
Kavli Institute for Particle Astrophysics and Cosmology, Stanford University,
Stanford, CA 94305, USA}
\begin{abstract}
A recently observed bump in the cosmic ray (CR) spectrum from 0.3--30 TV is likely
caused by a stellar bow shock that reaccelerates \emph{preexisting}
CRs, which further propagate to the Sun along the magnetic field lines.
Along their way, these particles generate an Iroshnikov-Kraichnan (I-K) turbulence
that controls their propagation and sustains the bump. {\it Ad hoc} fitting of the bump shape requires
six adjustable parameters. Our model requires none, merely depending on \emph{three physical
unknowns that we constrain using the fit.} These are the shock Mach number, $M$, its size,
$l_{\perp}$, and the distance to it, $\zeta_{\text{obs}}$. Altogether, they define the bump rigidity $R_{0}$. With $M$$\approx$1.5--1.6 and $R_{0}$$\approx$4.4 TV, the model fits the data with $\approx$$0.08\%$ accuracy. The fit
critically requires the I-K spectrum predicted by the model and rules
out the alternatives. These fit's attributes make an accidental agreement
highly unlikely. In turn, $R_{0}$ and $M$ derived from the fit impose the
distance-size 
relation on the shock:
$\zeta_{{\rm obs}}$(pc)$\sim$$10^{2}\sqrt{l_{\perp}(\text{pc})}$.
For sufficiently large bow shocks, $l_{\perp}$$=$$10^{-3}$$-$$10^{-2}$ pc, we
find the distance of $\zeta_{{\rm obs}}$$=$3--10 pc. Three promising stars in this range are: the Scholz's
Star at 6.8 pc, Epsilon Indi at 3.6 pc, and Epsilon Eridani
at 3.2 pc. Based on their current positions and velocities, we propose
that Epsilon Indi and Epsilon Eridani can produce the observed spectral
bump. Moreover, Epsilon Eridani's position is only $\sim$$6.7^{\circ}$ off
of the magnetic field direction in the solar neighborhood, which also changes the CR arrival direction distribution. Given the
proximity of these stars, the bump appearance may change in a relatively
short time.
\end{abstract}

\section{Introduction\label{sec:intro}}

Cosmic ray (CR) observations have much improved over the past decade
and encouraged a deep rethinking of ways in which CRs are accelerated
and transported. New features, such as breaks and significant differences
in the spectral indices of different species, are being discovered
in the energy range that deemed as well-studied (ATIC-2 -- \citealt{2009BRASP..73..564P},
CREAM -- \citealt{2010ApJ...714L..89A, 2011ApJ...728..122Y}, PAMELA
-- \citealt{2011Sci...332...69A}, \textit{Fermi}-LAT -- \citealt{2014PhRvL.112o1103A},
AMS-02 -- \citealt{2015PhRvL.114q1103A, 2015PhRvL.115u1101A, 2017PhRvL.119y1101A, 2018PhRvL.120b1101A, 2018PhRvL.121e1103A, PhysRevLett.124.211102},
NUCLEON -- \citealt{2018JETPL.108....5A, 2019ARep...63...66A,2019AdSpR..64.2546G, 2019AdSpR..64.2559G},
CALET -- \citealt{2019PhRvL.122r1102A}, DAMPE -- \citealt{eaax3793}).
These features bear the signatures of CR acceleration processes and
their propagation history.

A flattening in the spectra of CR $p$ and He at $\sim$300 GV was
discovered first \citep{ATIC06, 2010ApJ...714L..89A, 2011Sci...332...69A}
giving raise to a number of different interpretations. With new accurate
measurements of spectra of several primary ($p$, He, C, O, Ne, Mg,
Si) and secondary (Li, Be, B) species becoming available from AMS-02
\citep{2015PhRvL.114q1103A, 2015PhRvL.115u1101A, 2017PhRvL.119y1101A, 2018PhRvL.120b1101A, 2018PhRvL.121e1103A, PhysRevLett.124.211102},
only two main interpretations remain. These are (i) the intrinsic
spectral breaks in the injection spectra of typical CR sources or
two different types of sources with soft and hard spectra \citep[a so-called ``injection scenario,''][]{2012ApJ...752...68V},
and (ii) a break in the spectrum of interstellar turbulence resulting
in the break in the diffusion coefficient \citep[a so-called ``propagation scenario,''][]{2012ApJ...752...68V, 2012PhRvL.109f1101B, AloisioKink_2015}.
The latter has very few free parameters and predicts that the spectral
index change before/after the break in the spectra of secondary species
$\Delta\gamma_{{\rm sec}}$ should be twice as large as the spectral
index change in the spectra of primary species $\Delta\gamma_{{\rm sec}}\approx2\Delta\gamma_{{\rm pri}}$.
This scenario seems to be well-supported by the AMS-02 data
\citep[see fits and a discussion in][]{2020ApJ...889..167B, Boschini_2020}.


Another spectral feature, a softening in the spectrum of protons and
He at $\sim$10--30 TV, was also found and seems to be well established
now (\citealt{ATIC06,2018JETPL.108....5A, 2019ARep...63...66A, 2019AdSpR..64.2546G, 2019PhRvL.122r1102A, eaax3793},
see also parameterizations in \citealt{Boschini_2020}). Even though
the two mentioned scenarios work well for one break, it is hard to
imagine that the injection spectrum of an ensemble of CR sources and/or
the spectrum of interstellar turbulence conspire to produce two relatively
sharp breaks one decade in rigidity apart on a large Galactic scale.
More likely explanation is the presence of a local CR source, where
both breaks would form by a fresh primary component with a harder
spectrum and an appropriate cutoff produced by the local source. Meanwhile,
a scenario with a local supernova (SN) explosion \citep[e.g.,][]{Fang2020, Fornieri2020, Yuan2020}
predicts no breaks in the \textit{secondary} component \citep{2012ApJ...752...68V}
and thus is unfeasible. Moreover, the unprecedented accuracy with
which the sharp break in the spectrum is measured rules out a linear
composition of independent sources. The transition would be too smooth
unless the sources conspire to replace each other at the transition
point \citep{Niu2020}.

The fact that the spectra of secondaries are also flattening above the
first break similar to the primaries, albeit with different spectral
index, implies that the secondary species have already been present
in the CR mixture that the bump is made of. \textit{Therefore, the
bump has to be made out of the preexisting CRs with all their primaries
and secondaries that have spent millions of years in the Galaxy.}
There are two possibilities: either the bump was being formed in a
significant part of the Galaxy simultaneously with the secondaries
over the CR confinement time, or it has been created locally, relatively
short time before we started observing it. We will argue that the
second possibility is more likely.

Worth mentioning is the sharpness of the bump that may provide useful
restrictions on the distance to its formation site. Originating in
remote objects would result in a smoother appearance due to the diffusive
propagation and mixing with old CRs. An estimate of the maximum distance,
provided in Sect.~\ref{sec:Distance}, limits it to a 3--10 pc,
thus suggesting a field-aligned CR propagation from the bump formation
site directly to the observer. Assuming, still, that most of the CR
accelerators collectively produce a featureless power-law spectrum,
in this paper we are proposing a bow shock of a passing star and/or
a magnetosonic shock in the Local Bubble as a possible origin of the
feature recently observed in the sub-TeV--10 TeV energy range. 

{We build a relevant CR reacceleration-propagation model and argue that
it reproduces the data not coincidentally because:}
\begin{itemize}
\item {This model provides a fit to a complex high-fidelity spectrum
that contains two consecutive breaks with only a $0.08\%$ deviation
over nearly four decades in energy with no adjustable parameters (other
than the distance, size, and Mach number of the shock);}
\item {The fit critically requires specific wave turbulence, }{\emph{self-driven
by reaccelerated CR}}{, shown to be of an Iroshnikov-Kraichnan (I-K)
type. Other turbulence spectra, such as the more common for the background
interstellar medium Kolmogorov spectrum, do not fit the data;}
\item {There exists at least one suitable shock that is magnetically connected with the Sun well within
the uncertainties of the local magnetic field direction, it is a very powerful termination shock around
Epsilon Eridani star at 3.2 pc;
it is different from the bow shock, but may also reaccelerate CRs.}
\end{itemize}

The remainder of the paper is organized as follows. 
In Sect.~\ref{sec:bubble} we assess the local ISM environment for shocks suitable to our model.
In Sect.~\ref{sec:problem} we consider a modification of the preexisting CR spectrum by such a shock. 
Sect.~\ref{sec:Distance} describes the propagation of the modified CR spectrum to the observer through a self-driven turbulence. 
In Sect.~\ref{sec:fitting} we determine the unknown physical parameters of the shock by fitting the data to our model. 
In Sect.~\ref{sec:Passing-Stars} we discuss possible shock candidates around passing stars. 
In Sect.~\ref{sec:Anisotropy-and-Magnetic} we evaluate the CR anisotropy associated with the nearby shock. 
In Sect.~\ref{sec:LocalB-field} we describe the topology of the local magnetic field and discuss its perturbation mechanisms.
We conclude with a brief summary in Sect.~\ref{sec:Model-Summary} and a discussion in Sect.~\ref{sec:Discussion-and-Outlook}. 
Appendices \ref{sec:Wave-Generation} and \ref{sec:WaveEnTransf} describe the formalism of the generation of turbulence by the reaccelerated particles and the wave energy transformation. Appendix \ref{sec:EstimateDistance} details an estimate of the distance to the hypothetical bow-shock. Appendix \ref{sec:AnisApp} supplements the discussion of the large- and small-scale anisotropy provided in Sect.~\ref{sec:Anisotropy-and-Magnetic}.

\section{The Local Bubble\label{sec:bubble}}

The Local Bubble is a low density region of the size of $\sim$200
pc around the Sun filled with hot H\,\textsc{i} gas that was formed
in a series of SN explosions \citep[e.g.,][]{1999A&A...346..785S, 2011ARA&A..49..237F}.
Studies suggest 14--20 SN within a moving group, whose surviving
members are now in the Scorpius-Centaurus stellar association \citep{2011ARA&A..49..237F, 2016Natur.532...73B}.

An excess of radioactive $^{60}$Fe found in deep ocean core samples
of FeMn crust \citep{1999PhRvL..83...18K, 2004PhRvL..93q1103K, 2016Natur.532...69W},
in the Pacific Ocean sediments \citep{2016PNAS..113.9232L}, in lunar
regolith samples \citep{2009LPI....40.1129C, 2012LPI....43.1279F, 2014LPI....45.1778F},
and more recently in the Antarctic snow \citep{2019PhRvL.123g2701K}
indicates that it may be deposited by SN explosions in the solar neighborhood.
Observation of $^{60}$Fe \citep[the half-life $\tau_{1/2}$$\sim$2.6 Myr,][]{2009PhRvL.103g2502R}
in CRs by ACE-CRIS spacecraft \citep{2016Sci...352..677B}, while
only an upper limit for $^{59}$Ni ($\tau_{1/2}$$\sim$76 kyr) is established
\citep{1999ApJ...523L..61W}, suggests $\ga$100 kyr time delay between
the ejecta and the next SN, and thus SN events should be clustered
in space and time. A newly found excess at 1--2 GV in the spectrum of CR iron \citep{2021arXiv210112735B} derived from the combined  Voyager 1 \citep{2016ApJ...831...18C}, ACE-CRIS, and AMS-02 \citep{PhysRevLett.126.041104} data supports this picture.

Currently, there is no general agreement on the exact number of SNe
events and their precise timing, but it is clear that there could have been
several events during the last $\sim$10 Myr at distances of up to
100 pc \citep{2016Natur.532...69W}. The most recent SN events
in the solar neighborhood were 1.5--3.2 Myr and 6.5--8.7 Myr ago
\citep{2015ApJ...800...71F, 2016Natur.532...69W}. The measured spread
of the signal is $\sim$1.5 Myr \citep{2015ApJ...800...71F}. However,
each of these events could, in principle, consist of several consequent
SN explosions separated by some 100 kyr, as an estimated time spread
for a single SN, located at $\sim$100 pc from the Earth, is just
$\sim$100--400 kyr and the travel time is $\sim$200 kyr. A detailed
modeling by \citet{2016Natur.532...73B} indicates two SNe at distances
90--100 pc with the closest occurred 2.3 Myr ago at present-day Galactic
coordinates $(l,b)=(327^{\circ},11^{\circ})$, and the second-closest
exploded about 1.5 Myr ago at $(343^{\circ},25^{\circ})$, both with
stellar masses of $\sim$$9M_{\sun}$.

The consequent SN events generated multiple shocks that are likely
still present in the Local Bubble. The typical lifetime of such shocks
$\sim$2 Myr can be estimated assuming that the shock speed is equal to
the speed of sound in a $10^{6}$ K plasma and a distance of 200 pc
\citep{2002A&A...390..299B}. The same ballpark estimate can be obtained,
if one assumes the old shock velocity of $\sim$100 km s$^{-1}$ --
comparable to the typical velocity of the random motions in the interstellar
medium (ISM). The shock could have also emerged more recently from
a decaying turbulence left behind after the last SN events by wave
steepening and shock coalescence.

In fact, radio observations and sometimes X- and $\gamma$-rays reveal
structures, often referred to as ``radio loops,'' that cover a considerable
area of the sky. There are at least 17 known structures \citep[for details, see][and references therein]{2015MNRAS.452..656V}
with the radii of tens of degrees that can be as large as $\sim$$80^{\circ}$.
The best-known is Loop I, which has a prominent part of its shell
aligned with the North Polar Spur. The spectral indices of these structures
indicate a nonthermal (synchrotron) origin for the radio emission,
but the origin of the loops themselves remains unclear. One of the
major limitations is the lack of precise measurements of their distances. However, it is clear that some of them could be the old SNRs and their huge angular sizes hint at their proximity to the solar system.

That was the original motivation for the present paper. However, in the process of writing an interesting alternative emerged:
{\it a bow shock of a star rapidly moving in the solar neighborhood.} Such bow shocks are abundant in the Galaxy as clearly seen in images\footnote{See, e.g., a bow shock around LL Ori in the Orion Nebula: 
https://www.nasa.gov/multimedia/imagegallery/imagefeature1060.html
}
by the Hubble Space Telescope, and are able to reaccelerate CRs. We discuss such a possibility in Sect.~\ref{sec:Passing-Stars}.
A subset of fast moving stars is likely to be produced by the outer Lindblad resonant scattering \citep{Dehnen_1999}. 
{ The formalism described in the following sections is equally suitable for both scenarios, a magnetosonic shock and/or a passing star.}

\section{Spectrum Modification by a Weak Shock\label{sec:problem}}

Given the requirements for the observed sharp breaks in the CR spectrum,
we  approach the problem  of their  origin as follows. Consider a
weak, planar magnetosonic shock propagating through the rarefied plasma
in the Local Bubble. Assume the shock is moving at a speed $u_{1}$
in the positive $x$-direction, with which the ambient magnetic field
makes an angle $\vartheta_{nB}$. At first, we will describe the CR
distribution both upstream and downstream of the shock by  a stationary,
convection-diffusion equation in one dimension. In obtaining its solution
and analyzing whether it has a potential to reproduce the observed
bump at some distance upstream, we will follow the CRs reaccelerated
at the shock as they propagate along the magnetic flux tube, whose
foot is on the shock surface. The propagation problem is two-dimensional,
at a minimum, and we will address it in Sect.~\ref{sec:Distance}, including 
its connection to the one-dimensional reacceleration problem we consider in this section.

The convection-diffusion equation for the particle distribution function $f$ near the shock reads
\begin{equation}
u\frac{\partial f}{\partial x}-\frac{\partial}{\partial x}\kappa\frac{\partial f}{\partial x}=\frac{1}{3}\frac{du}{dx}p\frac{\partial f}{\partial p}.\label{eq:cd}
\end{equation}
Here $p$ is the particle momentum and the flow velocity has the following
values, respectively ahead and behind of the shock, 
\[
u\left(x\right)=\begin{cases}
-u_{1}, & x\ge0,\\
-u_{2}, & x<0,
\end{cases}
\]
\[
u_{1}>u_{2}>0.
\]
The particle diffusion coefficient in the direction along the shock
normal ($x$) is $\kappa=\kappa_{\parallel}\text{\ensuremath{\cos^{2}}\ensuremath{\vartheta_{nB}}}+\kappa_{\perp}\text{\ensuremath{\sin^{2}}\ensuremath{\vartheta_{nB}}}$,
where $\kappa_{\parallel}$  and $\kappa_{\perp}$ are the components of the diffusion tensor
along and accross the field, respectively.

The general solution of this problem is well known and we will use
it in the form of \citet{BlandOst78}: 
\begin{eqnarray}
 &  & f\left(x,p\right)=\label{eq:dsSol}\\
 &  & \begin{cases}
{\displaystyle f_{\infty}\left(p\right)+\left[F\left(p\right)-f_{\infty}\right]\exp{\left[-u_{1}\int_{0}^{x}\frac{dx^{\prime}}{\kappa\left(x^{\prime},p\right)}\right]},} & x\ge0;\\
F\left(p\right), & x<0.
\end{cases}\nonumber 
\end{eqnarray}
Here $f_{\infty}\left(p\right)$ is the background CR distribution
far upstream of the shock and $F\left(p\right)$ is that at its front
and downstream. The upstream part of the solution grows sharply with
momentum at sufficiently large $x$, owing to the growth of the diffusion coefficient $\kappa$
with $p$. This behavior can produce the required bump if the prefactor
$F-f_{\infty}$ properly depends on the momentum. The function $f_{\infty}$
is given by the background spectrum, while an equation for $F\left(p\right)$
may be straightforwardly obtained by integrating both parts of Eq.~(\ref{eq:cd})
across the flow discontinuity at $x=0$: 
\begin{equation}
p\frac{dF}{dp}+qF=qf_{\infty},\label{eq:ShSpec}
\end{equation}
where $q=3r/(r-1)$ with $r=u_{1}/u_{2}$ being the shock compression.

For the next step, it is convenient to write the solution as follows,
\begin{equation}
F=f_{\infty}-\int_{0}^{p}\left(\frac{p^{\prime}}{p}\right)^{q}\frac{\partial f_{\infty}}{\partial p^{\prime}}dp^{\prime}.\label{eq:SolAtShock}
\end{equation}
We have merely assumed here that $f_{\infty}$ grows slower than $p^{-q}$
as $p\to0$, where the singularity at low energies is naturally removed by the diminishing flux of ``aged'' CRs due to the fast ionization losses \citep{Strong_1998}. 

Now we need to specify $f_{\infty}$. If the solution in Eq.~(\ref{eq:dsSol})
correctly describes the observed CR spectrum, then at particle momenta
below the first break, the exponential term is very small and we can
derive $f_{\infty}$ from the observed spectrum in this energy range.
Of course, we need to know it also at higher momenta where the spectrum
is masked by the exponential term, presumably responsible for the
spectral feature. However, the spectrum below the first break is a
simple power-law, $\propto p^{-\sigma}$, with the index measured
to be close to $\sigma\approx4.85$ \citep[e.g.,][]{Boschini_2020}.
Assuming that it continues with the same slope $\sigma<q$, at least
to the second break, we substitute this power-law into the solution
for $F$ in Eq.~(\ref{eq:SolAtShock}) to obtain the following relation,
\[
\frac{F}{f_{\infty}}=1+\frac{\sigma}{q-\sigma}.
\]
Now we can express the solution upstream through $f_{\infty}$: 
\begin{equation}
f\left(x,p\right)=f_{\infty}\left\{ 1+\frac{\sigma}{q-\sigma}\exp\left[-u_{1}\int_{0}^{x}\frac{dx^{\prime}}{\kappa\left(x^{\prime},p\right)}\right]\right\} .\label{eq:SolFinal}
\end{equation}
Note, that assuming the shock intrinsic index $q$ being larger than
that of the background CRs, $q>\sigma\approx4.85$, implies a low-Mach
shock. For that reason, we have not added the freshly injected CRs
to the solution, but will discuss the implications of this omission
in Sect.~\ref{subsec:Spectrum-Beyond-the}. In addition to being dominated
by the background CRs at higher energies, as $q>\sigma$, they could
not be injected efficiently in the first place \citep[e.g.,][]{Hanusch2019ApJ},
provided that the Alfv\'enic shock Mach number, $M_{A}\lesssim2-3$.
The second term in the solution in Eq.~(\ref{eq:SolFinal}) represents
the shocked background spectrum.

In most of the CR transport regimes the diffusion coefficient $\kappa$ grows with momentum.
Therefore, at some distance ahead of the shock only particles with
higher momenta contribute to the second term in the above solution.
It is this growth with momentum that produces the spectrum upturn
at $\sim$0.5 TeV. The growth, however, saturates at $p\to\infty$,
and so the overall spectrum reestablishes its background profile, but
at an enhanced level, $f$$\approx q\left(q-\sigma\right)^{-1}f_{\infty}$.
This second transition is responsible for the spectrum softening at
$\sim$15--20 TV. The spectral shape thus has a required appearance
of the observed CR bump and it very much depends on the following
function of momentum, see Eq.~(\ref{eq:SolFinal}), 
\begin{equation}
\Phi\left(p\right)=u_{1}\int_{0}^{x}\frac{dx^{\prime}}{\kappa\left(x^{\prime},p\right)}\label{eq:FiOfp},
\end{equation}
where the variable $x$ is considered to be fixed, e.g., by the observer's
position. 

Clearly, the presence of the two breaks in the spectrum
requires the path integral $\Phi$ to vary between small and large
values with momentum $p$ (see Sect.~\ref{subsec:Possible-Time-Dependence} for details). This imposes a constraint on the distance $L\sim x/\cos\vartheta_{nB}$
from the shock to the observer. However, the integral along the particle
propagation path runs through regions with different propagation regimes,
between which $\kappa$ varies by orders of magnitude. For sufficiently
small particle momenta, $\Phi\sim1$ is already in the shock precursor.
If the CR intensity near the shock is sufficient to drive waves to
the Bohm regime, $\delta B\sim B$, that is, $\kappa\sim cr_{g}$, which
gives an estimate for the distance $L\sim r_{\text{g}}c/u_{1}$, where
$r_{\text{g}}$ is the proton gyroradius. For a TV rigidity of the
spectral hardening (first break) and $u_{1}\sim10^{-3}c$, the estimate
yields, as a lower bound, $L\sim$1 pc. There are two questions about
this estimate, though. First, if the Bohm regime is realistic for the
observed intensity of the CR excess. Second, if the $\kappa\propto p$
Bohm scaling fits its spectral shape. We address these questions in
Sect.~\ref{sec:fitting} and Appendix \ref{sec:Wave-Generation}, and the answer is negative for both questions. Therefore, we need to extend the integration in Eq.~(\ref{eq:FiOfp}) beyond the shock precursor.

Far from the shock, we may try to substitute a popular ISM value for
$\kappa_{\parallel}=10^{28}\left(p/mc\right)^{a}$, with $a=0.3-0.6$,
neglecting the above contribution to the path integral from the shock
precursor. We obtain the following upper bound on the distance to
the observer: 
\[
L\sim3\cdot10^{2}\,\frac{\cos\vartheta_{nB}}{u_{100}}\left(p/mc\right)^{a}\ \text{pc},
\]
where $u_{100}$ is the shock speed in units of 100 km s$^{-1}$,
and $p$ is the typical momentum of particles that form the bump. The
caveat in this estimate is that the plane-shock solution becomes invalid
upstream from the shock precursor. Even if we assume the most favorable
Kolmogorov scaling with $a=1/3$, the distance $L$ for the TeV particles
may reach a kpc scale. As we argue in Sect.~\ref{subsec:Momentum-Diffusion},
observing the sharp spectral breaks at such a long distance would
be untenable because of the momentum diffusion. In addition, particles
diffusing that far from the shock are unlikely to stay inside the
flux tube and never return to the shock which is required for the
solution in Eq.~(\ref{eq:SolFinal}). We, therefore, conclude that
neither lower nor upper bound on the distance is realistic.

A systematic approach to the path integral in Eq.~(\ref{eq:FiOfp})
must thus include the effects of waves that are self-generated by
the reaccelerated particles both near the shock and further along
their path to the observer. We will address these processes in the next section. However,
the rigidity dependence of the spectrum at the observation point can
be obtained from Eq.~(\ref{eq:SolFinal}) already at this point. As we will argue in
the next section, one may approximate the path-integrated rigidity dependence of  $\kappa(R)$ as
$\kappa\propto R^{a}$, with $a$ being the turbulence index less one.
By also equating $x$ to the observer's location the rigidity profile
of the solution in Eq.~(\ref{eq:SolFinal}) can be represented as follows:
\begin{equation}
f\left(R\right)=f_{\infty}\left(R\right)\left[1+Ke^{-\left(R_{0}/R\right)^{a}}\right]\label{eq:Sec3simplRes}
\end{equation}
Here $K\equiv$$\sigma/\left(q-\sigma\right)$, while the parameters
$R_{0}$ and $a$ will be formalized in the next two sections before
fitting the above equation to the data. We only note here that $K$
is determined by the background CR index $\sigma$ and the shock Mach
number $M$, while $R_{0}$ depends on the distance to the observer,
$\zeta_{\text{obs}}$. Neither $M$ nor $\zeta_{\text{obs}}$ are
known, but we will constrain them using the fit, as the data set contains
a sufficient amount of information for that.
{We emphasize that these are the only two parameters that are required by our model, while an ad hoc fitting of the bump spectrum requires six parameters: such as the two values of break rigidity, two widths of the breaks, and two changes of the spectral index at the breaks.}

\section{Distance to the Shock, Intervening Turbulence, and the Spectrum \label{sec:Distance}}

The goal of this section is to link the unknown shock distance,  
along with its other parameters, to the { rigidity} dependence of the
path integral $\Phi$ in Eq.~(\ref{eq:FiSimple}), which is observable.
While the shock speed can be reasonably estimated, the dependence
of $\Phi$ on the turbulence that scatters CRs is not known \emph{a
priori}. The level and the spectrum of this turbulence are the most
important quantities that determine both the observed spectral anomaly
and the distance to its source. The turbulence is generated by two
different mechanisms, depending on how far from the shock the
waves are excited. In the shock precursor, Alfv\'en waves are generated
by the ordinary ion-cyclotron instability, usually employed in diffusive
shock acceleration problems. Further out along the flux tube the turbulence is
driven by the lateral pressure gradient of the shocked CRs.

Self-generated waves strongly suppress the particle diffusivity $\kappa$
near the shock, but not necessarily down to the Bohm value, $\kappa_{B}=cr_{g}/3$,
assumed in Sect.~\ref{sec:problem} for a crude estimate. The wave generation can be related
to the pressure exerted by the shocked CRs along the tube through
the work done on the waves, e.g., \citep{Drury83}: 
\begin{equation}
\left(u+V_{A}\right)\frac{\partial E_{\text{w}}}{\partial \zeta}=V_{A}\frac{\partial P_{\text{CR}}}{\partial \zeta},\label{eq:WaveGen}
\end{equation}
where $E_{\text{w}}$ and $P_{\text{CR}}$ are the Alfv\'en wave energy
density and the CR pressure, respectively, $V_A$ is the Alfv\'en speed, $u=u_{1}/\cos\Theta_{nB}$,
and $\zeta$ is the coordinate running from the shock along the field
line. It is related to the coordinate $x$ used in Sect.~\ref{sec:problem}
as $x=\zeta\cos\vartheta_{nB}$. Within the quasilinear approximation,
the particle diffusion increases with $\zeta$ as $\kappa_{\parallel}\propto1/E_{w}\propto1/P_{\text{CR}}$,
which we apply in the shock precursor. At larger $\zeta$, where the cyclotron
instability subsides, the quasilinear treatment has to be abandoned.
Nevertheless, as long as the tube remains overpressured by the shocked
CRs, $\kappa_{\parallel}$ does not rise to the level of the background
ISM, $\kappa_{\text{ISM}}$, in contrast to the quasilinear predictions.
It remains at an intermediate level, $\kappa_{\text{int}}$, which is between
the diffusivity near the shock and in the ISM, $\kappa_{\text{ISM}}$. Using Eqs.~(\ref{eq:cd})
and (\ref{eq:WaveGen}), in Appendix \ref{sec:Wave-Generation} we
express the path integral in Eq.~(\ref{eq:FiOfp}) through $\kappa_{\text{int}}$:
\begin{eqnarray}
\Phi\left(p,\zeta\right) & = & \ln\frac{P_{\text{CR}}\left(0\right)}{P_{\text{CR}}\left(\zeta\right)}\nonumber \\
 & \simeq & \frac{u\zeta}{\kappa_{\text{int}}}+\ln\left[P_{\text{CR}}\left(0\right)\frac{\kappa_{\text{int}}}{\kappa_{B}}\left(1+\frac{u}{V_{A}}\right)^{-1}\right].\label{eq:FiOfpz}
\end{eqnarray}
Unlike in the shock precursor,
where the excited Alfv\'en waves are rapidly swept up by the shock,
the turbulence in the flux tube is long-lived and has enough time
for transforming to different modes and cascading in the wave number.
It supports the CR diffusivity at the level $\kappa_{\text{int}}$.
Its contribution is roughly characterized by the first term on the
r.h.s.\ in Eq.~(\ref{eq:FiOfpz}), but it requires a more accurate treatment.
We show in Appendix \ref{sec:WaveEnTransf} that the turbulence originates
from an acoustic instability driven by the excessive CR pressure in
the flux tube. 

\begin{figure}[tb!]
\includegraphics[scale=0.2,viewport=0bp 300bp 1210bp 935bp]{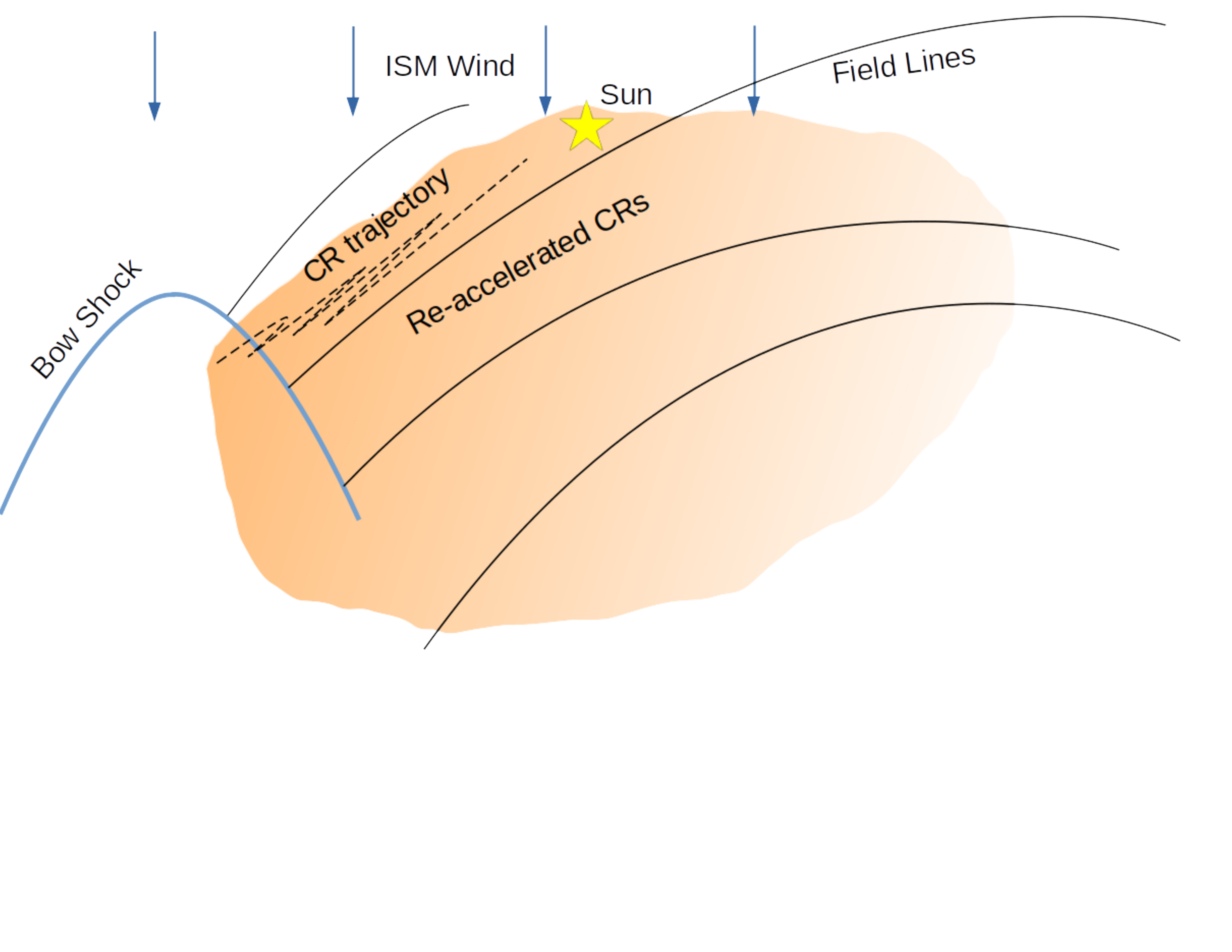}
\caption{Schematics of a bow-shock ahead of a moving star and reaccelerated
CRs, diffusing predominantly along the field lines. \label{fig:Bow-shock}}
\end{figure}

Up to now, there was no need to specify the shock geometry, as we
have used a one-dimensional model. To understand the excitation of
the acoustic instability requires at least a two-dimensional approach.
In Appendix \ref{sec:WaveEnTransf} we consider a bow shock ahead
of a moving star, Fig.~\ref{fig:Bow-shock}, as a simple 
two-dimensional setup for the flux tube. The acoustic instability inverse-cascades
to longer waves with a steep spectrum, $\propto$\,$k^{-3}$. It is, however,
intercepted by a forward Alfv\'en cascade that results in the I-K turbulence,
$k^{-3/2}$, thus yielding the $\kappa_{\text{int}}\propto \sqrt{p}$\,-scaling with momentum. One can see then that at large $\zeta$ the first term on the r.h.s.\ of Eq.~(\ref{eq:FiOfpz}) varies in a much broader range than the second one, when the particle rigidity varies
between the spectral breaks. Therefore, the main contribution to the
path integral in Eq.~(\ref{eq:FiOfp}) comes from the flux tube outside
of the shock precursor, which is consistent with our preliminary estimates. 
Since $\kappa_{\text{int}}\left(\zeta,p\right)$
maintains the $\sqrt{p}$\,-scaling throughout this region we can write
the value of $\Phi$ at the observation point $\zeta=\zeta_{\rm obs}$, as a
function of rigidity, $R$, as follows (cf. Eq. [\ref{eq:Sec3simplRes}]):
\begin{equation}
\Phi\left(\zeta_{\rm obs},R\right)=\left(\frac{R_{0}}{R}\right)^{1/2}.\label{eq:FiSimple}
\end{equation}
The quantity $R_{0}$  has a meaning of the bump rigidity, so that the spectral hardening
occurs at $R\approx0.1R_{0}$, while the softening -- at $R\approx4R_{0}$,
Sect.~\ref{subsec:Possible-Time-Dependence}. This quantity is calculated
in Appendix \ref{sec:EstimateDistance} and is given (in units of GV) as:
\begin{equation}
R_{0}=\left[ 3\left(\sigma-3\right)\xi\frac{u}{c}\sqrt{2\eta\Gamma\left(2\sigma-8\right)}\right] ^{2/\left(\sigma-3\right)}\ {\rm GV}.\label{eq:R0fromApp}
\end{equation}
Together with the shock index, $q$, in Eq.~(\ref{eq:SolFinal}), $R_{0}$
fully determines the shape of the spectrum. Apart from the measured
CR background index $\sigma\approx4.85$, and the shock velocity $u$,
it depends on the following two dimensionless parameters:
\begin{equation}
\xi\equiv\frac{\zeta_{\rm obs}}{\sqrt{r_{\text{GV}}l_{\perp}}}\quad\text{and}\quad\eta=\frac{KP_{\infty}}{\rho C_{s}V_{A}}\frac{\sigma-4}{R_{*}^{4-\sigma}},\label{eq:zeta-def}
\end{equation}
which {are} the normalized distance to the shock, and the pressure of
accelerated CRs $\eta$, normalized to the geometric mean of thermal
and magnetic pressure in the ISM ({ see} Eq.\ [\ref{eq:alf-eta-defs}]).
Here $l_{\perp}$ is the characteristic transverse scale of the flux
tube and $r_{\text{GV}}$ is the gyroradius of a GV proton. 

This concludes our theoretical consideration of reacceleration and propagation of
CRs to the observer. Our predictions thus depend on several unknown
parameters: the distance and the size of the shock (roughly equivalent
to $l_{\perp}$), the shock Mach number, $M$, and the shock speed,
$u$. The latter two can be related if the temperature of the ISM
near the shock is known. In fact, however, to make a precise comparison
of our model predictions with the observed spectrum, the following
two parameters suffice, the shock Mach number (to determine its index
$q$) and $R_{0}$. Courtesy of the I-K turbulence spectrum, derived
on theoretical grounds, our model thus contains no free parameters. 
Therefore, even though the physical parameters of the shock and
its geometry are not known, we can determine $M$ and constrain other
physical parameters by obtaining $R_{0}$ from the best fit to the data. 

\begin{figure}[tb!]
\includegraphics[scale=0.235,viewport=60bp 100bp 1210bp 935bp]{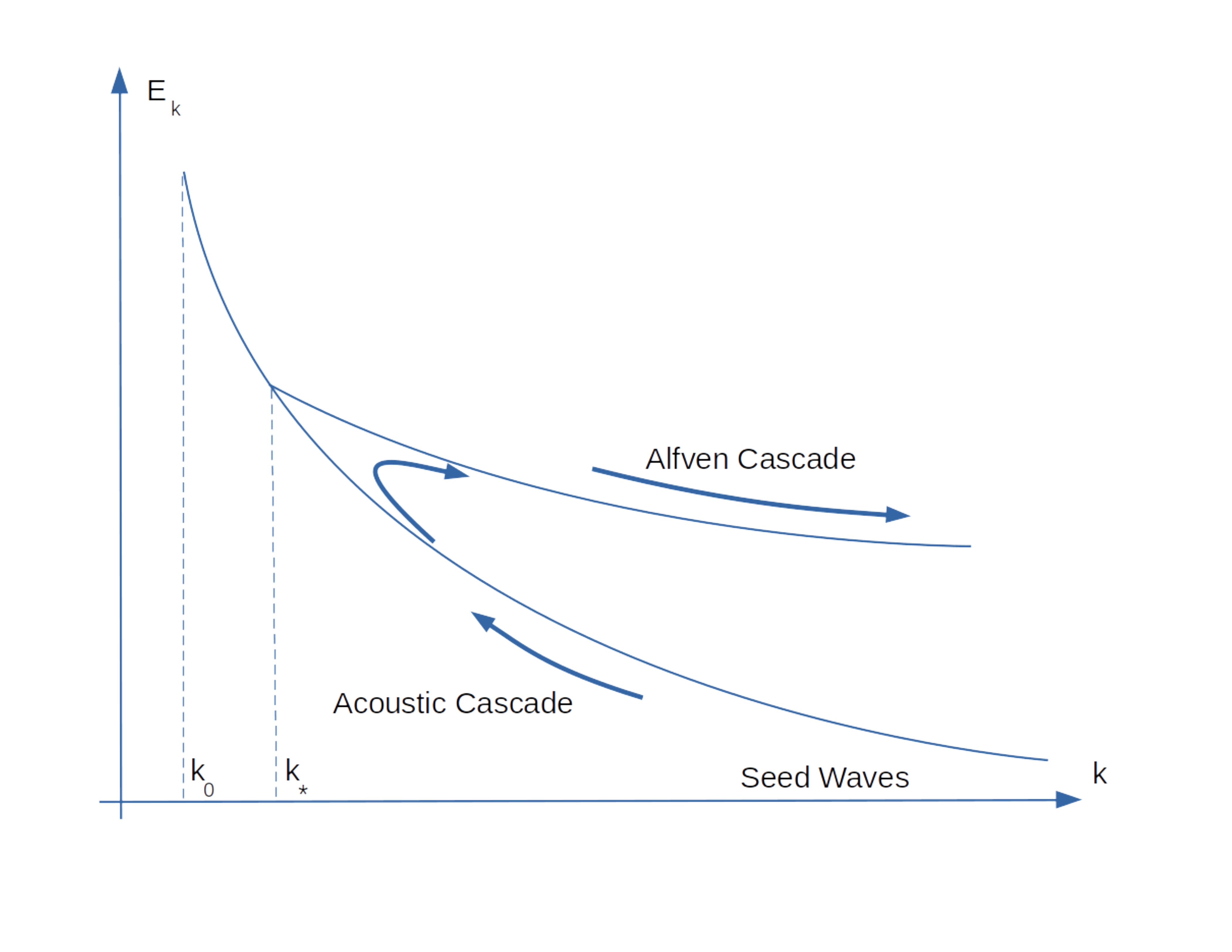}
\caption{Acoustic wave energy cascading to longer scales until they are intercepted
by the Alfv\'enic forward cascade. \label{fig:TurbSp}}
\end{figure}

\section{Fitting the Data\label{sec:fitting}}

\begin{figure}[tb!]
\includegraphics[scale=0.44]{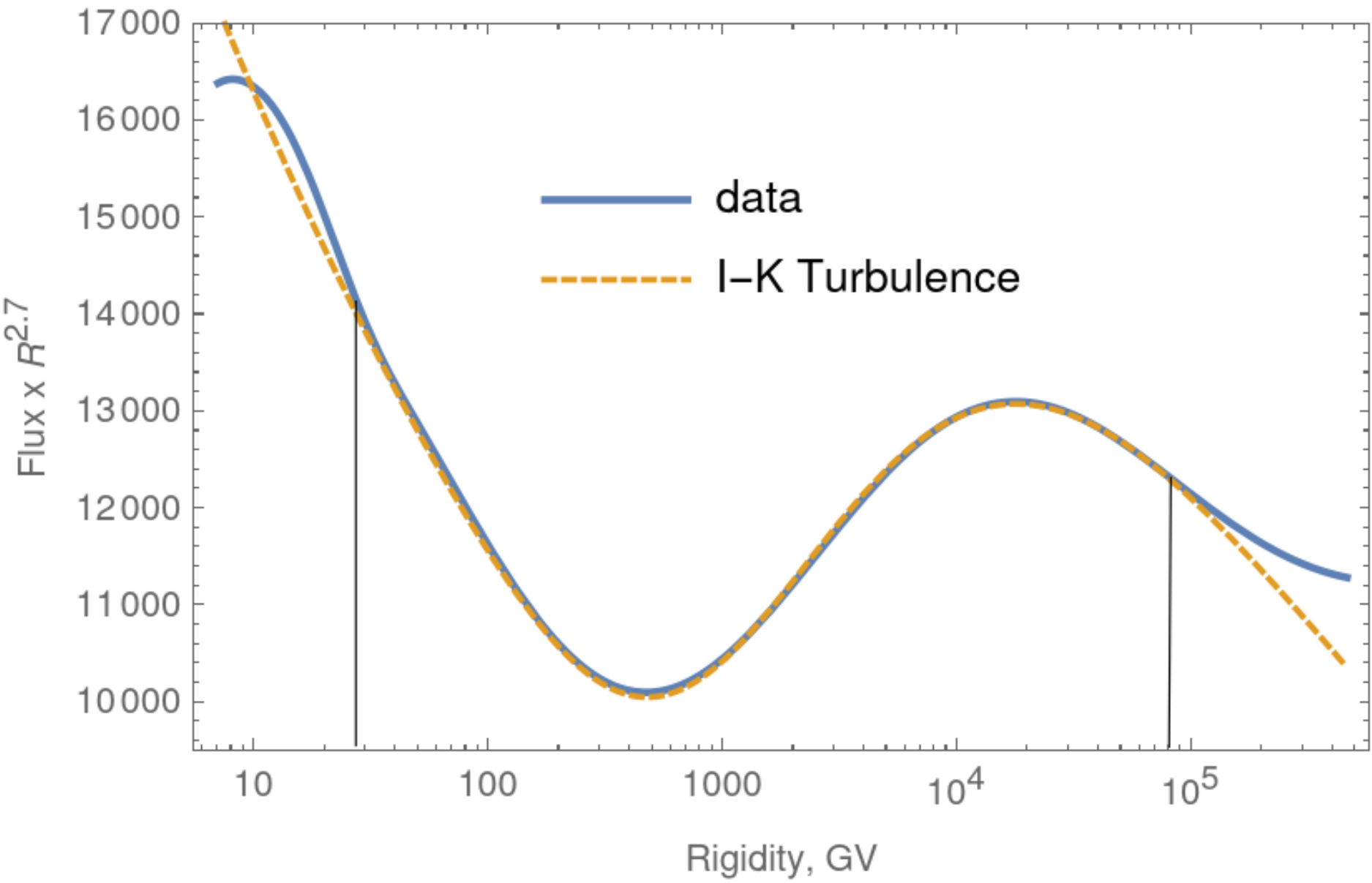} 
\caption{Fit of the diffusive shock acceleration result in Eq.~(\ref{eq:FitEq})
(dashed line) to the CR proton data compilation by \citet{Boschini_2020}
(solid line). The region between the vertical lines is physically
related to the shock acceleration in the Local Bubble and is used
to calculate the relative deviation between the two curves presented
in Eq.~(\ref{eq:STD}). Here we used $a=0.515$, $K=2.39$, $R_{0}=4434$ GV.
\label{fig:IKturb}}
\label{fig:Fit-of-the}
\end{figure}

Unlike the common practice to compare theoretical predictions with
discrete sets of data, we compare our prediction with a continuous
version of the observed proton spectrum. The latter has been provided
by \citet{Boschini_2020}, in a convenient form of analytic fit to
the actual multi-instrument data set. We will ignore individual instrumental
errors, as the cumulative data-set significantly diminishes them.
The analytic expression is rather cumbersome and contains ten fitting
parameters to accurately represent the local ISM CR spectrum; we show it plotted in Fig.~\ref{fig:IKturb}. We denote this spectrum by $f_{d}\left(R\right)$ and compare
with our two-parameter ($M$ and $R_{0}$) prediction in Eq.~(\ref{eq:FitEq}).
To this end, we rewrite  Eq.~(\ref{eq:Sec3simplRes}) in the following form: 
\begin{equation}
f\left(R\right)=2.3\times10^{4}R^{-0.15}\left[1+Ke^{-\left(R_{0}/R\right)^{a}}\right],\label{eq:FitEq}
\end{equation}
where $K(M)\equiv\sigma/\left(q-\sigma\right)$  (see below) and $R_{0}$ is defined
in Eq.~(\ref{eq:R0fromApp}). Here the CR density $f\left(R\right)$
is normalized to $fdR$, as opposed to $fp^{2}dp$ in Eq.~(\ref{eq:SolFinal}).
Also, following a common practice, we have multiplied $f$ by an additional
factor $R^{2.7}$ to make the spectral structure visible. The normalization
factor and the power-law index (0.15, stemming from the CR background
index $\sigma=4.85,$ introduced in the previous section) are fixed
by comparing the background CR spectrum $f_{\infty}$ in Eq.~(\ref{eq:SolFinal})
with the synthetic data, $f_{d}\left(R\right)$, from \citet{Boschini_2020}.
We match the two at lower rigidities, $R\ll R_{0},$ where the shock
does not perturb the spectrum near the heliosphere. 

Although we have calculated the CR diffusion coefficient, $\kappa\left(R\right)$\,$\propto$\,$R^{1/2}$
in Appendix \ref{sec:EstimateDistance}, we introduced an additional
parameter $a$ in Eq.~(\ref{eq:FitEq}) to investigate if other turbulence
models may also apply. The nominal value is thus $a=1/2$, but we
will explore the Kolmogorov
($a=1/3$) and the Bohm ($a=1$) models { as well.  The former is believed to be suitable
for the CR transport in the Galaxy, while the latter is expected to dominate near the shock front \citep{Bell78} and it  also occurs in the solar wind turbulence \citep{deWit2020}.}
We will scan the parameter
space by changing $a$ in the range $a=0.2-1.0$  and also determine
$K\equiv\sigma/\left(q-\sigma\right)$ and 
$R_{0}$ by minimizing the deviation of the model prediction from the
data. However, first we find the absolute minimum of this deviation which
happens to correspond to $a=0.515$ that is slightly larger than the nominal I-K value. 

Shown in Fig.~\ref{fig:Fit-of-the} is a nearly perfect match between
the data and our model, found for $K=2.39$ and $R_{0}=4434$. By
defining the relative deviation between the two curves as 
\begin{equation}
\Delta=\left.\int_{R_{1}}^{R_{2}}\left|f-f_{d}\right|dR\middle/\int_{R_{1}}^{R_{2}}f_{d}dR\right.,\label{eq:STD}
\end{equation}
where $f_{d}$ is the ``data,'' and $R_{1}=28$ GV and $R_{2}=8\times10^{4}$~GV
are chosen to contain the shock-related perturbation of the spectrum. 
The bump rigidity $R_{0}$
satisfies the relation $R_{1}\ll R_{0}\ll R_{2}$. Using the above
formula, we have computed that $\Delta\approx8.5\times10^{-4}$. We
discuss the incipient discrepancies between the two spectra outside
of this range in Sect.~\ref{subsec:Spectrum-Beyond-the}. An insignificant shift
in $a$ from $1/2$ might be produced by a minor contribution of the
Bohm scaling, $\propto p$, near the shock, not included into the path integral.
It can equally well be attributed to the observation errors, and should generally be ignored
in any realistic choice of the underlying turbulence model.  

\begin{figure}[tb!]
\includegraphics[scale=0.44]{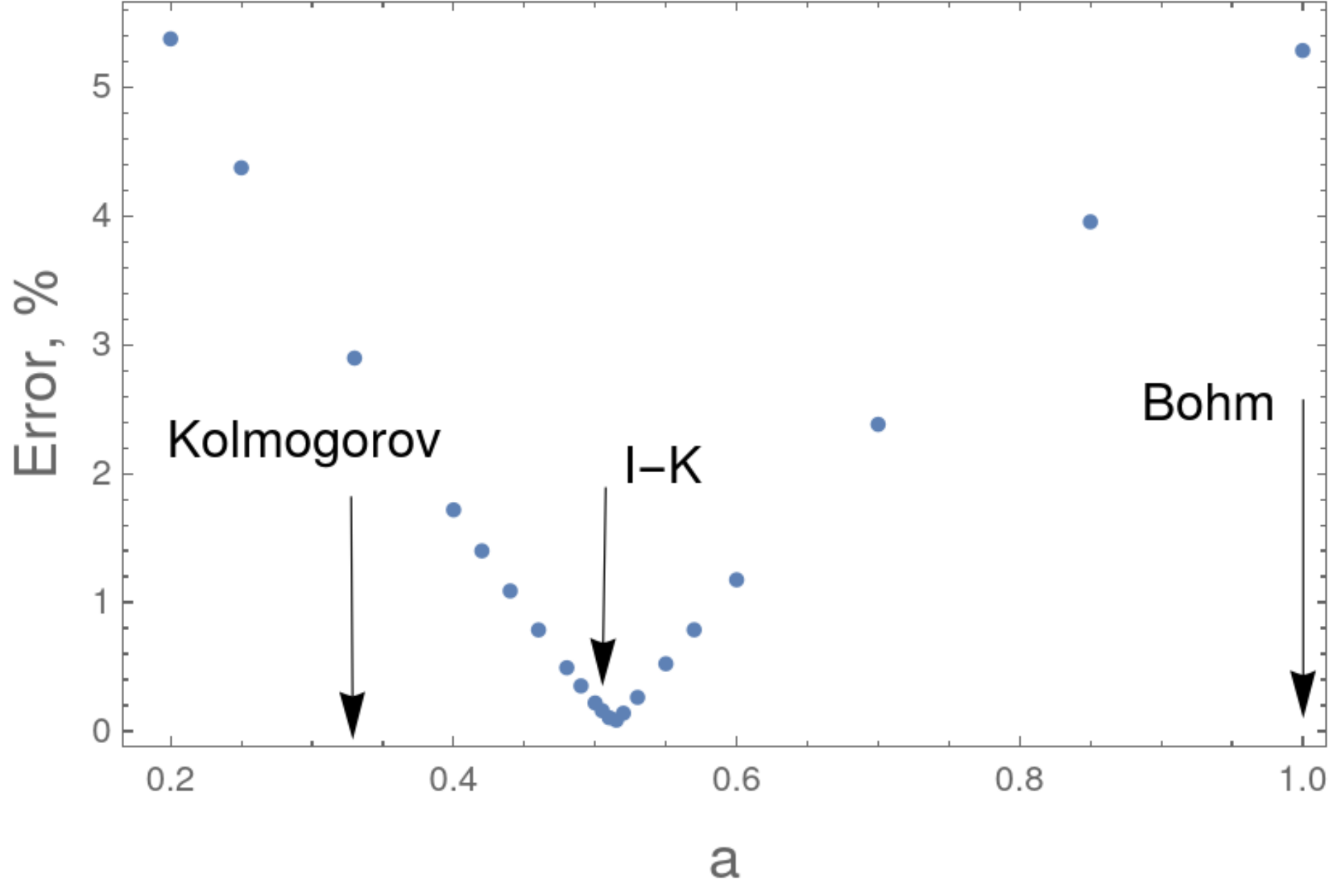}
\caption{The calculated deviation of the predicted spectrum from the the synthetic data vs.\ the index of the diffusion coefficient, Eq.~(\ref{eq:STD}). The indices corresponding to the Kolmogorov, I-K, and Bohm diffusion are shown by arrows.
\label{fig:Error-vs-turb}}
\end{figure}

\begin{figure}[tb!]
\includegraphics[scale=0.44]{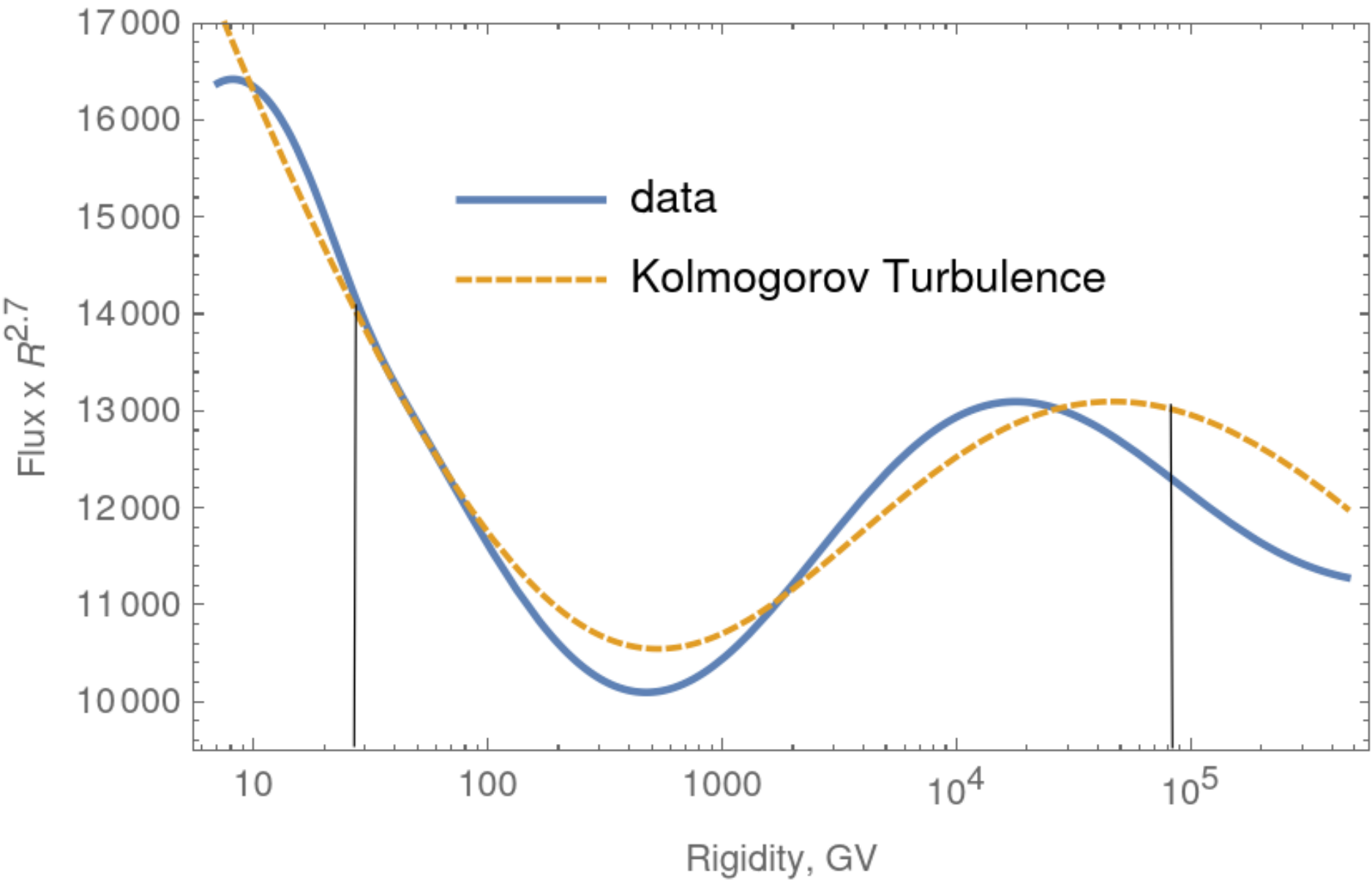}
\caption{Same as in Fig.~\ref{fig:IKturb}, but for Kolmogorov turbulence with
$a=1/3$. The best agreement is at $K=3.73$, $R_{0}=15665$ GV.\label{fig:Kolmogorov-Turb}}
\end{figure}

The sensitivity of our model predictions to the turbulence regime
is demonstrated in Fig.~\ref{fig:Error-vs-turb} that shows a scan of
the function $\Delta$ in the turbulence parameter $a$. Clearly,
the I-K model stands out reaching a sharp minimum of the error $\Delta$
very close to $a=1/2$. In fact, even at the exact I-K value $a=1/2$, the error
is about $0.2\%$, and, as we mentioned, even that minuscule deviation from the nominal
I-K index can still be consistent with the model. To demonstrate
that the I-K is by far the most likely turbulent regime in the flux
tube we plot in Fig.~\ref{fig:Kolmogorov-Turb} the model prediction
for the Kolmogorov turbulence with $a=1/3$. { This is the nearest to
the I-K spectral index (from the two shown in Fig.~\ref{fig:Error-vs-turb})}, but the disagreement with the data is significant.
In particular, it over-predicts the bump position by a factor 2-3
and is inconsistent with the newest data even if the error bars are
included. 

The CR bump amplitude $K\approx2.39$ implies the shock power-law
index $q\approx6.88$, which translates into the shock compression
$r\approx1.77$, and then into the Mach number $M\approx1.55$. Note
that these parameters formally relate to the scattering centers in
the shock precursor, which may move with the Alfv\'en speed $V_{A}$
backward with respect to the inflowing plasma. However, this is quite
uncertain as we do not know the shock angle, the turbulence level
and the plasma $\beta$. These factors can change the wave dispersive
properties. For example, if the turbulence level is sufficiently high,
the induced scattering of waves on thermal ions or other nonlinear
processes, such as interaction with magneto-acoustic waves, may symmetrize
the Alfv\'en waves in terms of the propagation direction, thus making
them effectively frozen into the flow. The latter was assumed in the
above calculations for simplicity.

Unlike $K$, which is fully determined by the shock compression, the
second model parameter, $R_{0}$, depends on the unknown distance
to the shock, its size, and speed. In order to constrain these quantities,
we first calculate the shock speed $u$ from the shock Mach number,
$M\approx1.55$, obtained from the best fit value of $K\approx2.4$.
Its upper bound may be placed by taking the temperature in the Local
Bubble $T\simeq10^{6}$ K \citep[e.g.,][]{Snowden2014}, which gives
$u_{1}\simeq150$ km s$^{-1}$. In denser and cooler regions the speed
is several times lower. Also, small density clumps in the ISM effectively reduce the shock speed. 

We can now substitute the parameters $K$ and
$R_{0}$, inferred from the data, in the expression in Eq.~(\ref{eq:zStarGeneral})
for the normalized distance $\xi$, while setting $R_{*}\approx1$: 
\begin{equation}
\frac{\zeta_{\rm obs}}{\sqrt{r_{\text{GV}}l_{\perp}}}\approx2\times10^{2}\ \frac{c}{u}\sqrt{\frac{\rho C_{s}V_{A}}{P_{\infty}}}.\label{eq:zStarFin}
\end{equation}
To get a crude estimate, we can substitute $r_{\text{GV}}=10^{12}$~cm, 
$c/u\sim3\times10^{3}$, and the normalized CR pressure $P_{\infty}/\rho C_{s}V_{A}\approx1$ into Eq.~(\ref{eq:zStarFin}). Then, we obtain $\zeta_{\rm obs}\sim$$10^{2}\sqrt{l_{\perp}}$, where the both lengths, $\zeta_{\rm obs}$ and $l_{\perp}$, are given in pc.

The value of $l_{\perp}$ is still uncertain. In contrast to $\zeta_{\rm obs}$,
however, we can constrain it by assuming that the shock is a bow shock
of a moving star, and its size is not much larger than our own heliosphere,
even if we leave a room for the reaccelerated CR to form a wake. Inserting
then $l_{\perp}\sim10^{-2}$ pc into the last estimate, we obtain
an upper bound on $\zeta_{\rm obs}\sim10$ pc. A lower bound on $l_{\perp}$
can be obtained if the Sun is just entering the flux tube, as shown
in Fig.~\ref{fig:Bow-shock}. { In this case, $l_\perp$ is the scale of  the edge of the flux tube rather than its full radius.} We will discuss the implications of
this scenario for the time-dependent CR spectrum, observed locally.
In any event, $l_{\perp}$ cannot be smaller than the gyroradius of
those reaccelerated CRs that make the main contribution to their pressure.
Formally this quantity is as small as $r_{\text{GV}}$, but particles
with such small gyroradii do not diffuse far from the shock and
a more plausible estimate is the gyroradius of particles with the
rigidity corresponding to the spectral upturn, that is $\sim$1 TV.
This estimate yields $l_{\perp}\sim10^{15}$ cm and, therefore, a
few pc for $\zeta_{\rm obs}$. Based on our estimates of $l_{\perp}$
in Eq.~(\ref{l-perp}), a moderate inverse cascade of Alfv\'en waves
might be required to allow for such a short $l_{\perp}$. We thus
conclude that the distance to the reaccelerating bow shock is likely
in the range 3--10 pc.


\subsection{Spectrum Beyond the Excess Region\label{subsec:Spectrum-Beyond-the}}

Our model spans the rigidity range $30$$<$$R$$<$$10^{5}$ GV, Fig.~\ref{fig:Fit-of-the}.
Beyond each end of this interval, the model under-predicts the data.
Both below $R=30$ GV and above $R=10^{5}$
GV  the spectrum is likely to be unrelated to the shock in question. Nonetheless, 
we briefly discuss how our model can be extended to these intervals. 

A moderate enhancement of the data above the model prediction
at the lower rigidities can be due to freshly injected primaries.
We have neglected their contribution to a broad bump in the spectrum
in the 10 TV range, because the shock is weak, thus producing much
steeper spectrum (index $q\approx6.9$) than the background spectrum
(index $\approx4.9$). For precisely that reason, however, the contribution
of these particles at lower energies can be increasingly visible.
They must have much higher density near the shock than the reaccelerated
particles. Therefore, by penetrating the magnetic flux tube, even
in a small number, these particles may partially survive the upstream
screening effect in Eq.~(\ref{eq:dsSol}) and reach the Sun.  
Apart from the injection, ionization energy losses may also distort the spectrum 
of aged CRs at lower rigidities. 

The deviation of the fit from the data at $R>10^{5}$ GV can be explained
by a possible slight concavity of the CR background spectrum $f_{\infty}$
that we have substituted in the shock solution as a straight power
law. In reality, however, while the spectrum between $30$$<$$R$$<$$10^{5}$
GV is dominated by the reacceleration at the bow shock, the underlying
background spectrum may flatten between these limits. At higher rigidities
the shock reacceleration stops working, because the acceleration time,
$\kappa/u^{2}$, approaches the shock convection time, $l_\perp/u$,
available for the reacceleration. At higher rigidities the background
spectrum reemerges with a flatter slope than it has below the first
break. The flattening is consistent with the CR spectrum behavior
at sub-knee rigidities.


\subsection{Momentum Diffusion\label{subsec:Momentum-Diffusion}}

Propagation of reaccelerated CRs to the Sun through an enhanced turbulence
comes with their enhanced diffusion in momentum. The required bi-directional
Alfv\'en wave spectrum is justified by the genesis of these waves in
the acoustic instability discussed in Appendix~\ref{sec:WaveEnTransf}.
Hence, relatively sharp kinks in the observed spectrum may constrain
the distance to their source, in addition to the estimate in Eq.~(\ref{eq:zStarFin}).
It is convenient to employ the following inverse relation between
the particle diffusivity in momentum, $D_{pp}$, and that along the
field, $\kappa_{\parallel}$, with a numerical factor justified by
\citet{DruryStrong2017}: 
\[
D_{pp}\simeq0.1\frac{V_{A}^{2}p^{2}}{\kappa_{\parallel}}.
\]
By noting that the spreading in momentum for the particles traversing
the distance $\zeta_{\rm obs}$ from the source is $\Delta p^{2}\sim D_{pp}\zeta_{\rm obs}^{2}/\kappa_{\parallel}$
and combining the above relation with Eq.~(\ref{eq:zStarFin})  and (\ref{eq:kap-int-fin}) with
$P_{\text{CR}}/P_{\infty}\sim1$ at large distances from the shock,
we find the relative rigidity spreading: 
\[
\frac{\Delta R}{R}\sim10^{2}\frac{V_{A}}{u}\sqrt{\frac{1\,\text{GV}}{R}}.
\]
This broadening of the peak at $R\sim10^{4}$ GV in Fig.~\ref{fig:Fit-of-the} is not large,
especially if the plasma $\beta$ is high, so $V_{A}\ll u$. It may
become closer to unity at the spectral minimum around $R\lesssim10^3$
GV. Nevertheless, the minimum breadth is not in tension with the
$\Delta R\sim R$ result. In addition, the above estimate gives an
upper bound on the momentum diffusion as it does not include the convective
screening of low-energy particles and a possibly asymmetric wave propagation.  

The above estimate of $\Delta R$
merely assumes that particles reaccelerated at the shock diffuse in
$\zeta$-direction, while also diffusing in momentum. To account for the
momentum diffusion more accurately, it should be included in the particle
transport in Eq.~(\ref{eq:cd}). If a sharper first break reported
by CALET prevails over other instruments, this modification of the
model may become desirable. Note, however, that other instruments,
particularly DAMPE that also claims high precision at the first break
rigidity, currently indicate a smoother than CALET transition (see,
e.g., \citealt{2020APh...12002441L} for the recent multi-instrument
data compilation).


\begin{deluxetable*}{rlccc}[t!]
	\tablewidth{0mm}
	\tablecaption{The model and problem parameters\label{tab:Model-and-problem}}
	\tablehead{
\colhead{}& \colhead{}& \colhead{Theoretical}& \colhead{Best fit}& \colhead{Measured}\\
\colhead{Notation}& \colhead{Description}& \colhead{prediction}& \colhead{value}& \colhead{value} 
	}
	\startdata
$s$ & $k^{-s}$ turbulence index & $3/2$ & 1.49 & \nodata \\
$\zeta_{\rm obs}$ & Distance to the bow shock & 3--10 pc &  \nodata &  \nodata \\
$\displaystyle \xi\equiv\frac{\zeta_{\rm obs}}{\sqrt{r_{\text{GV}}l_{\perp}}}$ & Normalized distance & Eq. (\ref{eq:zStarFin}) & \nodata & \nodata \medskip\\
$\displaystyle \eta=\frac{KP_{\infty}}{\rho C_{s}V_{A}}\frac{\sigma-4}{R_{*}^{4-\sigma}}$ & Normalized background CR pressure & \nodata &  \nodata & \nodata \medskip\\
$\displaystyle R_{0}/1\,\text{GV}=\left[ 3\left(\sigma-3\right)\xi\frac{u}{c}\sqrt{2\eta\Gamma\left(2\sigma-8\right)}\right] ^{2/\left(\sigma-3\right)}$ & CR Bump rigidity, GV &   \nodata & 4434 & \nodata \\
$\sigma$ & Index of background CR spectrum & 4.3--4.6\tablenotemark{a} &  \nodata & 4.85\tablenotemark{b}\\
$\displaystyle K\equiv\sigma/\left(q-\sigma\right)$ &  CR Bump magnitude&  \nodata & 2.4 & \nodata\\
	\enddata
	\tablenotetext{a}{Index, generally expected for CRs accelerated in SNRs and propagated in Kolmogorov turbulence with $a\approx$ 0.3 (see text).}
	\tablenotetext{b}{Index, extracted from multi-instrument data around $\sim$100 GV, presumably propagated through the flux tube in I-K turbulence.}
	\end{deluxetable*}

\subsection{Possible Time Dependence Due to the Spatial Gradient\label{subsec:Possible-Time-Dependence}}

For our estimate of the distance to the bow shock in Eq.~(\ref{eq:zStarFin})
to be plausible the CR pressure profile must be sufficiently steep
in the cross-field direction. Our analysis in Appendix~\ref{sec:EstimateDistance}
suggested that the lateral gradient scale $l_{\perp}$ must be in
the range $10^{15}$$-$$10^{16}$ cm. When the Sun crosses the flux tube
of that scale then the CR variation can be detectable. If the relative
velocity between the bow shock and the Sun is $\sim$100 km s$^{-1}$,
as the fit suggests, then the relevant crossing time is $\tau$$\sim$3--30
yr, which merits a closer look at the historical data.

Early reports on the spectral hardening date back to the ATIC-2 paper
\citep[data taken in 2002--2003,][]{ATIC06}, first two flights by CREAM
\citep[data taken in 2004-2005,][]{2010ApJ...714L..89A}, and first
two years of PAMELA data \citep[data taken in 2006--2008,][]{2011Sci...332...69A}.
The position of the first break (hardening) was significantly lower,
$\sim$200--240 GV vs.\ its current value of 450 GV from AMS-02
proton data \citep{2015PhRvL.114q1103A}, although the uncertainties are large. The situation with the second
break (softening) is much less clear. According to the current fit,
it is at 17 TV, which is, however, based on a multi-instrument compilation
with even larger uncertainties. There is no PAMELA data in this range, while
the ATIC-2 data (2002--2003) are more consistent with a $\sim$10
TV break, but the statistical errors are quite large.

Let us now turn to the solution parameters $K$ and $R_{0}$ in Eq.~(\ref{eq:FitEq})
(with $a=1/2$) that determine the two breaks whose rigidities we
denote by $R_{h}$ and $R_{s}$, where `h' and `s' stand for hardening
and softening, respectively. The $K$-value must change with time,
as it is proportional to the CR excess. Note that the one-dimensional
solution in Eq.~(\ref{eq:FitEq}) does not include the cross-field
variation of the CR flux, so $K$ formally depends only on the shock
index $q$ and the background CR index $\sigma$ that are constant.
The one-dimensional approximation applies only well inside the flux
tube, where the CR flux reaches its maximum. The second parameter,
$R_{0}$, clearly changes across the flux tube as it directly relates
to the turbulence enhancement. { By looking for the extrema in}  
Eq.~(\ref{eq:FitEq}) we can derive the following simple equation
for $R_{h}$ and $R_{s}$: 
\[
\frac{\sqrt{R_{0}/R}}{0.3}=1+e^{\sqrt{R_{0}/R}-\ln K}.
\]
We see that the both roots of this equation scale linearly with $R_{0}$,
but only logarithmically depend on $K$. The current value of $K=2.4$
is not big enough to neglect the variations in $\ln K$ completely.
Nevertheless, we justify this step by noting that had $K$ changed
strongly over the time, its value would be too close to the background
CR intensity back in 2002--2008 and the excess would have not been
detected.

The parameter $R_{0}$ varies across the tube with the turbulence
level, Eq.~(\ref{eq:SqrtR0}), which we expressed through the CR
pressure, $P_{\text{CR}}.$ After fixing $K=2.4$, from the equation
above we find: 
\[
R_{h}\approx0.1R_{0}\propto P_{\text{CR}},
\]
while $R_{s}\approx4R_{0}$. From here we conclude that, indeed, $R_{h}$
may have increased by a factor of two (from $\sim$230 GV measured
by ATIC and PAMELA) to the current value of 455 GV over 10--15 yrs,
provided that the Sun penetrated deeper into the flux tube and the
turbulence level increased by the same factor. The second break at
$R=R_{s}$ must have increased proportionally to $R_{h}$, unless
the turbulence index $a$ also changed over the time. This change
cannot be ruled out, since well outside of the tube a more likely
value, based on the B/C ratio, derives from the Kolmogorov turbulence
with $a\approx1/3$, rather than the I-K index that we have arrived at. On the
other hand, the argument about the detectability applied to the parameter
$K$ seems to be applicable to $a$ as well. 

{ We note that the drift in the opposite direction, i.e.\ the decrease of the break rigidity with time is also possible, and depends on the configuration of the magnetic tube and orientation of the shock relative to the Sun.}   
We shall return to the possible variability of the CR bump in Sect.~\ref{sec:LocalB-field}.

\section{Passing Stars\label{sec:Passing-Stars}}


{ However accurate the spectral fit may be, our model's success depends on identifying the object most likely responsible for the spectral bump.}
It did
not take long to find two passing stars, which perfectly match the
derived range of distances and velocities: the binary Scholz's Star
at 6.8 pc distance with the radial velocity 82.4 km s$^{-1}$, and
a triple system Epsilon Indi at 3.6 pc that has the radial velocity
--40.4 km s$^{-1}$ and a considerable proper motion. These are just
two examples indicating that such passings within a few pc distance
are not unusual. { We start with a brief discussion of their physical properties and turn to the critical question of their magnetic connectivity with the heliosphere afterward.}

The recently discovered Scholz's Star \citep{2014A&A...561A.113S}
is a system of red M9.5 and brown T5.5 dwarfs \citep{2015ApJ...800L..17M}
with masses of 0.095$M_{\sun}$ and 0.063$M_{\sun}$, correspondingly.
The system has a moderately eccentric orbit $e$ = 0.240 with a semi-major
axis of $\sim$2 au, and a period of 8 yr \citep{2019AJ....158..174D}
that amended the previous estimate of \citet{Burgasser_2015}. The
Scholz's Star has passed within the close proximity of the Sun, 0.25
pc, about 70--80 kyr ago, and is moving away from us. Therefore,
the Sun is located downstream in its system.

The peculiar proper motion of the Epsilon Indi system was noticed
already about two centuries ago \citep{1847MNRAS...8...16D}. The
main star A in the Epsilon Indi system is a K4.5V star\footnote{http://www.astro.gsu.edu/RECONS/TOP100.posted.htm
\label{f1}}
with mass $\sim$$0.77M_{\sun}$. The recently discovered two smaller
stars B and C belong to the brown dwarf family with classes T1.5 and
T6 and masses of 0.072$M_{\sun}$ and 0.067$M_{\sun}$, correspondingly
\citep{2018ApJ...865...28D}. The main star and the binary brown dwarf
system are separated by $\sim$1500 au. The Epsilon Indi star is moving
toward us with the Sun located upstream.

Both star systems can generate multiple colliding bow shocks with
rich opportunities for particle acceleration \citep[e.g.,][]{Bykov2019}: (i) Each companion star
produces its individual bow shock, (ii) The shock velocities are modulated
with the orbital motion, and (iii) the tight separation in each system
ensures that the shocks are colliding and amplifying each other. In
the case of the Scholtz's star, a combination of the directed and orbital 
motion (with period $\tau$) produces downstream magnetic and kinetic perturbations with
the wavelengths $\lambda_{B}\sim V_{A}\tau\sim10^{14}$--$10^{15}$
cm and $\sim$2 au $\approx3\times10^{13}$ cm, respectively. They
are relevant for the scattering and confinement of CRs with up to
TV rigidity. Moreover, the orbital motion injects magnetic and kinetic
helicity into the turbulence past the star systems, which is necessary
for an inverse MHD cascade \citep{Pouquet76,pouquet2018helicity}.
Even though we have considered the propagation of reaccelerated CRs
upstream of a shock, the turbulence left behind by the Scholtz's star
may have important implications for our model. It can provide an enhanced
scattering and confinement of particles at the rigidities beyond the
TV range.

The model we discussed in the paper can be applicable to the
Epsilon Indi system. Epsilon Indi must have a bow shock that is very
much different from the ordinary one formed by a single star. A 1500
AU separation, 11-yr orbital period, and colliding stellar winds inject
much more turbulent power than the Scholtz's star does. All this happens
downstream of the bow shock and has no direct impact on the reaccelerated
CRs upstream. However, the reacceleration process may be enhanced,
and the bow shock is likely to be rippled due to the stellar dynamics
behind it. An exciting, but speculative, aspect of this dynamics is
a possible 11-yr variations in the CR bump parameters. Here the coincidence
with the solar cycle period is accidental, as our Sun maintains this
period for hundreds of millions of years \citep{Luthardt2017FossilFR}.


{ 
Meanwhile, even slower moving stars, such as the Epsilon Eridani\textsuperscript{\ref{f1}} ($\sim$20 km s$^{-1}$), may be viable candidates. Epsilon Eridani star at a distance of 3.2 pc is the closest of the three and fits well within the derived range of distances. It is a K2 dwarf with a mass of 0.82 $M_\sun$, a radius of 0.74 $R_\sun$, and an effective temperature of $\approx$5000 K \citep{2012ApJ...744..138B}. Nevertheless, it has a high mass loss rate of 30$\dot{M}_\sun$ and a full width of its astrosphere reaches about 8000 au with the corresponding angular size $\sim$42\arcmin{} (larger than the Moon!) as viewed from the Earth \citep{Wood2002}.

Higher speed through the ISM implies faster particle acceleration and a longer flux tube with reaccelerated CRs. These CRs may then reach the Sun before they dissolve in the ISM. Alternatively, a slower shock in a cooler phase of the ISM (e.g., warm ionized medium, or WIM with $T$$\sim$$10^{4}$~K) may produce a similar spectrum, albeit at a slower pace and suppressed CR self-confinement due to ion-neutral collisions. A large shock, similar to that of the Epsilon Eridani star, may compensate for these drawbacks. These considerations widen the choice of suitable objects in the solar vicinity.
}

\section{Anisotropy from Magnetically Connected CR Source\label{sec:Anisotropy-and-Magnetic} }

Perhaps the most compelling evidence for the CR bump association with
a nearby star might come from the star fortuitously found in the same
flux tube with the heliosphere. There are only a handful of stars
with suitable characteristics within 10 pc around the Sun. Therefore,
the chances are very much against its existence. Nevertheless, Epsilon
Eridani 
appears to be the \emph{one}. 
It is only a small turn of $\approx$$6.7^{\circ}$ away from the local magnetic field
direction, and as already mentioned, it is located at 3.2 pc away from the Sun and possesses a huge astrosphere.
 
In Sect.~\ref{sec:LocalB-field}, we discuss the topology of the local magnetic field and its perturbation mechanisms, whereby may increase the probability for other stars to be
connected to the Sun. However, the Epsilon Eridani stands out in that all it
takes for it to be magnetically connected to the Sun is just a straight
field line. If we assume this simple field configuration, two implications
follow: the obvious one is that the star is indeed exceptional. The
chance to find a suitable candidate star no further than $6.7^{\circ}$ off of the local
field direction is only $\approx$0.3\%, assuming that its position
on the sky is random. To evaluate the total probability of this fortunate
outcome, we sample 3-4 stars that have detectable astrospheres (bow
shocks) of a size $>$$1000$ au (to exceed a 10-TeV CR Larmor radius) within
5 pc of the Sun listed by \citet{Wood2002}. The probability
of at least one of them to be found in such a narrow solid angle
is not larger than one percent. This observation makes the bump no
longer surprising. Indeed, the \textquotedblleft standard\textquotedblright{}
CR acceleration and propagation models that predict featureless spectra
do not typically include such sources.

The second implication is that the source proximity is likely to produce
imprints in the CR arrival directions other than a generally expected
large-scale anisotropy. The latter can result from uneven distribution
of sources in the Galaxy, for example, unrelated to the rigidity bump
phenomenon studied in this paper. Historically, the discovery of a
small-scale CR anisotropy in 1-10 TeV range, or ``hot
spots,'' predated the rigidity bump discovery. The
Milagro observatory \citep{Milagro08PRL} reported a surprisingly
sharp, $\sim$10$^{\circ}$ anisotropy at a $10^{-4}$ level of the
dominant isotropic CR component, atop the large-scale anisotropy at
the $10^{-3}$ level. The observatory was decommissioned in 2008. Among its results, was also a hardening of the CR proton spectrum associated with the sharp anisotropy spots. We retrospectively link these data with
other early indications of the 10-TeV rigidity bump, discussed in
Sect.~\ref{subsec:Possible-Time-Dependence}. 

The anisotropy spots  have been observed later by several other instruments. An
update of post-Milagro results can be found, e.g.,  in \citet{Ahlers2017}.
They are in reasonable agreement with the original Milagro findings.
However, the difference in the width of the small-scale structures
and their relative position on the large-scale angular profile are
noticeable. A number of factors may have caused the difference, such
as different types of instruments, data processing, or even a putative
time dependence of the CR spectrum in the TeV range, discussed in
Sect.~\ref{subsec:Possible-Time-Dependence}. 

Unlike in our study of CR reacceleration in Sect.~\ref{sec:problem},
we need to include the pitch-angle dependence of reaccelerated CRs, which
makes the problem more difficult. Therefore, we split it in two parts: we consider the acceleration
part of the problem as solved, and study the CR propagation along the
source's flux tube. We take a step back from the rigidity
spectrum obtained in Sect.~\ref{sec:problem} and assume
a generic source with given momentum and pitch-angle distribution.
For a high mass-loss but slowly moving star, its wind termination
shock might reaccelerate CRs more efficiently than its bow shock.
Having the Epsilon Eridani in mind, we further simplify the problem
using its relative motion  to the local gas at $76^{\circ}$ to
the line of sight \citep{Wood2002}. Since the latter is close to
the local field direction, we may neglect the source motion along the
field. With this additional symmetry, the following Fokker-Planck
equation is suitable for the CR propagation along the field:
\begin{eqnarray}
\frac{\partial f}{\partial t}+c\mu\frac{\partial f}{\partial\zeta}+\nu_{\perp}^{\prime}f&=& \label{eq:FPinit}\\ 
\frac{\partial}{\partial\mu}\left(1-\mu^{2}\right)&D&(p,\mu)\frac{\partial f}{\partial\mu}+Q(p,\mu)\delta\left(\zeta\right).
\nonumber
\end{eqnarray}
Here $\mu$ is the pitch angle cosine of a particle with momentum
$p$. The spatial coordinate $\zeta$ is directed along the field,
and the source is at the origin. The other two coordinates are removed
by averaging the distribution function across the flux tube. The particle
flux through its boundary (arising from an integral endpoint in averaging)
is simplified to the form of $\nu_{\perp}^{\prime}f$ on the l.h.s.\
\citep[e.g., ][]{MalkovAharonian2019}. The CR scattering frequency $D$ can be related to
the Alfven fluctuation energy density $E_{k}^{A}$, determined in
Eq.~(\ref{eq:AcousAlfSpectraFinal}), using the standard quasi-linear
formula \citep{KulsrNeutr69}: 
\begin{equation}
D=\left.\frac{\pi^{2}c}{\left|\mu\right|r_{g}^{2}B_{0}^{2}}\rho_{0}E_{k}^{A}\right|_{k=1/r_{g}\left|\mu\right|}\equiv4\nu_{\text{IK}}\left(p\right)\left|\mu\right|^{1/2},
\label{eq:PAdiff}
\end{equation}
where we have specified the scattering frequency using the I-K
spectrum obtained in Appendix~\ref{sec:WaveEnTransf}, Eq.~(\ref{eq:AcousAlfSpectraFinal}):
\[
\nu_{\text{IK}}=\frac{\pi}{16}\sqrt{\frac{\gamma_{D}}{V_{A}r_{g}}}.
\]

We represent the source term in Eq.~(\ref{eq:FPinit}) as follows:
$Q=\nu_{\text{IK}}Q_{0}\left(p\right)Q_{1}\left(\mu\right)$, where
$Q_{1}\left(\mu\right)=Q_{1}\left(-\mu\right)\approx const$, using
the problem reflection symmetry, $\zeta\to-\zeta$. Most of the above
restrictions can be relaxed at the expense of more algebra in treating
the transport problem below, but largely without new insight. We,
therefore, continue to transform Eq.~(\ref{eq:FPinit}) by introducing
dimensionless time and coordinate, and absorbing the source momentum
dependence $Q_{0}\left(p\right)$ into $f$ as follows:
\begin{equation}
\nu_{\text{IK}}t\to t,\quad\frac{\nu_{\text{IK}}}{c}\zeta\equiv\frac{\zeta}{\lambda_{\text{CR}}}\to\zeta,\quad\frac{f}{Q_{0}}\to f.\label{eq:NonDimVar}
\end{equation}
Next, we remove the $\left|\mu\right|^{1/2}$-singularity from $D$
by changing the variables:
\[
\chi=\left|\mu\right|^{1/2}\text{sgn}(\mu).
\]
However, we will demonstrate that the $\left|\mu\right|^{1/2}$-suppression
of particle scattering at the $90^{\circ}$ pitch angle has a profound
observable effect on the solution of Eq.~(\ref{eq:FPinit}). Assuming
a steady state we transform Eq.~(\ref{eq:FPinit}) to the following
equation:
\begin{equation}
\frac{\partial}{\partial\chi}\left(1-\chi^{4}\right)\frac{\partial f}{\partial\chi}-\chi^{3}\frac{\partial f}{\partial\zeta}-\nu_{\perp}f=Q_{1}\left(\chi\right)\delta\left(\zeta\right)\label{eq:FPndim}
\end{equation}
where $\nu_{\perp}\equiv\left|\chi\right|\nu_{\perp}^{\prime}$/$\nu_{\text{IK}}$.
It is also reasonable to assume that the lateral particle escape from
the tube is symmetric, $\nu_{\perp}\left(-\chi\right)=\nu_{\perp}\left(\chi\right).$
The solution of the last equation has then the following cross-symmetry,
$f\left(-\zeta,\chi\right)=f\left(\zeta,-\chi\right)$. By making
use of it, the solution can be written in the following compact form
for $\zeta\neq0$:
\begin{equation}
f\left(\zeta,\chi\right)=\sum_{n=0}^{\infty}C_{n}f_{n}\left[\chi\cdot\text{sgn}(\zeta)\right]e^{-\lambda_{n}\left|\zeta\right|}.
\label{eq:fexpans}
\end{equation}
Further details of this solution can be found in Appendix~\ref{sec:AnisApp}.   Below,  
we summarize the salient features of it. 

The lowest eigenvalue $\lambda_{0}\to0$,
for $\nu_{\perp}\to0$, thus yielding a fully isotropic solution at
$\left|\zeta\right|=\infty$. Furthermore, for any $\nu_{\perp}>0$,
$f\to0$ for $\left|\zeta\right|\to\infty$. The eigenvalues $\lambda_{n}$
with $n\ge1$ are almost independent of $\nu_{\perp}\ll1$. We assume
this inequality to be strong. Were it not so, CRs from the star would
not reach the Sun because of a strong lateral escape from the flux
tube. While $\lambda_{0}\ll1$ (for vanishingly small $\nu_{\perp}$),
the next eigenvalues $\lambda_{1}\simeq45$, $\lambda_{2}\simeq130,\dots$
continue to grow rapidly with $n$. The solution of Eq.~(\ref{eq:FPndim})
is thus determined by the first $n$ terms of the series in Eq.~(\ref{eq:fexpans}),
where $\lambda_{n}\sim1/\left|\zeta\right|$. 

According to Sturm's oscillation theorem, the last condition limits the CR anisotropy scale
from below, since the number of zeros of $f_{n}$ is $n$. It is also
worth noting that the zeros are on the positive side of $\chi$-axis
for positive eigenvalues $\lambda_{n}$ for $n\ge1$ and on its negative
side for negative eigenvalues, $-\lambda_{n}$. Because of this difference,
the full solution in Eq.~(\ref{eq:fexpans}) possesses the following
observable property. It develops the small-scale anisotropy only for
the particles moving away from the source. The returning particles
exhibit only a large scale anisotropy. If they are a few mean free
paths away from the source, then they have already scattered by a
large angle, so their angular distribution is smooth. An exception
to this rule are particles moving close to the $90^{\circ}$ pitch
angle, which merits a separate discussion later in this section and in Appendix~\ref{sec:AnisApp}. 

\begin{figure}[tb!]
\includegraphics[scale=0.48]{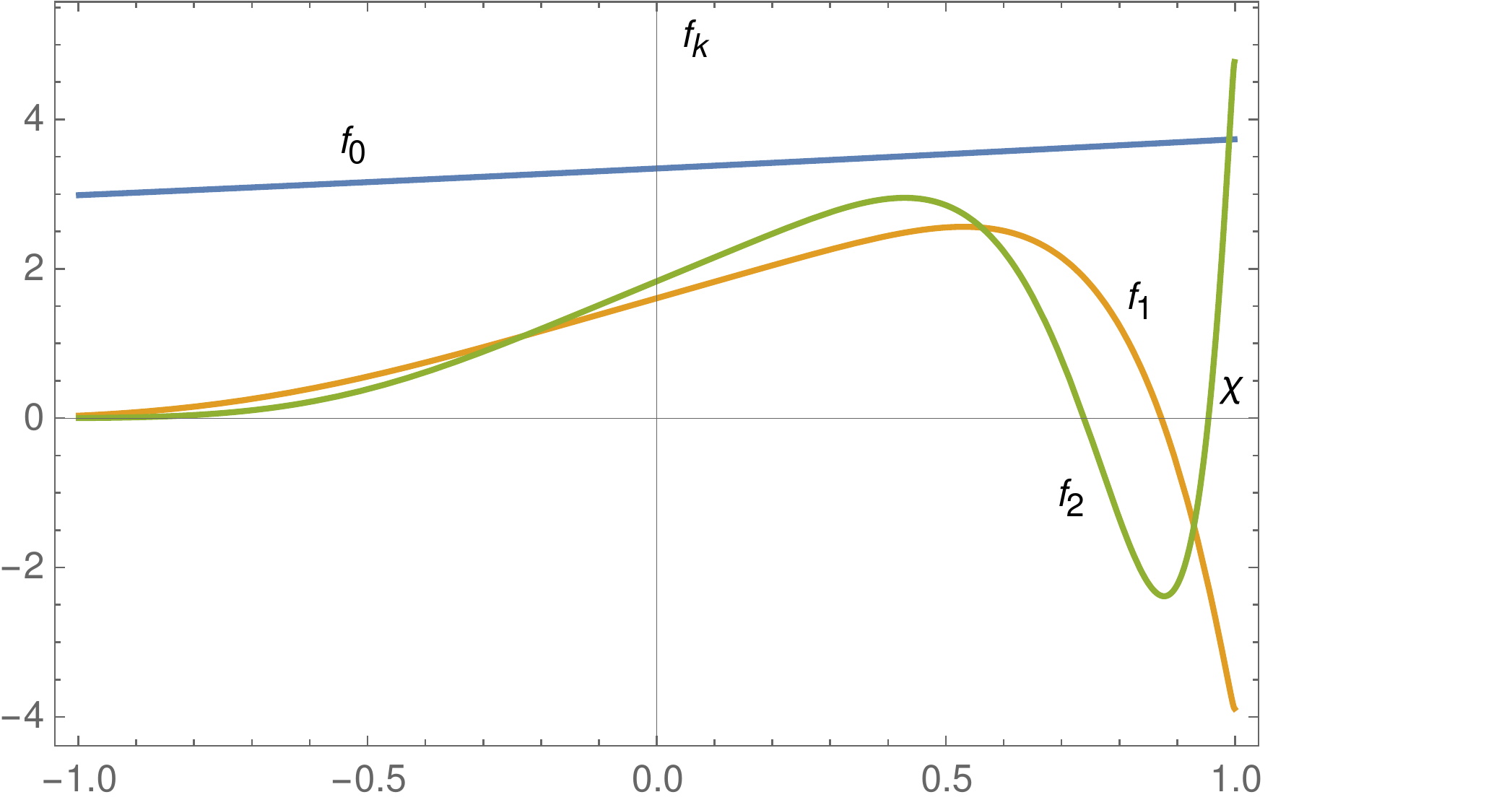}
\caption{First three eigenfunctions $f_{n}\left(\chi\right)$, $n=0,1,2$,
of the spectral problem in Eq.~(\ref{eq:SpProblem}) for $\nu_{\perp}=0.01$.
They correspond to the three positive eigenvalues $\lambda_{0}$$\approx2\sqrt{5\nu_{\perp}}\approx0.45$,
$\lambda_{1}\approx45.0$, $\lambda_{2}\approx132$. 
\label{fig:The-first-three}}
\end{figure}

We show the first three eigenfunctions in Eq.~(\ref{eq:fexpans})
with $n=0,1,2$ in Fig.~\ref{fig:The-first-three}. The ``ground-state''
$f_{0}$ carries the dominant isotropic part of the solution in the
series in Eq.~(\ref{eq:fexpans}) and the bulk of its large scale anisotropy.
It can be decomposed approximately as follows $f_{0}\left(\chi\right)\approx A+A^{\prime}\chi=A+A^{\prime}\left|\mu\right|^{1/2}\text{sgn}(\mu)$.
The constants $A$ and $A^{\prime}$ may be related using Frobenius
expansions at the endpoints $\chi=\pm1$. Neglecting small terms $\sim$$\nu_{\perp}\ll1$,
to the first order in $\lambda_{0}<1$, the expansions yield $A^{\prime}/A\approx\lambda_{0}/4$
at the both endpoints. To this order, the constant $A$ comes from
the normalization condition, $\parallel$$f_{n}$$\parallel^{2}\equiv\left(f_{n},f_{n}\right)=1$,
that results in $A\approx\sqrt{5/\lambda_{0}}$ and then $f_{0}\approx\sqrt{5/\lambda_{0}}+\sqrt{5\lambda_{0}}\chi/4$.
(See Appendix \ref{sec:AnisApp}, as needed, for notation.)

\begin{figure}[tb!]
\includegraphics[scale=0.45]{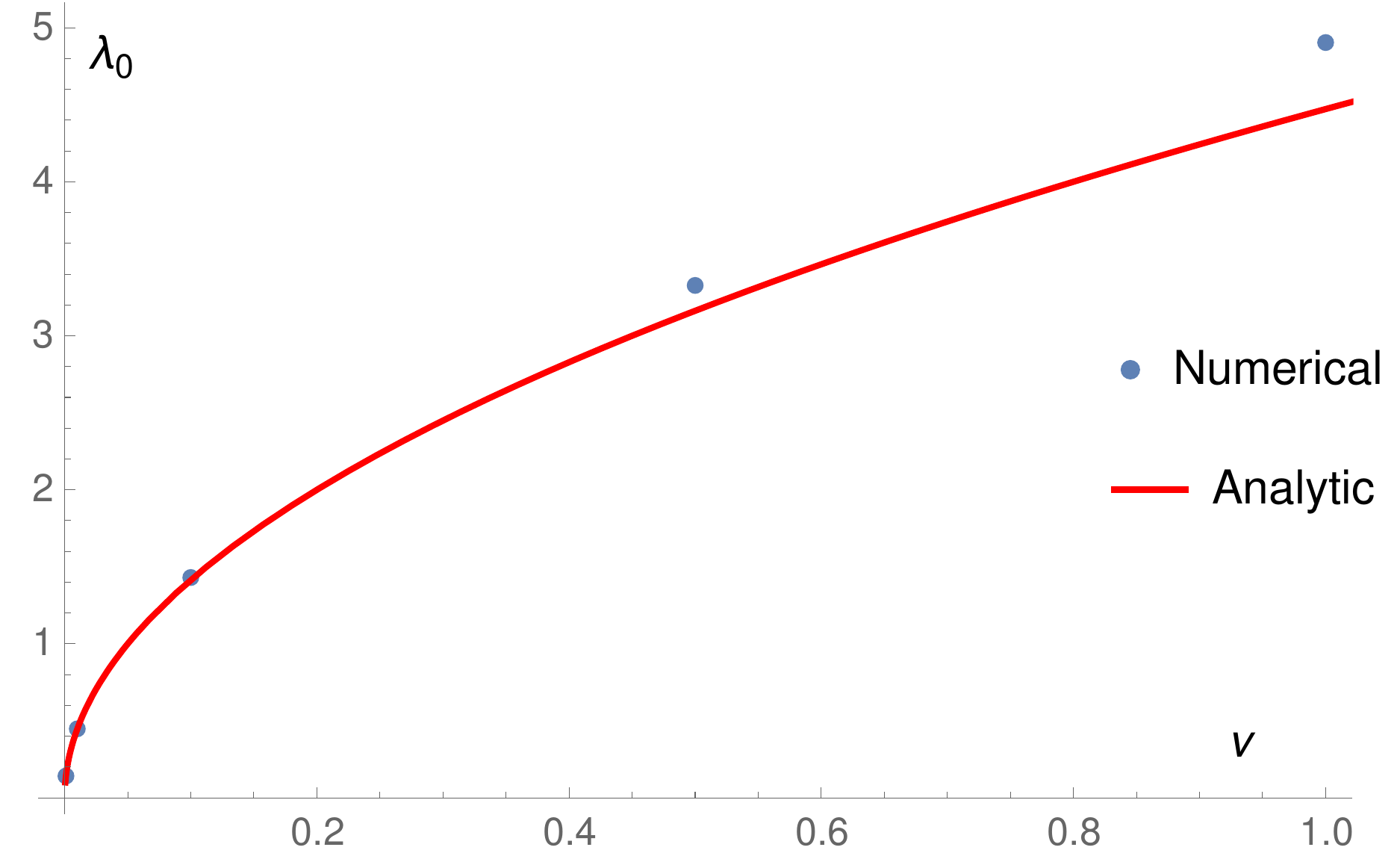}
\caption{The ``ground state'' eigenvalue $\lambda_{0}$ as a function $\nu_{\perp}$
given in Eq.~(\ref{eq:Lam0OfNu}) (line) and direct numerical integration
of Eq.~(\ref{eq:SpProblem}) (points).
\label{fig:The-ground-state}}
\end{figure}

From these considerations, we see that the anisotropy at some distance
from the source, where $\lambda_{1}\left|\zeta\right|\gtrsim3$, approaches $\approx$$\lambda_{0}/4$,
regardless of its origin: be it the source, the CR losses from the
flux tube, or both. The linear (in $\chi$) anisotropy component should
be detectable in the flux tube as long as $\lambda_{0}\left|\zeta\right|\lesssim2-3$,
depending on the strength of the source. More elaborate algebra with
the Frobenius series at the endpoints and matching them inside of
the interval $\chi\in\left(-1,1\right)$ leads to the following dependence
of $\lambda_{0}\left(\nu_{\perp}\right)$:
\begin{equation}
\lambda_{0}\approx2\sqrt{5\nu_{\perp}}.
\label{eq:Lam0OfNu}
\end{equation}
It is formally valid for $\nu_{\perp}\ll1$, but remains accurate
even when $\lambda_{0}$ exceeds unity, which we illustrate in Fig.~\ref{fig:The-ground-state}.
This simple relation establishes the range of source visibility and
anisotropy with no need to solve the full propagation problem.

\begin{figure*}[tb!]
\includegraphics[width=0.49\textwidth, height=0.23\textheight]{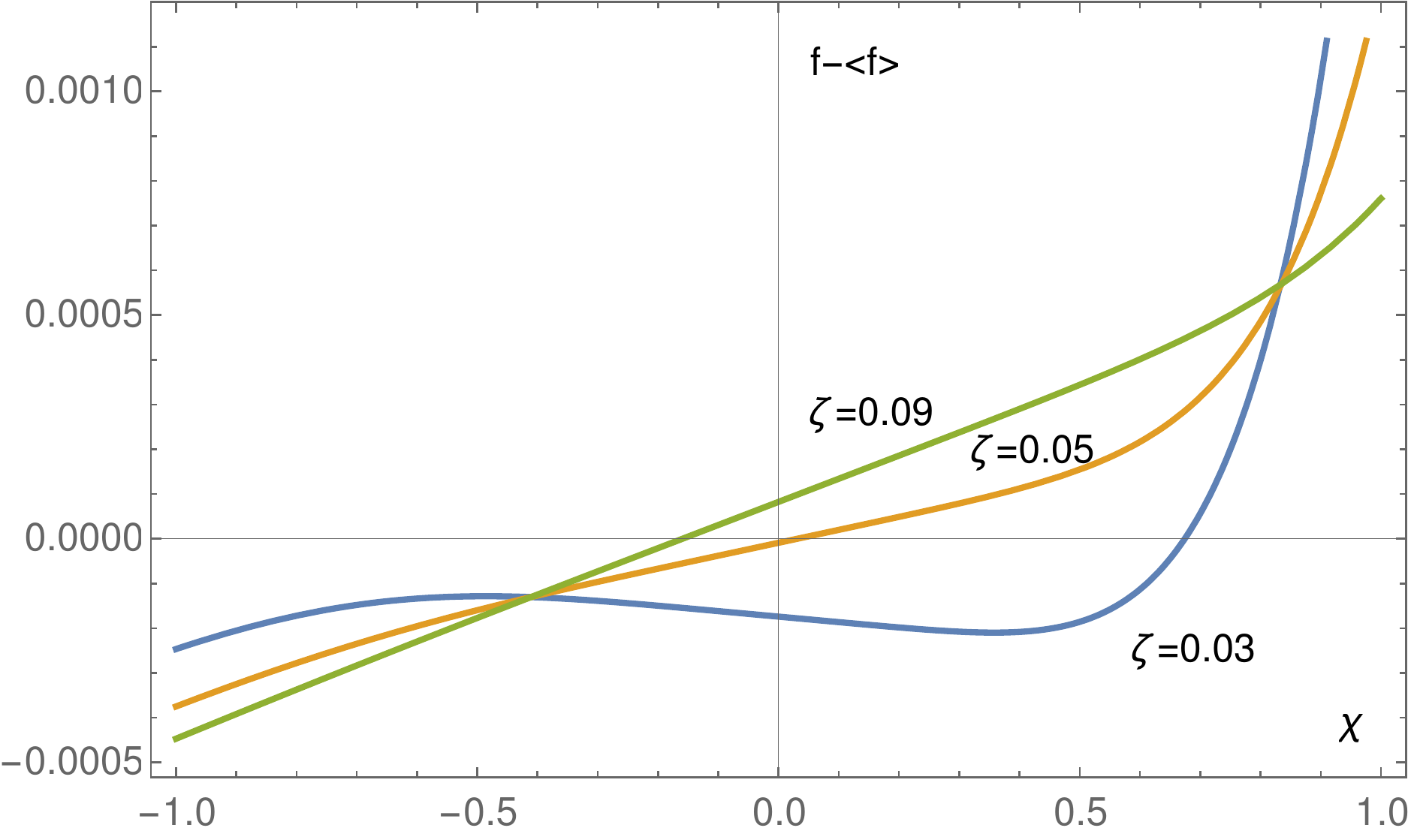}\hfill
\includegraphics[width=0.49\textwidth, height=0.23\textheight]{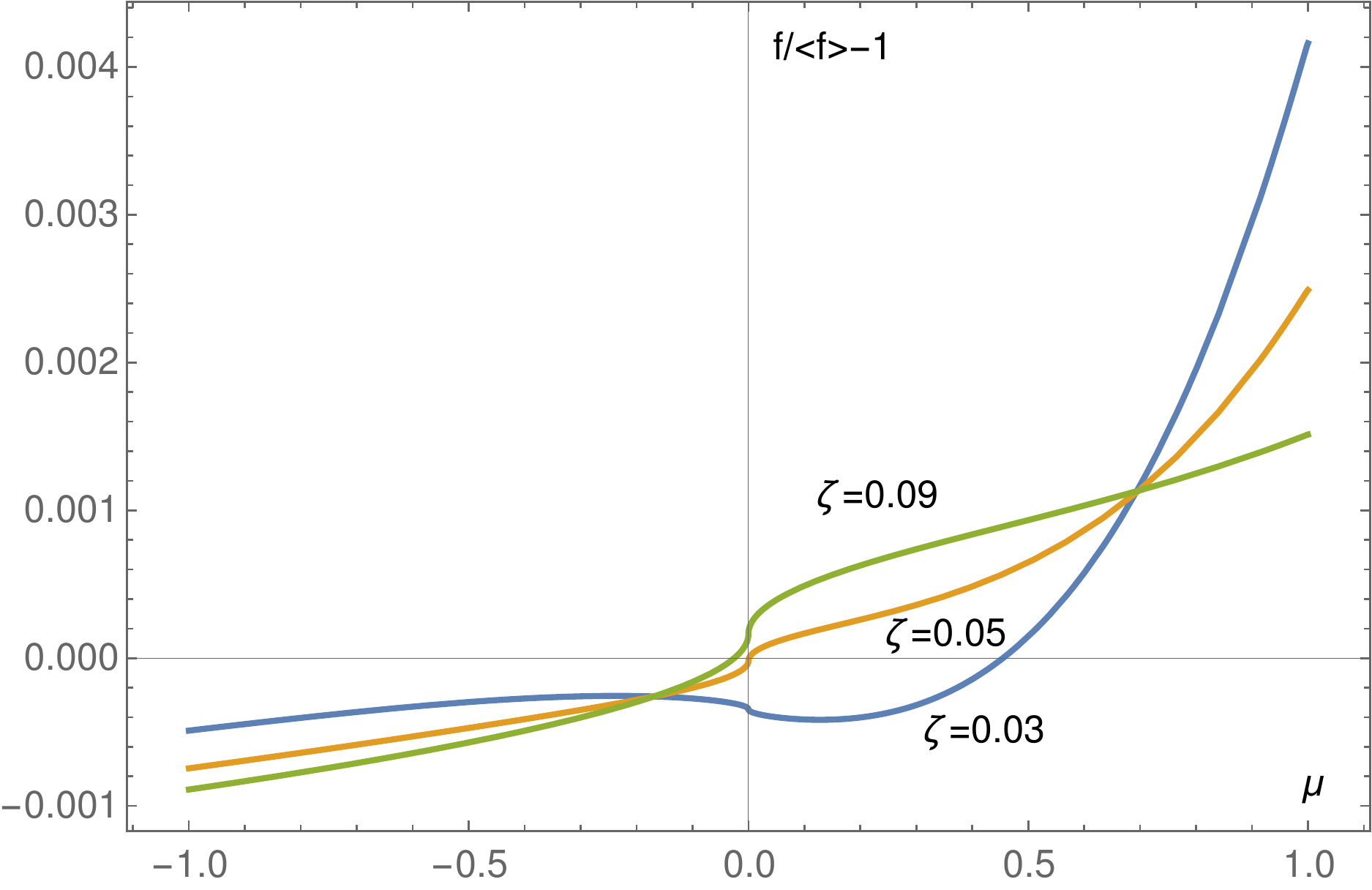}
\caption{Left panel: particle angular distribution $f-\left\langle f\right\rangle $,
using variable $\chi$ that shows no enhanced anisotropy at $\chi=0$.
Right panel: the same three solutions as a function of $\mu$ shown
as a relative anisotropy, $\left(f-\left\langle f\right\rangle \right)/\left\langle f\right\rangle $.
The isotropic part $\left\langle f\right\rangle \approx0.5$0 for
all three values of $\zeta$, and decreases with $\zeta$ insignificantly,
since $\nu_{\perp}=10^{-6}$ and $\lambda_{0}\approx$ 0.00447, $\lambda_{1}\approx45.0$,
$\lambda_{2}\approx132$. The source anisotropy is $Q_{1}\left(\chi\right)=1+0.01\left|\chi\right|^{3}$. See Eq.~(\ref{eq:NonDimVar}) for dimensionless redefinitions of $f$ and $\zeta$.
\label{fig:AnisInChiMu}}
\end{figure*}

The small-scale anisotropy ($f_{n\ge1}$) can be observed at the dimensionless
distance to the source $\lambda_{1}\left|\zeta\right|\lesssim1$.
To illustrate the variation of small- and large-scale anisotropy with
the distance to the source, we show in Fig.~\ref{fig:AnisInChiMu}
the angular distributions of CRs at three different $\left|\zeta\right|$.
For $\zeta$ varying by only a factor of three, the CRs change their smooth
dipolar anisotropy at longer distances to a sharper anisotropy when they are
closer to the source. An observer approaching the source would also see progressively field-aligned CRs.

A more robust feature of the solution can be seen by using the conventional
angular variable $\mu$ instead of $\chi$. This feature becomes
visible at distances $\left|\zeta\right|>\lambda_{1}^{-1}\approx0.05$,
where the sharp anisotropy near the field direction decays. Note,
that $\lambda_{1}$ is practically independent of the lateral losses
$\nu_{\perp}$. The solution is dominated by $f_{0}$ (smooth at $\chi=0$), but changes
at $\mu=0$ very sharply. This is a direct consequence of the I-K
turbulence, already confirmed by the fit of the rigidity spectrum, independent
of the anisotropy. The effect would be absent in the Bohm diffusion
case, characterized by a $\mu$-independent pitch-angle scattering
frequency in Eq.~(\ref{eq:PAdiff}). By contrast, it should be slightly
more pronounced in the Kolmogorov turbulence, in which the equivalent
of our $\chi$-variable would scale with $\mu$ as $\chi=\mu^{1/3}$
rather than $\chi=\mu^{1/2}$. We conclude that the I-K turbulence
confirmed by the rigidity spectrum must also manifest itself in a
sharp increase of the CR intensity across the magnetic horizon. The
increase is predicted to have the following angular dependence:
\begin{equation}
f_{0}\approx\sqrt{5/\lambda_{0}}\left[1+\frac{\lambda_{0}}{4}\left|\mu\right|^{1/2}\text{sgn}(\mu)\right].
\label{eq:f0OfMu}
\end{equation}

Remarkably, a sharp increase across the presumed magnetic horizon
($\mu=0$) in the CR intensity map was indeed observed by the HAWC
and IceCube instruments. The best fit borderline between the global
CR excess and deficit areas on the intensity map (black crossed curve
in Fig.~11 of \citealp{Abeysekara2019}) does not seem to deviate from
$\mu=0$ line (black curve) by more than several degrees anywhere
on the sky. This borderline position is predicted by Eq.~(\ref{eq:f0OfMu}),
including the intensity profile across it, shown in Fig.~8 of the same
reference. It shows a slice of the intensity map along the declination
$\delta\approx-20^{\circ}.$ The angular dependence in Eq.~(\ref{eq:f0OfMu})
can be recognized on the plot as a steep rise of the CR intensity near the RA
angle $\Theta\approx\Theta_{0}=150^{\circ}$. Note that it is sufficient
to replace our variable $\mu$ in Eq.~(\ref{eq:f0OfMu}) by $\mu\approx\pi\left(\Theta-\Theta_{0}\right)/180$
for this consideration.

However, the excess region spills over the magnetic horizon in some areas
in the Northern Hemisphere, for example, at the Milagro Region B.
Nevertheless, the spillovers are not in conflict with Eq.~(\ref{eq:f0OfMu})
since the latter represents only a gyrotropic (gyro-phase averaged)
component of the full angular distribution of CRs. The non-gyrotropic
correction to it can be significant due to, e.g., a nonlinear gyrophase
bunching that occurs when particle gyromotion falls in resonance with
a strong Alfven wave \citep[e.g.,][]{Lutom66,m98}. Unlike the robust
sharp increase across the $\mu=0$ line, the gyrophase bunching is
likely to be transient. It may change due to the wave propagation
or spreading of the particle bunch over the pitch-angle-gyrophase
plane. Such spreading occurs due to the nonlinearity of oscillations
of particles trapped into the resonance, augmented with other wave
harmonics. However heuristic the above arguments, the former Milagro
Region B now looks quite different on the Milagro successors HAWC
and IceCube's intensity maps.

The energy dependence of the CR anisotropy must also be affected by
a nearby source. Fig.~5 in \citet{Amenomori2017} shows an enhanced
anisotropy in a limited range between 0.1 TeV and 100 TeV. It is this range
where the spectral anomaly studied in this paper was observed. After
the enhancement, the anisotropy briefly declines at $\sim$100 TV,
and then it starts growing again. We note that CRs reaccelerated at a nearby shock
would merge into the background below this rigidity. 

Meanwhile, the
overall rigidity dependence of the CR anisotropy, considered in an
extended range from 0.1--1000 TV, was interpreted as being approximately
flat \citep[e.g.,][]{cowsik2016spectral}. However, the flatness is inconsistent
with the expected density gradient of CRs in the Galaxy combined with the rigidity
dependence of their scattering rate \citep[e.g.,][]{Giacinti2018}. Hence, the anisotropy of Galactic
CRs must grow with rigidity, as the pitch-angle scattering rate generally decreases with it. 

Our model resolves the paradox of flat anisotropy. It is sufficient
to subtract the contribution of CRs in the 10-TeV bump area, reaccelerated
at a nearby source, and the CR anisotropy will grow with rigidity, as expected. 
An interesting question is how exactly it grows.
Although the CRs reaccelerated nearby are more anisotropic than the background CRs, we can remove
them from the data set by their rigidity range. We then take a somewhat
broader range from the data in Fig.~5 of \citet{Amenomori2017} that
show the anisotropy increase from $\sim10^{-4}$ to $10^{-2}$ between
$10^{-1}$  and $10^{3}$ TV. This interval contains the bump. The critical point
is that the background CRs propagate their final 3--5 pc through the
same flux tube as the reaccelerated CRs. According to our model, they
are also scattered by the I-K turbulence. Assuming then that the residual
background CR anisotropy grows as $R^{\delta}$, we indeed recover
the I-K index $\delta\simeq0.5$ from the above anisotropy data. This
is the third independent indication of the I-K turbulence in the flux
tube, in addition to the theoretical consideration in Appendix \ref{sec:WaveEnTransf}
and the rigidity data fit in Sect.~\ref{sec:fitting}.

Another interesting signature of the I-K turbulence is
a sharp variation of the CR intensity across the magnetic horizon.
According to Eq.~(\ref{eq:f0OfMu}), the relative anisotropy of the
CR intensity, $I$, near the magnetic horizon at $\Theta=\Theta_{0}$ can be represented as follows
\[
\frac{I-\bar{I}}{\bar{I}}=\frac{1}{12}\left(\pi\nu_{\perp}\left|\Theta-\Theta_{0}\right|\right)^{1/2} \text{sgn\ensuremath{\left(\Theta-\Theta_{0}\right)}} ,
\]
where $\bar{I}$ is the all-sky average, the angle $\Theta$ is given in degrees, and the value of $\Theta$ is assumed to be close to $\Theta_{0}$. 
Unlike the rigidity fit obtained earlier, the above formula requires the fitting parameter
$\nu_{\perp}$ that we cannot determine ``from the first principles''
at this point. However, the characteristic square root profile is recognizable
in Fig.~8 of the paper by \citet{Abeysekara2019}.

At the outset of this section, we emphasized the role of a nearby
star magnetically connected with the heliosphere. 
If a better candidate is not found, the Epsilon Eridani will likely be the first object on the sky directly
linked to a significant part of the Galactic CR spectrum in the relevant energy range between $\sim$1 TeV and 100 TeV.

\section{Topology of the magnetic field in the solar neighborhood}\label{sec:LocalB-field}

It is clear that the structure of the local magnetic field may facilitate CR transport from one direction and impede it from another. 
To ascertain how stringent the magnetic connectivity requirement is, we extend the discussion of the Local Bubble in Sect.~\ref{sec:bubble} by zooming into several pc around the Sun.

Both the matter distribution and magnetic topology in solar proximity are challenging to describe. Most of the objects (other stars, pulsars, and quasars) suitable to probe the medium within 3--10 pc of the Sun are located too far away, and thus any such probe has to be corrected for the properties of the medium along the line of sight.  Nevertheless, much has been learned over the recent decade, although there are two somewhat different views on the local environment. Each of them might impact our model in different ways. Therefore, we are discussing them separately. 

\subsection{Complex Local Interstellar Clouds (CLIC)\label{subsec:CLIC}}

The CLIC model describes the local ISM environment as a cluster of isolated clouds \citep{Redfield2008, 2011ARA&A..49..237F, Linsky2019}.
Using 157 stars within 100 pc around the Sun, \citet{Redfield2008} were able to identify 15 partially ionized WIM clouds with temperatures
ranging between 3900 K and 9900 K.
These clouds fill
a volume within 15 pc of the Sun. According to these studies,
four such clouds are currently interacting with the heliosphere \citep{Linsky2019}. More
precisely, the Sun is currently located at the very edge of the Local Interstellar Cloud (LIC) and will leave it in
3000 years, if not earlier, 
transiting to the G Cloud. This region
has a higher temperature and lower gas density than the WIM. While
the inter-cloud gas temperature turned out to be lower than $10^{6}$~K
(earlier estimates), it remains ionized, due to, e.g., past supernova
events \citep{Breitschwerdt1999}. 

The local magnetic field direction, poorly known a decade ago, is now measured
using independent techniques producing consistent results \citep{Zirnstein2016}.
However, given the complexity of matter distribution under the CLIC
concept, the local field direction probably does not persist across
the CLIC. It shall then be naive to determine whether a nearby star
is magnetically connected with the Sun by extending the local field
line. The relative cloud motion perturbs the large-scale magnetic
field in the inter-cloud space next to the Sun. Since the inter-cloud
field is frozen into the hot plasma, not mixing with the cloud gas,
the field is likely to align with the adjacent cloud boundaries. 
The results of the Solar Wind ANisotropies (SWAN) experiment on board the Solar and Heliospheric Observatory (SOHO) satellite and the Interstellar Boundary EXplorer (IBEX) point to the same direction of the magnetic field just outside of the heliosphere, i.e.\  it is roughly tangent to the boundary of the Local Cloud in the direction of the G-cloud, as expected if field lines are compressed between both clouds
\citep{ferriere2015interstellar}.
The heliosphere perturbs the local field by draping the ISM magnetic field and the solar wind outflow (so-called Axford-Cranfill effect, \citealp{1972NASSP.308..609A}). 

Therefore, the star responsible for
the bump is not required to have its angular coordinates tightly aligned
with the local field direction because extraneous cloud motions bend
the CR flux tube. The latter makes its way to the Sun between the clouds. In this case, our task is to ascertain whether, given star coordinates and
the local field direction, the field can change its direction in the local ISM to connect to the star. To this end, we will consider
both the sharp local changes in the field direction, e.g.,  caused by 
perturbations of the ISM by the heliosphere itself and gradual ones associated
with the motions of isolated clouds and turbulence. 

The flux tube may also twist and bend intrinsically. Since the CRs
reaccelerated at a bow shock are mostly protons, they will carry a
current along the flux tube that needs to be compensated by the plasma
return current to maintain the medium's charge neutrality. This current
may drive a kink instability that will bend and twist the flux tube.
Rough requirements for its onset come from the classical Kruskal-Shafranov
(K-S) threshold. It is usually expressed in terms of a critical current
flowing through a plasma column, $I_{\text{KS}}=\pi l_{\perp}^{2}cB/L$,
where $l_{\perp}$ is the radius of the plasma column and $L$ -- its
length. The plasma column here is floating between two conducting
electrodes. 

More analogous to the CR flux tube is a plasma column
insulated from anode and its respective end can move freely \citep{furno2006current,Ryutov2006}, similar to that in a well-known plasma lamp (aka Tesla ball).
This setting would substantially decreases the instability threshold. Let
us rewrite the conventional K-S threshold in terms of a number density of reaccelerated CRs:
\begin{equation}
\frac{n_{\text{CR}}}{n_{0}}>4\pi\frac{V_{A}^{2}}{cu}\frac{r_{\text{GV}}}{\zeta}.\label{eq:KScrit}
\end{equation}
Here $n_{0}$ is the background plasma density, $V_{A}$ is the Alfv\'en
velocity, $u=U_{\text{sh}}/\cos\Theta_{nB}$ (with $U_{\text{sh}}$
and $\Theta_{nB}$ being the shock velocity and the obliquity angle,
respectively), and $r_{\text{GV}}$ is the Larmor radius of a GV proton.
For an estimate, we have replaced the plasma column length $L$ in the K-S threshold
with a local distance from the shock, $\zeta$. Indeed, $n_{\text{CR}}\left(\zeta\right)$
decreases along the tube from its maximum at the shock, $n_{\text{CR}}\left(0\right)$,
that exceeds the background CR density by the factor $K\approx2.4$,
according to the fit in Sect.~\ref{sec:fitting}. Assuming the inter-cloud density $n_{0}\lesssim0.01$
\citep{Linsky2019}, we estimate $n_{\text{CR}}\left(0\right)/n_{0}\sim10^{-6}$.
The l.h.s.\ of Eq.~(\ref{eq:KScrit}) decreases with $\zeta$ faster
than its r.h.s. The $n_{\text{CR}}\left(\zeta\right)$ can be estimated
from Eq.~(\ref{eq:SolFinal}) assuming an I-K turbulence spectrum and
the power-law index of CR, $q\approx4.5$, for simplicity. This yields
a simple scaling $n_{\text{CR}}\propto l_{\perp}^{3}/\zeta^{3}$,
so that we obtain the effective length of the kink-unstable part of
the flux tube, 
\begin{equation}
\frac{L_{\text{unst}}}{l_{\perp}}\sim\left(\frac{l_{\perp}}{r_{\text{GV}}}\frac{n_{\text{CR}}\left(0\right)}{n_{0}}\frac{cu}{V_{A}^{2}}\right)^{1/2}.\label{eq:KinkLength}
\end{equation}

Given the parameters estimated above, we find $L_{\text{unst}}/l_{\perp}\gtrsim10$
being perhaps insufficient to dislocate the source's magnetic coordinates
by more than a few degrees for $l_{\perp}$$\sim$$10^{16}$~cm. However,
the above $L_{\text{unst}}$ gives only a conservative lower bound
to this quantity. First, the Kruskal-Shafranov threshold is significantly
lower in the case of a plasma column with a ``loose end,'' as we
already mentioned. Second, we estimated $n_{\text{CR}}\left(0\right)$
from the fit in Sect.~\ref{sec:problem}, neglecting lateral losses from the tube, which
is unrealistic. Given $\zeta_{\text{obs}}\gg l_{\perp}$, the losses
must be very significant. We note that the fit remains accurate since
changes in $n_{\text{CR}}\left(0\right)$ are compensated by changes
in the shock compression ratio, Sect.~\ref{sec:fitting}. The reason is that the fit
contains a combination of these parameters, entering the spectrum
normalization parameter $K\approx2.4$, which in reality is higher
at the base of the flux tube due to the neglected losses. 

The magnetic connectivity criterion may promote the Epsilon Eridani
star to the leading candidate for the CR bump as this star is close
to the local field direction. Its Galactic J2000 coordinates\footnote{http://simbad.u-strasbg.fr/simbad\label{simbad}} are $(l_{\varepsilon\text{Eri}}, b_{\varepsilon\text{Eri}})=(196^{\circ},-48^{\circ})$,
while the local field direction at $>$1000 au from the Sun is evaluated by \citet{Zirnstein2016} based on IBEX data: $(l_{B_\text{loc}},b_{B_\text{loc}})=(206^{\circ},-50{}^{\circ})$. Here we converted their coordinates $(l'_{B_\text{loc}},b'_{B_\text{loc}})=(25.98^{\circ}\pm0.70^{\circ},50.09{}^{\circ}\pm0.57^{\circ})$ using an expression: 
\begin{equation} \label{eq-conversion}
(l_{B_\text{loc}},b_{B_\text{loc}})=(l'_{B_\text{loc}}+180^{\circ},-b'_{B_\text{loc}}). 
\end{equation}
The kink instability of the flux tube discussed above can compensate for the difference between the direction of the magnetic field and the location of the star. Note also a huge size of the astrosphere of Epsilon Eridani that is approaching $1^{\circ}$ and an even larger bow shock it creates. More importantly, for
other passing star candidates, such as Epsilon Indi, the local field direction
is unlikely to persist beyond a distance comparable to the size of
local clouds in the CLIC, as we stated earlier. Since the heliosphere
is believed to be practically in the inter-cloud space, the local field
most probably aligns with the cloud boundary. 


The next set of data about the field direction concerns a larger volume
around the Sun, about 40 pc across. The field orientation
is determined here with a significantly lesser accuracy than locally.
\citet{ferriere2015interstellar} summarizes the results of \citet{Lallement2005}
as $(l_{B_{40}},b_{B_{40}})=(227^{\circ}\pm20^{\circ},-25^{\circ}\pm20^{\circ})$,
with the errors being only roughly estimated. A more recent systematic
study by \citet{Frisch2015} provides $(l_{\text{F}},b_{\text{F}})=(206^{\circ},-49^{\circ})$
with a somewhat lesser uncertainty $\pm16^{\circ}$. (Note that these magnetic field directions are also converted using Eq.~[\ref{eq-conversion}].) This direction is consistent with the local direction discussed in the
previous paragraph but, given the significant uncertainty, may be
distinct from it. The difference is expected since, in addition
to the field perturbations caused by the local clouds, the scales
of energy injection into the ISM turbulence are likely to include
the 1--5 pc interval \citep{Haverkorn2008}. The above data constitute
an average field direction that may fluctuate on this
scale, which is just in the range of distances to the stars we are
interested in. Minimal fluctuation would place Epsilon Eridani on the same magnetic line with the Sun since the field direction indicated above is only $\approx$$6.7^\circ$ away from the star's position on the plane of the sky. Alternatively, curvature and diamagnetic drifts, as well as a lateral spreading of CRs, can compensate for this minor mismatch.

While the Epsilon Eridani star can seamlessly align
with the local (and the 40-pc averaged) field, the Epsilon Indi requires
a significant field line deflection from the local direction to connect
the reaccelerated CRs to the Sun. The Epsilon Indi star has Galactic
coordinates\textsuperscript{\ref{simbad}} $(l_{\varepsilon\text{Ind}},b_{\varepsilon\text{Ind}})=(336^{\circ},-48^{\circ})$,
that is, $\approx$$70^{\circ}$ away from the local field direction.
In part, the deflection might occur locally due to the heliospheric distortion of the ISM field mentioned earlier (see the previous paragraph, as well).
However, additional gradual deflection by as much as $\sim$$90^\circ$
over the
3.6 pc distance between the Sun and the star might be necessary. It does not seem to be impossible if
the field is largely wired through the inter-cloud space, especially
with some assistance from the kink instability. We conclude that the magnetic connection with the Epsilon Eridani star is practically guaranteed if the field maintains its local direction over a distance $\sim$3 pc. The connection to other possible sources that are not aligned with the local field is strongly affected by the CLIC cloudlets.

\subsection{Single Cloud Morphology}

An alternative account of the ISM within $\sim$10 pc of the
Sun is proposed by \citet{Gry2014,Gry2017}. In their interpretation
of the UV absorption data, the heliosphere is located inside of a
single monolithic cloud. Its bulk velocity is perturbed at different
locations so that the sightline velocities are perceived as coming
from independently moving clouds. \citet{Gry2014} provide two motivations
for the revision of the \citet{Redfield2008} model, outlined above.
First, for isolated clouds, one would expect some sightlines to pass
between them, which was not observed. This objection does not seem
to be crucial, given the limited number of sightlines and uncertain
filling factor of the clouds. The second objection regards the lack
of evidence in the Ca-II K data to support the G Cloud existence \citep{Crawford1998}.
However, this objection does not logically rule out other clouds as
separate entities, with a hot ionized medium between them. The \citet{Gry2014}
model has the advantage of a fewer number of free parameters, but
the \citet{Redfield2008} model fits the data more accurately. The
mono-cloud configuration might have a strong impact on the Epsilon
Indi scenario and would probably be less important for the Epsilon
Eridani scenario, since its direction is close to the local field. 

\subsection{Implications for Cosmic Rays}

While not making a definitive selection between the two morphologies,
we slightly incline to the \citet{Redfield2008} model, as it implicates
a sizable volume of highly ionized inter-cloud plasma through which
the CR flux tube with self-generated turbulence can be channeled.
Switching to the \citet{Gry2014} concept would require a reconsideration
of the CR propagation through the WIM. However, one aspect of the
\citet{Gry2014} interpretation of the data might have important implications
for our CR reacceleration model. Namely, they infer from the data
a shock wave slowly moving through the Local Cloud towards the heliosphere.
They further hypothesize that this is an imploding shock caused by
a pressure imbalance between the Local Cloud and a hot Local Bubble
plasma. While the shock is subsonic, it might be in a radiative state,
thus having a significant compression ratio $\sim$1.5, despite a
relatively low speed of 20--26 km s$^{-1}$. These shock parameters are not
very promising for efficient CR acceleration. However, if the shock
is quasi-perpendicular, as suggested, it may accelerate more efficiently
operating in a shock-drift acceleration regime. 

The single cloud scenario, suggested by \citet{Gry2014}, has an advantage
for our reacceleration model of not being constrained by magnetic
connectivity requirement since the proposed imploding shock should
cover half of the sky. We also note that its effect on the possible
time variability of the CR bump might be the opposite of a nearby
bow shock. Although the observer is still upstream of the shock, we
do not expect a strong suppression of particle diffusivity associated
with the flux tube crossing discussed in Sect.~\ref{sec:fitting}. In this case, the
lower-energy break (hardening) will likely decrease with time, as
particles with progressively lower energies can diffuse to the observer
from the shock, Eq.~(\ref{eq:SolFinal}).

\section{Model Summary\label{sec:Model-Summary}}

We have proposed an astrophysical explanation of the excess in the
Galactic CR proton spectrum, recently observed with an unprecedented
accuracy. The key elements of our model are as follows: 
\begin{enumerate}
\item The unexpected bump in the 10 TV rigidity range is entirely a product
of reacceleration of the pre-existing CRs in the Local Bubble (both
primary and secondary) by a weak shock wave, such as a bow shock of
a passing star or an imploding shock moving through the Local Cloud. We exclude a source of primary freshly accelerated
CRs, such as a nearby SN remnant (SNR), as the main contributor to
the bump, because of the co-presence of secondaries in it which cannot
be ``fresh''. 
\item In the current epoch, the bow shock is magnetically connected with the
Sun. The reaccelerated CRs reaching the Sun propagate through a magnetic
flux tube with an enhanced scattering turbulence generated by these
CRs. It is primarily driven by CR pressure gradient across the flux
tube. A turbulent cascade to shorter scales maintains the I-K Alfv\'en wave spectrum, $\propto k^{-3/2}$. It leads to the particle mean-free path that scales as $\lambda\propto\sqrt{l_{\perp}r_{g}}$
with the flux tube characteristic scale $l_{\perp}$ and particle
gyroradius $r_{g}$. This scaling is robust, as it is required
to fit the entire rigidity profile of the bump with correct amounts
of hardening and softening. 
\item Propagation through the I-K turbulent flux tube leaves imprints in
the CR arrival directions. The intensity map has
a narrow excess near the magnetic field direction for particles
coming from the source and a shallow deficit in the opposite field
direction. It also has a sharp increase across the magnetic horizon.
These model predictions are qualitatively consistent with observations. 
\end{enumerate}

We have solved the model equations, initially using the unknown shock
Mach number, $M$, and the nominal distance from the shock to the Sun,
$\zeta_{\rm obs}$, as fitting parameters. We have fitted the solution to
the most recent data including the TeV CR excess with a $\sim$0.1\% accuracy.
Capturing the complexity of the CR excess requires six ad hoc parameters. 
{ These are the two  break rigidities, two widths of the breaks, and two changes of the spectral index at the breaks}.
Our model solution is determined by only {\it two physical quantities.}  
These are the Mach number $M$ 
and the bump rigidity $R_0$, whereas $R_0$  
depends on the distance to the shock and its size.
This significant reduction in the number of fitting parameters makes a 
coincidental agreement highly unlikely. As the model has no free parameters,
we were able to constrain unknown physical parameters.
Our findings are: 
\begin{enumerate}
\item The distance to the bow shock along the flux tube $\zeta_{\rm obs}=$ 3--10
pc, assuming that the characteristic scale of the flux tube (bow shock) across the field is $l_{\perp}\sim10^{15}-10^{16}$
cm.  
\item The characteristic sound Mach number on the part of the shock that
efficiently reaccelerates CRs is $M$$\approx$1.5--1.6.
\end{enumerate}
The scales are obtained assuming that the shock is propagating in
a hot ISM with the temperature $T\sim10^{6}$ K, which for the calculated
Mach number corresponds to the shock speed $u\sim$ 100 km s$^{-1}$.
Our model parameters are summarized in Table~\ref{tab:Model-and-problem}.

\section{Discussion and Outlook\label{sec:Discussion-and-Outlook}}

It is, perhaps, too early to embark on quantitative modeling of the
secondaries Li, Be, B, as well as on heavier primaries, He, C, and
O. In the relevant rigidity range $\ga200$ GV they still have significant
error bars \citep{2008APh....30..133A, 2017PhRvL.119y1101A, 2018PhRvL.120b1101A}.
However, it is clear that the spectral hardening above $\ga$200 GV
is more pronounced in the spectra of secondaries. One can roughly reproduce this difference through the injection of
pre-existing CR species 
to the shock solution and limiting their acceleration in the region
above the spectral hardening because of the shock weakness. While
the resulting softening of the overall primary spectrum is not pertinent
to the reaccelerated secondaries, the latter will have a stronger
break than the primaries.

Worth mentioning is that the proposed scenario of the local magnetosonic
or bow shock predicts the same rigidity for the spectral breaks for
all CR species. Though very different in underlying physics, the implications
are similar to the ``propagation scenario'' \citep{2012ApJ...752...68V} mentioned in the introduction
and require very few free parameters. However, instead of the ``propagation
scenario'' pertinent to the whole Galaxy, our proposed scenario is
local. Observations of the diffuse emission above $\sim$30 GeV from
the interstellar gas at different distances from the Sun may be able
to discriminate between several scenarios of the observed bump, such
as the local disturbance or the global properties of the ISM.

We found, at least, two closeby fast-moving
stars that may drive shocks capable of CR reacceleration, the double Scholtz's star and triple Epsilon Indi, 
{ and a slower star, Epsilon Eridani, that has a mass loss rate of $30\dot{M}_\sun$ and a huge astrosphere extending over 42\arcmin{} on the sky. Our model is directly
applicable to both the Epsilon Indi and Epsilon Eridani systems. The former should have a bow shock
that is very much different from the shock formed by a single star.
There isi also a possibility of 11-yr variations in the CR bump properties
due to the 11-yr orbital period of its widely separated A and B+C
components. The latter, Epsilon Eridani, is the closest of the three and is aligned with the direction of the local magnetic field that makes it a primary candidate.} The turbulence left behind the double Scholtz's Star,
which is moving away from the Sun, may also provide an enhanced scattering
and confinement of particles at the rigidities beyond the TV range.

Another possibility is a weak imploding shock moving through the Local Cloud or a magnetosonic shock propagating through
the rarefied plasma in the Local Bubble. In ordinary gas dynamics
an arbitrary initial motion generally steepens into shocks, unless
dissipative effects come into play earlier \citep{Whitham2011}. In
the collisionless plasma of the Local Bubble wave dispersion and ion
reflection off the shock front may damp the shock by enabling anomalous
(i.e.\ collisionless) dissipation mechanisms \citep{Sagdeev66}.
However, for this to happen the shock needs to be supercritical \citep{Kennel1988},
i.e., typically having the Mach number $M\gtrsim2$. This is not the
case for the shock of $M$$\approx$1.5--1.6 that we inferred by fitting the
observed spectrum. We conclude that such shocks may propagate in the
Local Bubble even long after the last SN explosions we discussed in
Sect.~\ref{sec:bubble}. Other sustainable sources of mechanical
energy capable of driving weak shocks have also been suggested \citep[e.g.,][]{Cox_2003}.

{ Additional support to the local origin of the TV bump comes from its enhanced anisotropy associated with the source proximity. The background CR anisotropy is then much lower in the TV range. However,
with the bump-CRs added, the total observed anisotropy appears as an almost flat function of rigidity, as observed. 
The angular structure predicted by the model contains a robust step-like feature near the magnetic horizon, which also has been detected by HAWC and IceCube.
We defer the quantitative comparison of the anisotropic component of the model with the data to a separate study.} 

\acknowledgments 
We thank the anonymous referee for useful comments. Mikhail A.\ Malkov acknowledges support from NASA
ATP-program within grant 80NSSC17K0255 and from the National Science
Foundation under grant No.~NSF PHY-1748958. Igor V.\ Moskalenko
acknowledges support from NASA Grant No.~NNX17AB48G.

\appendix

\section{Wave Generation\label{sec:Wave-Generation}}

Although macroscopic in nature, Eq.~(\ref{eq:WaveGen}) approximately
applies to the partial CR pressure, $P_{\text{CR}}\left(R\right)$,
and wave spectral energy, $E_{\text{w}}\left(k\right),$ if the particle
distribution is nearly isotropic. In this case, a simplified resonance
condition for the wave-particle interaction, $k\approx r_{g}^{-1}\left(R\right),$
may be adopted instead of the more accurate cyclotron resonance condition,
$k_{\parallel}v_{\parallel}\approx\omega_{c}\left(R\right)$ \citep{Skill75a}.
The partial pressure of reaccelerated CRs and the wave energy density
in Eq.~(\ref{eq:WaveGen}) are normalized as follows 
\[
P_{\text{CR}}\left(p\right)=\frac{8\pi c}{3\rho V_{A}^{2}}p^{4}\left(f-f_{\infty}\right),
\]
\[
\frac{\left\langle \delta B^{2}\right\rangle }{B_{0}^{2}}=\int E_{\text{w}}\left(R\right)d(\ln R).
\]
Since we already know the CR enhancement from the fitting parameter
$K,$ we can quantify their suppression effect on $\kappa$, as compared
to its large background value, $\kappa_{\text{ISM}}\sim10^{28}$ cm~s$^{-1}$
(for GV particles). 

Being controlled by the wave generation, the suppression
factor is limited by a nonlinear saturation, cascading, and damping
of the self-generated waves. These phenomena are not included in Eq.~(\ref{eq:WaveGen}).
A plausible expectation is that the waves saturate at $\delta B\sim B$
for $k\sim r_{g}^{-1}$ under a sufficient pressure gradient ($f\gg f_{\infty}$)
that drives these waves \citep{McKVlk82,Voelk84}. The diffusion coefficient
of the resonant particles is then close to the Bohm value\footnote{Shorter waves may saturate at higher amplitudes, $\delta B\gg B_{0}$
\citep{Bell04}, close to strong shocks, but their impact on the particle
confinement may even be negative. Indeed, $\kappa\left(R\right)$
grows too steeply with $R$, $\kappa\propto R^{2}$, when these short
waves dominate the spectrum. Besides, the shock is weak in our case.}, $\kappa_{B}=cr_{g}/3$. However, unlike the CRs that are freshly
accelerated at strong shocks, the pressure of reaccelerated CRs only
moderately exceeds the background CR pressure (by a factor of $K\approx2.4)$.

We, therefore, determine $\kappa_{\parallel}$ using a quasi-linear
approximation, according to which it remains reasonably large compared
to $\kappa_{B}$, but much smaller than the background value $\kappa_{\text{ISM}}$.
As the amplitude of the waves decreases outside of the shock precursor,
$\kappa_{\parallel}$ increases and approaches an intermediate value
$\kappa_{\text{int}}$, such as $\kappa_{B}$$\ll\kappa_{\text{int}}\ll\kappa_{\text{ISM}}$.
This value is typical for such distances from the shock where an enhanced turbulence
generated by the reaccelerated CRs persists while the CRs begin to
escape the shock surroundings along the field, albeit with the suppressed
particle diffusivity. Their diffusion is impeded by self-generated
waves that propagate at a low speed, $V_{A}$. Hence, an enhanced
turbulence persists even after the CRs largely diffuse away. We incorporate
this effect into the integration constant of Eq.~(\ref{eq:WaveGen}).
The quasilinear result then reads 
\begin{equation}
\kappa_{\parallel}=\frac{\kappa_{B}}{E_{w}}=\left(1+\frac{u}{V_{A}}\right)\frac{\kappa_{B}}{P_{\text{CR}}+P_{\text{int}}},\label{eq:kappaPar}
\end{equation}
where we have defined an arbitrary constant $P_{\text{int}}=\left(\kappa_{B}/\kappa_{\text{int}}\right)\left(1+u/V_{A}\right)$
that comes from the integration of Eq.~(\ref{eq:WaveGen}). It satisfies
the boundary condition $\kappa_{\parallel}\to\kappa_{\text{int}}$
for $P_{\text{CR}}\to0$, and represents the diffusivity suppression
ahead of the shock at $\zeta\sim \zeta_{\text{int}}$, where the reaccelerated
particles diffuse away, but the wave turbulence is still significant,
$\kappa_{\text{int }}\ll\kappa_{\text{ISM}}$. 

To find $P_{\text{CR }}$,
we close Eq.~(\ref{eq:WaveGen}) by a convection-diffusion balance
for $P_{\text{CR}}$ along the flux tube, 
\begin{equation}
uP_{\text{CR}}+\kappa_{\parallel}\frac{\partial P_{\text{CR}}}{\partial \zeta}=0.\label{eq:CRtranspBal}
\end{equation}
This balance extends Eq.~(\ref{eq:cd}) to the area further ahead
of the shock precursor by applying it along the flux tube. 

Near the
shock the particle transport turns towards the shock normal, as $\kappa_{\perp}\approx\kappa_{B}^{2}/\kappa_{\parallel}$
increases. This part of the precursor does not contribute significantly
(see Sect.~\ref{sec:problem}) to the path integral in Eq.~(\ref{eq:FiOfp}).
We place it after the main part on the r.h.s.\ of the result below.
From Eq.~(\ref{eq:FiOfp}), using (\ref{eq:CRtranspBal}), we thus obtain: 
\begin{eqnarray}
\Phi\left(R,\zeta_{\rm obs}\right) & = & \ln\frac{P_{\text{CR}}\left(0\right)}{P_{\text{CR}}\left(\zeta_{\rm obs}\right)}\\
 & \simeq & \frac{u\zeta_{\rm obs}}{\kappa_{\text{int}}}+\ln\left[P_{\text{CR}}\left(0\right)\frac{\kappa_{\text{int}}}{\kappa_{B}}\left(1+\frac{u}{V_{A}}\right)^{-1}\right].\nonumber \label{eq:FiOfRzsun}
\end{eqnarray}
Here we have also assumed that $\zeta_{\rm obs}\gg \zeta_{\text{int}}$. To be
consistent with the fit, this expression must vary with $R$ as $R^{-1/2}$,
according to Eq.~(\ref{eq:FiSimple}). Assuming then that $\kappa_{\text{int}}\propto R^{1/2}$,
which we will justify in Appendix \ref{sec:WaveEnTransf}, we find
that only the first term on the r.h.s.\ behaves this way. The second
term varies as $\ln R$. Besides, it does not contribute to the range
of $\Phi\left(R\right)$ significantly. Indeed, according to the fit
in Sect.~\ref{sec:fitting}, $\Phi$ varies by a factor $\sim$100
in the interval 30 GV $<R<100$ TV which is well above the range of
the second term on the r.h.s.\ of Eq.~(\ref{eq:FiOfRzsun}). It
varies by less than an order of magnitude in the same rigidity range.

The main contribution to $\Phi$ thus comes from an extended region
outside of the shock precursor, which gives us an access to the distance
to the source. Unlike the second term, this contribution is directly
proportional to $\zeta_{\rm obs}$. However, this is a quasilinear result
that needs to be modified if the wave-wave interaction, not included
in Eq.~(\ref{eq:WaveGen}), is essential. The other obstacle to the
estimate of $\zeta_{\rm obs}$ is the unknown $\kappa_{\text{int}}$. We
will also consider these aspects in Appendix \ref{sec:WaveEnTransf}.

It follows that a cloud of reaccelerated CRs must be self-confined
in the lateral and wake areas of the bow-shock before they escape
the star surroundings. The CR cloud must also be elongated in the
field direction, because $\kappa_{\perp}\approx\kappa_{B}^{2}/\kappa_{\parallel}\ll\kappa_{\parallel}$.
Since the CR pressure gradient is then directed largely across the
field, the resonant cyclotron and the nonresonant CR current, or \citet{Bell04}
instabilities that develop most rapidly along the field are less efficient
than the CR pressure-driven acoustic instability. The latter, also
called Drury instability \citep{Dorfi84, DruryFal86, KangJR92}, has
been compared with the nonresonant instability by \citet{MDS10PPCF}.

Although only a specific case of CR modified shock precursor has been considered in the comparison,
in which both the CR current and pressure gradient instabilities stem
from the same CR population balanced by the ram pressure of the inflowing
plasma, we can apply some of those results to the present study. For
the plasma-$\beta\sim1$, the growth rates of Drury and Bell instabilities
are similar, even assuming the Bell instability operating under the
most favorable conditions of the field aligned current and wave vector.
Drury instability, however, is not sensitive to its wave vector orientation
with respect to the field, particularly if $\beta>1$, which is likely
to be the case in the Local Bubble \citep{Spangler2009}. Therefore,
Drury instability should drive waves across the field more efficiently.

Now we focus on the saturated spectrum of this instability and its
implications for the CR scattering.
Unlike in the resonant cyclotron and Bell instabilities the magnetic
perturbations of the acoustic instability can be considered as passively
frozen into the fluid perturbations. Assuming that $\beta>1$, this
approximation is sufficient for our purposes. Later we will take magnetic
perturbations pertaining to the particle scattering into account.
First, we adopt a reduced equation for unidirectional acoustic perturbations
driven by the CR pressure gradient in the form presented by \citet{MD09}.
In a CR cloud elongated along the field line in $\zeta$-direction, the
CR pressure is in $r$-direction, in a cylindrical symmetry, for example.
We neglect $\zeta$-dependence for now, also considering $r$ as a locally
Cartesian coordinate. The equation for unstable acoustic waves can
be written as a 1-D evolution equation: 
\begin{equation}
\frac{\partial\tilde{\rho}}{\partial t}-C_{s}\frac{\partial\tilde{\rho}}{\partial r}-\frac{\gamma+1}{2\rho_{0}}C_{s}\tilde{\rho}\frac{\partial\tilde{\rho}}{\partial r}-\mu\frac{\partial^{2}\tilde{\rho}}{\partial r^{2}}=\gamma_{D}\tilde{\rho},\label{eq:Burg}
\end{equation}
where $C_{s}$ is the sound velocity, $\tilde{\rho}$ is the plasma
density perturbation, $\rho_{0}$ is its background density, and $\gamma$
is the adiabatic index. The l.h.s.\ is essentially a Burgers equation
that describes unstable waves (driven by the instability term on the
r.h.s.)~aligned with the CR gradient and propagating along the characteristics
$r=-C_{s}t+r_{0}$. The characteristics $r=C_{s}t+r_{0}$, associated
with the damped waves, have already been eliminated from Eq.~(\ref{eq:Burg}).

The small viscosity $\mu$ is only important at discontinuities, resulting
from the nonlinear steepening of unstable waves. The growth rate for these waves is 
\begin{equation}
\gamma_{D}=-\frac{1}{2\rho_{0}C_{s}}\frac{\partial\overline{P}_{\text{CR}}}{\partial r},\label{eq:GammaDrury}
\end{equation}
where $\overline{P}_{\text{CR}}$ is the full (momentum integrated)
CR pressure. The growth rate has also a stabilizing component owing
to the CR diffusion. We have omitted it since it is small compared
to $\gamma_{D}$, assuming that $\kappa/l_{\perp}>C_{s}$. Here $l_{\perp}$
is the characteristic scale of the pressure gradient ($l_{\perp}^{-1}\sim\overline{P}_{\text{CR}}^{\,-1}\partial\overline{P}_{\text{CR}}/\partial r$).

A handle on the dynamics and saturated state of the unstably driven
Burgers turbulence can be obtained by transforming to the reference
frame moving with unstable sound waves, $r\to r+C_{s}t$. This boost
along $r$ transforms away the second term on the l.h.s.\ of Eq.~(\ref{eq:Burg}).
Small-amplitude seed waves will then grow exponentially until the
third (nonlinear) term balances the driver on the r.h.s.\ Before
it happens, the initial, say sinusoidal, profile of the seed perturbation
steepens into a periodic shock sequence with the seed
wave period. In this saturated state, the following balance is maintained
across the most part of the period: 
\begin{equation}
\frac{\partial\tilde{\rho}}{\partial r}\approx\frac{1}{\left(\gamma+1\right)C_{s}^{2}}
\frac{\partial\overline{P}_{\text{CR}}}{\partial r}.\label{eq:NLinstBal}
\end{equation}
As the viscose term in Eq.~(\ref{eq:Burg}) remains small everywhere
except in the shock transition, the density $\tilde{\rho}$ has a
saw-tooth profile. It consists of smooth curves with slopes proportional
to the local slope of $\overline{P}_{\text{CR}},$ connected by the
shocks. For this reason it is called a shock-train solution. The latter
is a generic consequence of nonlinear and instability terms present
in many evolution equations studied in the past (see, e.g., \citealp{MD09}
and references therein).

The shock-train solution is in compliance with a natural requirement
that the period averaged $\left\langle \tilde{\rho}\right\rangle \approx0$,
while, according to Eq.~(\ref{eq:NLinstBal}) it must remain monotonic
almost everywhere, except for the thin shock transitions. It also
proved to be a strong attractor to which a broad class of initial
profiles converge. Meanwhile, in the shock transitions the last term
on the l.h.s.\ of Eq.~(\ref{eq:Burg}) is of the same order as the
nonlinear and instability terms. The fact that the slope of $\tilde{\rho}$
between the shocks closely follows that of the $\overline{P}_{\text{CR}},$
immediately relates the amplitude of the density jumps to the distance
between the shocks, a property that we will use in the sequel.

\section{Wave Energy Transformation\label{sec:WaveEnTransf}}

Unlike in an idealized periodic solution discussed earlier, in any
realistic shocktrain structure shocks have different strengths. Therefore,
they move at different speeds and coalesce, so their strengths increase
while fewer shocks remain in the system. However, this shock dynamics
is typical for a one-dimensional system evolving, for example, in
a periodic box, with a time-asymptotic state of only one strong shock
per period. 

The system considered here has at least three important
differences. First, unlike the initial value problem outlined in the
previous subsection, the seed waves are continuously fed to the Drury
instability from the resonance instability of CRs diffusing away from
the bow shock. Hence, the asymptotic single-shock state of a maximum
strength is unrealistic. Second, the shock formation occurs in an
elongated region being two-dimensional, at a minimum. The shocks propagate
then towards its axis at different, albeit close angles. Therefore,
their coalescence proceeds along intersection lines, not along the
entire shock surface simultaneously. As a result, a random web-like
shock network emerges. Third, the inverse cascade associated with
the shock merger events is arrested by the forward Alfv\'en cascade at a scale 
that plays a role of the Rhines's scale.
This effect is akin to the Rhines's phenomenon where an inverse eddy cascade
is intercepted by a wave turbulence cascade \citep{Rhines1975}. We will discuss
this scenario below for the sake of CR scattering by the Alfv\'en waves.

Let us first estimate the density of the shocks and their average strength
while they merge. These quantities will determine the spectrum of
Alfv\'en waves that are generated by the shock merger. The particle
diffusivity along the field, $\kappa_{\text{int}}$, depends on the
Alfv\'en spectrum. We recall that we need $\kappa_{\text{int}}$ to
derive the distance to the bow-shock using Eq.~(\ref{eq:FiOfp}).

To describe the shock merger, we return for a while to the simple
picture of a periodic 1-D shocktrain. Let its initial period be $2b.$
The energy density of the shock train moving at a speed $C_{s}$,
as in ordinary acoustic waves, has two quadratic contributions. They
come from the density and velocity perturbations and are equal in
a plane acoustic wave \citep{Landau_Fluid}, thus carrying the specific
energy density $C_{s}^{2}\tilde{\rho}^{2}/\rho_{0}^{2}$. Using Eq.~(\ref{eq:NLinstBal}),
we can write this quantity within one shocktrain period between the
points $r=\pm b$, as 
\begin{equation}
E=\frac{C_{s}^{2}}{2b\rho_{0}^{2}}\int_{-b}^{b}\tilde{\rho}^{2}dr=\frac{b^{2}}{3\left(\gamma+1\right)^{2}\rho_{0}^{2}C_{s}^{2}}\left[\frac{\partial\overline{P}_{\text{CR}}}{\partial r}\right]^{2}.\label{eq:AcEnergy}
\end{equation}
We have used an approximation of linear $\overline{P}_{c}\left(r\right)$
on a short interval $b\ll l_{\perp}$. Introducing the shocktrain
amplitude parameter $A$, 
\[
A^{2}=\frac{2\pi^{2}}{3\left(\gamma+1\right)^{2}\rho_{0}^{2}C_{s}^{2}}\left[\frac{\partial\overline{P}_{\text{CR}}}{\partial r}\right]^{2},
\]
and the shocktrain wave number $k_{0}\equiv\pi/b$, we can express
the above energy density through the spectral density as follows:
\[
\frac{A^{2}}{2k_{0}^{2}}=\int_{k_{0}}^{\infty}E_{k}^{s}dk.
\]
From here we obtain the shocktrain spectral density $E_{k}^{s}$:
\begin{equation}
E_{k}^{s}=\begin{cases}
A^{2}k^{-3}, & k\ge k_{0};\\
0, & k<k_{0}.
\end{cases}\label{eq:ShTrSpec}
\end{equation}

It is seen that as the shocks merge, that is $k_{0}$ decreases, the
total energy density increases as $k_{0}^{-2}$. The spectrum at $k>k_{0}$
is unaffected by the mergers and remains \emph{independent of} $k_{0}$,
as we consider the amplitude parameter $A$, that is the $\partial\overline{P}_{\text{CR}}/\partial r$
profile, to remain constant. This convenient feature of the acoustic
inverse cascade allows us to evade a difficult question of how to
determine $k_{0}$ for the purpose of obtaining a magnetic counterpart
of the acoustic shocktrain turbulence. Since the inverse cascade proceeds
to a sufficiently small value $k_{0}\gtrsim l_{\perp}^{-1}$, we may
assume that $k_{0}r_{g,\text{max}}<1$ and the longest waves will
not influence the CR scattering significantly. As we know from the
fit, the maximum rigidity of reaccelerated CRs is in the range $10-100$
TV, so the gyroradius cannot significantly exceed $10^{16}$ cm in
a few $\mu$G field. Further constraints on $k_{0}$ would require
a consideration of the lateral profile of the CR flux tube, which is beyond
the scope of this paper.

In the zeroth approximation, magnetic perturbations associated with
the shocktrain generation passively follow the gas motion in acoustic
waves. In the Local Bubble plasma, presumably with $\beta>1$, oblique
magnetosonic perturbations would be strongly damped, were they not
be driven by the CR pressure gradient. In contrast to the driven magnetoacoustic
waves, Alfv\'en waves generated by the shocktrain turbulence propagate
along the field essentially undamped and cascade to shorter scales,
thus interacting with all reaccelerated CRs. Alfv\'en wave generation
is quite similar to what is known in the MHD turbulence studies as
``Alfv\'enization'' of the Kolmogorov hydrodynamic cascade. Although
the excitation of waves with longer scales is also expected due to
the helicity conservation, we focus on the direct Alfv\'en cascade as
it is relevant to the CR scattering. It proceeds via interaction of
counter-propagating Alfv\'en wave packets, as described by, e.g., \citet{goldr97}.

The wave packet interaction time is $1/kV_{A}$, so that the relative
change in velocity perturbation can be written as 
\[
\frac{\delta v_{k}}{v_{k}}\sim\frac{1}{v_{k}}\frac{dv_{k}}{dt}\frac{1}{kV_{A}}\sim\frac{v_{k}}{V_{A}},
\]
where the time derivative is estimated from the nonlinear interaction
term $\sim\boldsymbol{v}\cdot\nabla\boldsymbol{v}\sim kv_{k}^{2}$
of the MHD equations. Since $\delta v_{k}\left(t\right)$ is a random
process with zero mean, it takes $N_{k}\sim\left(V_{A}/v_{k}\right)^{2}$
interactions before the velocity perturbations increase to $\sim$$v_{k}$.
The time needed for that is $\tau_{k}=V_{A}/kv_{k}^{2}$. The energy
flux across the spectrum is 
\[
\varepsilon\sim\frac{v_{k}^{2}}{\tau_{k}}\sim k\frac{v_{k}^{4}}{V_{A}}.
\]
Assuming this flux to be constant in a steady state, we obtain the
power spectrum of Alfv\'en waves 
\begin{equation}
E_{k}^{A}\sim v_{k}^{2}/k\sim\frac{\sqrt{\varepsilon V_{A}}}{k^{3/2}},\label{eq:AlfvSpec}
\end{equation}
which is, of course, the I-K spectrum. 

Now we need to relate the energy
flux $\varepsilon$ to its source in the acoustic turbulence of merging
shocks. One possibility is that three-wave interactions, such as decays
of magnetoacoustic waves to other magnetoacoustic and Alfv\'en waves
\citep{Livshits1970, Kuznetsov2001}, $M\to M+A$ or $M\to A+A$, couple
sufficiently broad ranges of the two spectra. On the other hand, the
decay's growth rate scales unfavorably with the wave number in the
long wave limit, $\gamma_{d}\sim kV_{A}\delta B^{2}/B^{2}$, where
the acoustic turbulence condensates. Besides, the weak turbulence
scenario associated with the three-wave processes is not viable for
a strong CR driver that seems to be present by virtue of the fit.

In strong turbulence regimes, the energy flux is usually obtained
from a prescribed energy ``injection'' rate at an outer scale. In
our case the Alfv\'en waves are fed in from a continuous acoustic spectrum
rather than a determinate scale. However, the spectral shapes of these
two spectra are formed by strong individual cascades without interaction,
so they have markedly different indices (3 for acoustic- and 3/2 for
Alfv\'en waves). Therefore, they must intersect, say at $k=k_{*}$ (cf. the Rhines scale).
At this scale the turbulent kinetic energy of Alfv\'en waves is in equipartition
with the acoustic energy (shocks), $E_{k_{*}}^{A}\sim E_{k_{*}}^{s}$.
Since only one of them dominates for all  $k$ not too close to $k_{*}$, the
interaction between them falls off with the growing $\left|k-k_{*}\right|$.
The above equipartition requirement alone does not determine $k_{*}$
and $\varepsilon$. The second condition that we need is that the
energy transfer rates in the two cascades coincide at $k\sim k_{*}$.
In the acoustic cascade this rate is the same as the shock coalescence
rate, which is $\sim$$\gamma_{D}$ and is scale-invariant. Therefore,
the second condition is $\varepsilon\left(k_{*}\right)\sim\gamma_{D}v_{k_{*}}^{2}$.

From these two conditions we find 
\begin{equation}
k_{*}\sim\frac{\gamma_{D}}{V_{A}},\;\;\;\varepsilon\sim\gamma_{D}V_{A}^{2},\label{eq:kstarANDeps}
\end{equation}
that is $v_{k_{*}}\sim V_{A}$. In the range $k>k_{*}$ the two spectra
separate from each other as follows 
\begin{equation}
E_{k}^{s}\sim\frac{\gamma_{D}^{2}}{k^{3}},\quad E_{k}^{A}\sim\frac{\sqrt{\gamma_{D}}\,V_{A}^{3/2}}{k^{3/2}}.\label{eq:AcousAlfSpectraFinal}
\end{equation}
The behavior of the spectra at $k<k_{*}$ may deviate from these scalings,
since the inverse acoustic cascade looses a significant part of its
energy flux to the Alfv\'en forward cascade at $k\sim k_{*}$. In addition,
the inverse Alfv\'en cascade may also proceed from $k_{*}$ to smaller
wave numbers at a different slope. Meanwhile, using the first of the
two relations in Eq.~(\ref{eq:kstarANDeps}) we can express $k_{*}$
through the ``observable'' quantity $P_{\text{CR}}$, which can
be inferred from the fit, 
\begin{equation}
k_{*}l_{\perp}\sim\frac{P_{\text{CR}}}{\rho C_{s}^{2}}\sqrt{\beta}.\label{l-perp}
\end{equation}

Assuming that the background CR pressure is similar to the gas pressure,
and it is enhanced by the reacceleration only by a factor of a few
(according to the fit, $K=2.4$, Sect.~\ref{sec:fitting}), the
above ratio cannot be much larger than unity. Since for the highest
energy reaccelerated CRs we assume $r_{g}\lesssim l_{\perp}$, the
region $k<k_{*}$ is not important for the particle scattering. At
the same time, the boundary in momentum space of self-confined particle
at $r_{g}\left(p\right)\sim k_{*}^{-1}$ shifts to lower momenta with
growing $P_{\text{CR}}$, provided that $l_{\perp}$ is constant.
This shift limits the maximum energy of reaccelerated CRs that can
be observed, unless there is a significant cascading of Alfv\'en waves
to lower $k<k_{*}.$

\section{Estimate of the Distance to the hypothetical bow-shock \label{sec:EstimateDistance}}

Now we can estimate the distance to the shock, $\zeta_{\rm obs}$, using
an expression for the Alfv\'en wave spectrum in Eq.~(\ref{eq:AcousAlfSpectraFinal}). 
Since we use the nonlinearly
transformed spectrum, the quasilinear phase integral in Eq.~(A3)
is no longer applicable
and we turn to its original form in Eq.~(\ref{eq:FiOfp}). We know
the function of particle rigidity, $\Phi\left(R\right)$, on the l.h.s.\ of
this equation from the fit while the particle diffusivity, $\kappa_{\text{int}}$,
can be estimated as follows 
\begin{equation}
\kappa_{\text{int}}\sim r_{g}c\frac{B^{2}}{\delta B_{k}^{2}}\sim r_{g}c\left.\frac{V_{A}^{2}}{v_{k}^{2}}\right|_{kr_{g}\sim1}.\label{eq:kap-int-est}
\end{equation}

Introducing a dimensionless parameter, characterizing the CR pressure
of reaccelerated particles, 
\[
\Pi=\frac{P_{\text{CR}}}{\rho C_{s}V_{A}},
\]
so that $\gamma_{D}\sim\Pi V_{A}/l_{\perp}$, and using Eqs.~(\ref{eq:AlfvSpec})
with (\ref{eq:AcousAlfSpectraFinal}), for $\kappa_{\text{int}}$
we obtain 
\begin{equation}
\kappa_{\text{int}}\sim\frac{c}{3}\sqrt{r_{g}l_{\perp}/\Pi}.\label{eq:kap-int-fin}
\end{equation}
The base-line m.f.p.\ of reaccelerated particles propagating along
the flux tube is thus 
\[
\lambda_{CR}\sim\sqrt{r_{g}l_{\perp}}.
\]
Substituting
$\kappa_{\parallel}=\kappa_{\text{int}}$ and $u_{1}=\cos\vartheta_{nB}u$
in Eq.~(\ref{eq:FiOfp}) and comparing it with Eq.~(\ref{eq:FiSimple}), we may write 
\begin{equation}
\sqrt{R_{0}}\simeq\frac{3u}{c\sqrt{r_{\text{GV}}l_{\perp}}}\Psi(\zeta_{\rm obs}),\label{eq:SqrtR0}
\end{equation}
where 
\[
\Psi(\zeta)\equiv\int_{0}^{\zeta}\sqrt{\Pi}d\zeta,
\]
and $r_{\text{GV}}$ is the gyro-radius of a GV-proton. Note that, 
the rigidity $R$ canceled out from the above relation
and $R_{0}\approx4400$ can now be considered as a dimensionless number.

Next, we need to calculate the normalized CR pressure $\Pi\left(\zeta\right)$.
Since it derives from the background CRs with a distribution $f_{\infty}$,
it is convenient to relate $\Pi$ to the background CR pressure $P_{\infty}$.
From the solution of acceleration problem in Eq.~(\ref{eq:SolFinal})
we have 
\begin{equation}
\Pi=\frac{P_{\infty}}{\rho C_{s}V_{A}}\left.\int_{0}^{\infty}R^{3}
\left(f-f_{\infty}\right)dR\right/\int_{0}^{\infty}f_{\infty}\frac{R^{4}dR}{\sqrt{1+R^{2}}},
\end{equation}
where the integral in the denominator represents the background CR
pressure $P_{\infty}$. The integration variable $R$ is normalized
here to the rest mass proton rigidity $mc/e$. In representing the
pressure of reaccelerated CRs, characterized by $f-f_{\infty}$, we
have used the approximation $R\gg1$. This approximation is not accurate
near the shock, but as we have seen from crude estimates of $\zeta_{\rm obs}$
in Sect.~\ref{sec:problem}, the main contribution to the path
integral comes from large $\zeta$ ahead of the shock precursor, where
$f\approx f_{\infty}$ at $R\sim1$. The main contribution to $P_{\infty}$
comes from $R\sim1$, but the exact behavior of $f_{\infty}$ along
the path is unknown (see however the discussion of the Voyager 1 data
by \citealp{DruryStrong2017}). We attach this uncertainty to a new
parameter $R_{*}\sim1$, by assigning the lower limit of the integral
in the denominator to $R_{*}$ and replacing the square root expression
by $R$.

The above expression for $\Pi$ rewrites: 
\begin{equation}
\Pi=\eta\int_{0}^{\infty}R^{3-\sigma}dR\exp\left[-\frac{\alpha}{\sqrt{R}}\Psi(\zeta_{\rm obs})\right],\label{eq:PiIntExpr}
\end{equation}
where 
\begin{equation}
\alpha=\frac{3u}{c\sqrt{r_{\text{GV}}l_{\perp}}}\quad\text{and}\quad\eta=\frac{KP_{\infty}}{\rho C_{s}V_{A}}\frac{\sigma-4}{R_{*}^{4-\sigma}}.\label{eq:alf-eta-defs}
\end{equation}
Using Eq.~(\ref{eq:PiIntExpr}), we have the following differential
equation for the path integral $\Psi$,
\[
\frac{d\Psi}{dz}=B\Psi^{4-\sigma},
\]
where 
\[
B=\alpha^{4-\sigma}\sqrt{2\eta\Gamma\left(2\sigma-8\right)},
\]
and $\Gamma$ is the gamma-function. After solving it and using Eq.~(\ref{eq:SqrtR0}),
we obtain the following result for $\zeta_{\rm obs}$: 
\begin{equation}
\frac{\zeta_{\rm obs}}{\sqrt{r_{\text{GV}}l_{\perp}}}\!=\!\frac{(c/3u)R_{0}^{\left(\sigma-3\right)/2}R_{*}^{2-\sigma/2}\sqrt{\rho C_{s}V_{A}}}{\sqrt{2\left(\sigma-4\right)}\sqrt{\Gamma\left(2\sigma-8\right)}\left(\sigma-3\right)\sqrt{KP_{\infty}}}.\label{eq:zStarGeneral}
\end{equation}
By solving the last expression for $R_{0}$, we obtain the result
in Eq.~(\ref{eq:R0fromApp}).

\section{Large- and Small-Scale anisotropy\label{sec:AnisApp}}

The purpose of this Appendix is to supplement Sect.~\ref{sec:Anisotropy-and-Magnetic}
with further discussion of observations and interpretations of the
CR anisotropy at different scales, as well as less essential details
of the methods we used there. 

Since the discovery of a narrow component in the CR arrival direction
by Milagro observatory \citep{Milagro08PRL}, much have been learned
about CRs in the same rigidity range 1--20 TV. Based on this learning,
we have concluded Sect.~\ref{sec:Anisotropy-and-Magnetic} by linking
together the recently detected rigidity bump with the sharp field-aligned
CR excess, and the step-like increase in the CR intensity across the
magnetic horizon. Note, that these two groups of particles are well
separated in the cosine of pitch angle, $\mu$, as they concentrate
near $\mu=1$ and $\mu=0$, respectively. The ``glue'' for these
three ingredients is the I-K turbulence spectrum required by the fit
of the CR bump rigidity profile. 

The I-K turbulence is also consistent
with the step-like increase across the magnetic horizon. The effect
of the turbulence spectrum on the sharp angular component, concentrated
near $\mu=1$, is weaker than that on the other two ingredients. The
reason for that is the pitch angle dependence of the scattering frequency
that vanishes at $\mu=1$ regardless of the scattering turbulence.
The causally linked new observational findings give us a fresh glimpse
of the Milagro hot spots, discovered more than a decade ago.

The Milagro sharp excess was found to be aligned with the heliotail
direction, which is now believed to be not far from the local magnetic
field direction (see, e.g., \citealt{Abeysekara2019} and Sect.~\ref{sec:LocalB-field}).
Though it is unlikely to be a heliotail phenomenon. Thanks to the
combined action of the solar wind and the Sun's motion through the
ISM at $\approx$26 km s$^{-1}$, the heliosphere is a strongly elongated
magnetic cavity \citep{Pogorelov2015} with an average field being
significantly lower inside than outside. Since the Larmor radius of
a 10-TeV proton in a 1 $\mu$G field is $\approx$2000 au, while the
heliosphere is only 200-300 au across and shrinks towards the tail,
these particles cannot be \emph{accelerated} there or even significantly
deflected upon entrance. The field inside the cavity is just too weak,
or its coherence length is too small, or both. If we still assume
that a strong deflection somehow occurs, we will need to accept that
the rigidity bump and the enhanced anisotropy in the same 1-10 TV
range are not causally associated. Indeed, since the magnetic field
at a local scale can be considered as static, it could not create
the energy bump in the isotropic CR component, no matter how much
of a CR deflection it might produce. It is therefore more likely that
the anisotropy have been built up externally.

Several such mechanisms have been suggested and in part reviewed critically,
e.g., by \citet{M_PoP2015}, and more extensively by \citet{Ahlers2017}.
Some of these mechanisms employ the CR propagation in a particular
realization of the ISM turbulence (\emph{not ensemble-averaged}).
In principle, magnetic perturbations could produce a magnetic lensing
effect, similar to the light ray caustics in refractive media. A simple
example is a highly variable light pattern seen on the bottom of a
swimming pool in a sunny day. However, such a particle focusing strongly
depends on their rigidity. The 1 TV and 10 TV protons cannot be collimated
simultaneously to produce the same hot spot in the sky, which, however,
was observed by Milagro. In fact, they found that the spectrum of
the collimated particles is flatter than the background. This behavior
can hardly be consistent with the magnetic lensing effect, but suggests
that the hot spot in the angular distribution and the rigidity
bump are two sides of the same phenomenon. 

An \emph{ensemble-averaged, }and thereby more\emph{ }robust
collimation mechanism, suggested by \citet{Milagro10}, correctly
predicts the Milagro sharp component's parameters, including the spectral
hardening between 1--10 TV. It assumes magnetic connectivity of the
source and heliosphere, as is also proposed in the present paper,
but the Milagro-related model \emph{requires} the source to be at
a few 100 pc, most likely being an SNR. The similarity with the mechanism
described in Sect.~\ref{sec:Anisotropy-and-Magnetic} is in that the
small scale anisotropy is also built on a larger-scale anisotropy,
near its minimum at $\mu=1$ of the first mode of the scattering operator,
$f_{1}\left(\mu\right)$. One difference, however, is in that the
CRs from the distant source are pitch-angle scattered on an anisotropic
Goldreich-Shridhar wave spectrum towards the minimum of $f_{1}\left(\mu\right)$
at $\mu=1$. In the I-K turbulence, inferred earlier in this paper,
the small scale anisotropy would be an addition of the second eigenfunction
$f_{2}$, to the larger-scale component $f_{1}$. This composition
requires a much closer source than any of the known SNRs, since the
$f_{2}$ component would have decayed at such long distance.

Nevertheless, both in the nearby shock and SNR scenarios, the small scale anisotropy
forms in the source's loss-cone, that is at the minimum of $f_{1}\left(\mu\right)$.
(Observe a decrease in $f_{1}$ at $\mu=\chi=1$, in Fig.~\ref{fig:The-first-three}).
The loss cone in momentum space develops at the dimensionless distance
$\zeta\lesssim1/\lambda_{1}$ \emph{from the source}, because particles
moving close to the magnetic axis are poorly scattered and escape
along the field. So, apart from the almost isotropic $f_{0},$ the
$f_{1}$ component dominates the solution at $\zeta\gtrsim1/\lambda_{1}$.
In the range $1/\lambda_{2}<\zeta<1/\lambda_{1}$, both $f_{1}$ and
$f_{2}$ may be visible, depending on the source angular profile.
Adding them up we obtain a narrow excess in a broader deficit area
on the intensity map, around the field direction. This appears to
be consistent with the intensity map shown in Fig.~11 of \citet{Abeysekara2019}.
But should it always be? 

Indeed, the minimum of $f_{1}$ is not necessarily the minimum on
the overall intensity map. Evidently, $f_{1}$ and $f_{2}$ have a
common isolated source, while the overall large-scale anisotropy is
likely to originate from a stronger background CR component. It may
have a minimum on the opposite side of the field line, for example.
This is possible if the background CRs come from the Galactic center
direction, which is roughly opposite to the heliotail. The background
CR may also be formed by distributed CR sources, each of which having
an individual loss cone, not seen at the heliosphere position for
the reasons described above. Diamagnetic and curvature drifts in a
possibly complicated magnetic environment around the heliosphere may
further distort and mix up the CR background particles. 

Although the above straightforward considerations speak for not to
require the coincidence of the global intensity minimum with the sharp
excess, it has been brought up as an issue by \citet{Ahlers2017}
in their discussion of the model of \citet{Milagro10}. Note that
this issue has already been addressed in Sect.~6 of this reference.
More interestingly, as we noted, the intensity map in Fig.~11 presented
by \citet{Abeysekara2019} can be interpreted as a narrow excess in
a broad deficit near the point $-\boldsymbol{B}_{\text{IMF}}$ (relatively
narrow red ``island'' surrounded by the blue ``sea''). However, the slice through
this map near longitude $\delta=-20^{\circ}$, shown in Fig.~8 of their
paper, seems to be at a flat top of the intensity curve. It is also
worth noting that a clear excess in the slice is seen in the HAWC,
but not in the IceCube data set. Going back to the Milagro results,
a possibly related sharp excess is between the minimum and the maximum,
though closer to the latter. To make matters even more complicated,
the deficit around an excess may also be a data processing effect
\citep{Milagro08PRL}. 

To conclude this discussion, the available small-scale anisotropy
data are not sufficient to scrutinize the models proposed in \citet{Milagro10}
and in the present paper. At the same time, the sharp intensity growth
across the magnetic equator is consistent with the effect of a nearby
source combined with the I-K propagation regime. An interesting implication
of this behavior is that there should be very little mirroring across
the magnetic horizon. The turbulence should be largely incompressible
at the relevant scales in the heliospheric environment. The present model also 
constrains  the source distances required to observe 
both the intensity enhancement inside the loss-cone at $\mu=1$ and the step-wise 
increase across the magnetic horizon at $\mu=0$.   

Turning to the details omitted in Sect.~\ref{sec:Anisotropy-and-Magnetic},
the coefficients entering Eq.~(\ref{eq:fexpans}) are defined in a
standard fashion: 
\begin{equation}
C_{n}=\int_{-1}^{1}\chi^{3}Q_{1}\left(\chi\right)f_{n}\left(\chi\right)d\chi,\label{eq:Cn}
\end{equation}
Here $\left\{ f_{n}\left(\chi\right)\right\} _{n=0}^{\infty}$ is
an orthonormal system of eigenfunctions obtained from the following
spectral problem 
\begin{equation}
\frac{\partial}{\partial\chi}\left(1-\chi^{4}\right)\frac{\partial f}{\partial\chi}-\nu_{\perp}f=-\lambda\chi^{3}f\label{eq:SpProblem}
\end{equation}
with the inner product defined as follows 
\[
\left(f_{n},f_{m}\right)\equiv\int_{-1}^{1}f_{n}\left(\chi\right)f_{m}\left(\chi\right)\chi^{3}d\chi=\delta_{nm},
\]
while 
\[
\int_{-1}^{1}f_{n}\left(-\chi\right)f_{m}\left(\chi\right)\chi^{3}d\chi=0
\]
for all $n$ and $m$.

This spectral problem is set by a condition of regularity of $f_{n}$
at the singular points $\chi=\pm1$ in Eq.~(\ref{eq:SpProblem}).
No specific boundary condition is needed, which corresponds to the
so-called Weyl's limit point case for this Sturm-Liouville problem.
When obtaining the $\nu_{\perp}f$ term in Eq.~(\ref{eq:SpProblem})
by integrating across the flux tube from its axis to the boundary,
we make a simple assumption $\nu_{\perp}=const$, even though it must
depend on $\chi$. The flux tube boundary is where the pitch angle
scattering frequency $D$ drops, as the particle self-driven turbulence
drops beyond this point. Particles that cross this boundary are assumed
to escape along the field line, so that the turbulence is not driven
outside of the tube, as long as $\nu_{\perp}\ll1.$ A more realistic
determination of this quantity is not worthwhile for our purpose,
as it is likely to be small indeed, as we argued in Sect.~\ref{sec:Anisotropy-and-Magnetic}.
By virtue of the symmetry of Eq.~(\ref{eq:FPndim}), $f\left(-\zeta,\chi\right)=f\left(\zeta,-\chi\right)$
and the condition $\nu_{\perp}>0$, the discrete spectrum is also
symmetric and consists of an infinite sequence, $\pm\lambda_{0},\pm\lambda_{1},\pm\lambda_{2},\dots$
of eigenvalues ordered as follows: $0<\lambda_{0}<\lambda_{1}<\lambda_{2},\dots$
\citep{Richardson1918}. Our sign convention is that each $f_{n}\left(\chi\right)$
corresponds to $\lambda_{n}>0$, while the negative eigenvalues $-\lambda_{n}$
correspond to $f_{n}\left(-\chi\right)$.

\bibliography{}

\begin{thebibliography}{}
\expandafter\ifx\csname natexlab\endcsname\relax\def\natexlab#1{#1}\fi
\providecommand{\url}[1]{\href{#1}{#1}}
\providecommand{\dodoi}[1]{doi:~\href{http://doi.org/#1}{\nolinkurl{#1}}}
\providecommand{\doeprint}[1]{\href{http://ascl.net/#1}{\nolinkurl{http://ascl.net/#1}}}
\providecommand{\doarXiv}[1]{\href{https://arxiv.org/abs/#1}{\nolinkurl{https://arxiv.org/abs/#1}}}

\bibitem[{{Abdo} {et~al.}(2008){Abdo}, {Allen}, {Aune}, {Berley}, {Blaufuss},
  {Casanova}, {Chen}, {Dingus}, {Ellsworth}, {Fleysher}, {Fleysher},
  {Gonzalez}, {Goodman}, {Hoffman}, {H{\"u}ntemeyer}, {Kolterman}, {Lansdell},
  {Linnemann}, {McEnery}, {Mincer}, {Nemethy}, {Noyes}, {Pretz}, {Ryan},
  {Parkinson}, {Shoup}, {Sinnis}, {Smith}, {Sullivan}, {Vasileiou}, {Walker},
  {Williams}, \& {Yodh}}]{Milagro08PRL}
{Abdo}, A.~A., {Allen}, B., {Aune}, T., {et~al.} 2008, Physical Review Letters,
  101, 221101, \dodoi{10.1103/PhysRevLett.101.221101}

\bibitem[{Abeysekara {et~al.}(2019)Abeysekara, Alfaro, Alvarez, Arceo,
  Arteaga-Vel{\'a}zquez, Rojas, Belmont-Moreno, BenZvi, Brisbois,
  Capistr{\'a}n, {et~al.}}]{Abeysekara2019}
Abeysekara, A., Alfaro, R., Alvarez, C., {et~al.} 2019, The Astrophysical
  Journal, 871, 96

\bibitem[{{Ackermann} {et~al.}(2014){Ackermann}, {Ajello}, {Albert},
  {Allafort}, {Baldini}, {Barbiellini}, {Bastieri}, {Bechtol}, {Bellazzini},
  {Blandford}, {Bloom}, {Bonamente}, {Bottacini}, {Bouvier}, {Brandt},
  {Brigida}, {Bruel}, {Buehler}, {Buson}, {Caliandro}, {Cameron}, {Caraveo},
  {Cecchi}, {Charles}, {Chaves}, {Chekhtman}, {Chiang}, {Chiaro}, {Ciprini},
  {Claus}, {Cohen-Tanugi}, {Conrad}, {Cutini}, {Dalton}, {D'Ammando}, {de
  Angelis}, {de Palma}, {Dermer}, {Digel}, {Di Venere}, {do Couto e Silva},
  {Drell}, {Drlica-Wagner}, {Favuzzi}, {Fegan}, {Ferrara}, {Focke},
  {Franckowiak}, {Fukazawa}, {Funk}, {Fusco}, {Gargano}, {Gasparrini},
  {Germani}, {Giglietto}, {Giordano}, {Giroletti}, {Glanzman}, {Godfrey},
  {Gomez-Vargas}, {Grenier}, {Grove}, {Guiriec}, {Gustafsson}, {Hadasch},
  {Hanabata}, {Harding}, {Hayashida}, {Hayashi}, {Hewitt}, {Horan}, {Hou},
  {Hughes}, {Inoue}, {Jackson}, {Jogler}, {J{\'o}hannesson}, {Johnson},
  {Kamae}, {Kawano}, {Kn{\"o}dlseder}, {Kuss}, {Lande}, {Larsson}, {Latronico},
  {Longo}, {Loparco}, {Lovellette}, {Lubrano}, {Mayer}, {Mazziotta}, {McEnery},
  {Mehault}, {Michelson}, {Mitthumsiri}, {Mizuno}, {Moiseev}, {Monte},
  {Monzani}, {Morselli}, {Moskalenko}, {Murgia}, {Nemmen}, {Nuss}, {Ohsugi},
  {Okumura}, {Orienti}, {Orlando}, {Ormes}, {Paneque}, {Panetta}, {Perkins},
  {Pesce-Rollins}, {Piron}, {Pivato}, {Porter}, {Rain{\`o}}, {Rando},
  {Razzano}, {Razzaque}, {Reimer}, {Reimer}, {Ritz}, {Roth}, {Schaal},
  {Schulz}, {Sgr{\`o}}, {Siskind}, {Spandre}, {Spinelli}, {Strong},
  {Takahashi}, {Takeuchi}, {Thayer}, {Thayer}, {Thompson}, {Tibaldo},
  {Tinivella}, {Torres}, {Tosti}, {Troja}, {Tronconi}, {Usher},
  {Vandenbroucke}, {Vasileiou}, {Vianello}, {Vitale}, {Werner}, {Winer},
  {Wood}, {Wood}, {Yang}, \& {Fermi LAT Collaboration}}]{2014PhRvL.112o1103A}
{Ackermann}, M., {Ajello}, M., {Albert}, A., {et~al.} 2014, \prl, 112, 151103,
  \dodoi{10.1103/PhysRevLett.112.151103}

\bibitem[{{Adriani} {et~al.}(2011){Adriani}, {Barbarino}, {Bazilevskaya},
  {Bellotti}, {Boezio}, {Bogomolov}, {Bonechi}, {Bongi}, {Bonvicini},
  {Borisov}, {Bottai}, {Bruno}, {Cafagna}, {Campana}, {Carbone}, {Carlson},
  {Casolino}, {Castellini}, {Consiglio}, {De Pascale}, {De Santis}, {De
  Simone}, {Di Felice}, {Galper}, {Gillard}, {Grishantseva}, {Jerse},
  {Karelin}, {Koldashov}, {Krutkov}, {Kvashnin}, {Leonov}, {Malakhov},
  {Malvezzi}, {Marcelli}, {Mayorov}, {Menn}, {Mikhailov}, {Mocchiutti},
  {Monaco}, {Mori}, {Nikonov}, {Osteria}, {Palma}, {Papini}, {Pearce},
  {Picozza}, {Pizzolotto}, {Ricci}, {Ricciarini}, {Rossetto}, {Sarkar},
  {Simon}, {Sparvoli}, {Spillantini}, {Stozhkov}, {Vacchi}, {Vannuccini},
  {Vasilyev}, {Voronov}, {Yurkin}, {Wu}, {Zampa}, {Zampa}, \&
  {Zverev}}]{2011Sci...332...69A}
{Adriani}, O., {Barbarino}, G.~C., {Bazilevskaya}, G.~A., {et~al.} 2011, \sci,
  332, 69, \dodoi{10.1126/science.1199172}

\bibitem[{{Adriani} {et~al.}(2019){Adriani}, {Akaike}, {Asano}, {Asaoka},
  {Bagliesi}, {Berti}, {Bigongiari}, {Binns}, {Bonechi}, {Bongi}, {Brogi},
  {Bruno}, {Buckley}, {Cannady}, {Castellini}, {Checchia}, {Cherry},
  {Collazuol}, {di Felice}, {Ebisawa}, {Fuke}, {Guzik}, {Hams}, {Hasebe},
  {Hibino}, {Ichimura}, {Ioka}, {Ishizaki}, {Israel}, {Kasahara}, {Kataoka},
  {Kataoka}, {Katayose}, {Kato}, {Kawanaka}, {Kawakubo}, {Kohri},
  {Krawczynski}, {Krizmanic}, {Lomtadze}, {Maestro}, {Marrocchesi}, {Messineo},
  {Mitchell}, {Miyake}, {Moiseev}, {Mori}, {Mori}, {Mori}, {Motz}, {Munakata},
  {Murakami}, {Nakahira}, {Nishimura}, {de Nolfo}, {Okuno}, {Ormes}, {Ozawa},
  {Pacini}, {Palma}, {Papini}, {Penacchioni}, {Rauch}, {Ricciarini}, {Sakai},
  {Sakamoto}, {Sasaki}, {Shimizu}, {Shiomi}, {Sparvoli}, {Spillantini},
  {Stolzi}, {Suh}, {Sulaj}, {Takahashi}, {Takayanagi}, {Takita}, {Tamura},
  {Terasawa}, {Tomida}, {Torii}, {Tsunesada}, {Uchihori}, {Ueno}, {Vannuccini},
  {Wefel}, {Yamaoka}, {Yanagita}, {Yoshida}, {Yoshida}, \& {Calet
  Collaboration}}]{2019PhRvL.122r1102A}
{Adriani}, O., {Akaike}, Y., {Asano}, K., {et~al.} 2019, \prl, 122, 181102,
  \dodoi{10.1103/PhysRevLett.122.181102}

\bibitem[{{Aguilar} {et~al.}(2015{\natexlab{a}}){Aguilar}, {Aisa}, {Alpat},
  {Alvino}, {Ambrosi}, {Andeen}, {Arruda}, {Attig}, {Azzarello}, {Bachlechner},
  \& et~al.}]{2015PhRvL.114q1103A}
{Aguilar}, M., {Aisa}, D., {Alpat}, B., {et~al.} 2015{\natexlab{a}}, \prl, 114,
  171103, \dodoi{10.1103/PhysRevLett.114.171103}

\bibitem[{{Aguilar} {et~al.}(2015{\natexlab{b}}){Aguilar}, {Aisa}, {Alpat},
  {Alvino}, {Ambrosi}, {Andeen}, {Arruda}, {Attig}, {Azzarello}, {Bachlechner},
  \& et~al.}]{2015PhRvL.115u1101A}
---. 2015{\natexlab{b}}, \prl, 115, 211101,
  \dodoi{10.1103/PhysRevLett.115.211101}

\bibitem[{{Aguilar} {et~al.}(2017){Aguilar}, {Ali Cavasonza}, {Alpat},
  {Ambrosi}, {Arruda}, {Attig}, {Aupetit}, {Azzarello}, {Bachlechner}, {Barao},
  {Barrau}, {Barrin}, {Bartoloni}, {Basara}, {Ba{\textcommabelow
  s}e{\v{g}}mez-du Pree}, {Battarbee}, {Battiston}, {Becker}, {Behlmann},
  {Beischer}, {Berdugo}, {Bertucci}, {Bindel}, {Bindi}, {de Boer}, {Bollweg},
  {Bonnivard}, {Borgia}, {Boschini}, {Bourquin}, {Bueno}, {Burger}, {Burger},
  {Cadoux}, {Cai}, {Capell}, {Caroff}, {Casaus}, {Castellini}, {Cervelli},
  {Chae}, {Chang}, {Chen}, {Chen}, {Chen}, {Cheng}, {Chou}, {Choumilov},
  {Choutko}, {Chung}, {Clark}, {Clavero}, {Coignet}, {Consolandi}, {Contin},
  {Corti}, {Creus}, {Crispoltoni}, {Cui}, {Dadzie}, {Dai}, {Datta}, {Delgado},
  {Della Torre}, {Demakov}, {Demirk{\"o}z}, {Derome}, {Di Falco}, {Dimiccoli},
  {D{\'\i}az}, {von Doetinchem}, {Dong}, {Donnini}, {Duranti}, {D'Urso},
  {Egorov}, {Eline}, {Eronen}, {Feng}, {Fiandrini}, {Fisher}, {Formato},
  {Galaktionov}, {Gallucci}, {Garc{\'\i}a-L{\'o}pez}, {Gargiulo}, {Gast},
  {Gebauer}, {Gervasi}, {Ghelfi}, {Giovacchini}, {G{\'o}mez-Coral}, {Gong},
  {Goy}, {Grabski}, {Grandi}, {Graziani}, {Guo}, {Haino}, {Han}, {He}, {Heil},
  {Hoffman}, {Hsieh}, {Huang}, {Huang}, {Huh}, {Incagli}, {Ionica}, {Jang},
  {Jia}, {Jinchi}, {Kang}, {Kanishev}, {Khiali}, {Kim}, {Kim}, {Kirn}, {Konak},
  {Kounina}, {Kounine}, {Koutsenko}, {Kulemzin}, {La Vacca}, {Laudi},
  {Laurenti}, {Lazzizzera}, {Lebedev}, {Lee}, {Lee}, {Leluc}, {Li}, {Li}, {Li},
  {Li}, {Li}, {Li}, {Li}, {Lim}, {Lin}, {Lipari}, {Lippert}, {Liu}, {Liu},
  {Lordello}, {Lu}, {Lu}, {Luebelsmeyer}, {Luo}, {Luo}, {Lyu}, {Machate},
  {Ma{\~n}{\'a}}, {Mar{\'\i}n}, {Martin}, {Mart{\'\i}nez}, {Masi}, {Maurin},
  {Menchaca-Rocha}, {Meng}, {Mikuni}, {Mo}, {Mott}, {Nelson}, {Ni}, {Nikonov},
  {Nozzoli}, {Oliva}, {Orcinha}, {Palmonari}, {Palomares}, {Paniccia},
  {Pauluzzi}, {Pensotti}, {Perrina}, {Phan}, {Picot-Clemente}, {Pilo},
  {Pizzolotto}, {Plyaskin}, {Pohl}, {Poireau}, {Quadrani}, {Qi}, {Qin}, {Qu},
  {R{\"a}ih{\"a}}, {Rancoita}, {Rapin}, {Ricol}, {Rosier-Lees}, {Rozhkov},
  {Rozza}, {Sagdeev}, {Schael}, {Schmidt}, {Schulz von Dratzig}, {Schwering},
  {Seo}, {Shan}, {Shi}, {Siedenburg}, {Son}, {Song}, {Tacconi}, {Tang}, {Tang},
  {Tescaro}, {Ting}, {Ting}, {Tomassetti}, {Torsti}, {T{\"u}rko{\v{g}}lu},
  {Urban}, {Vagelli}, {Valente}, {Valtonen}, {V{\'a}zquez Acosta}, {Vecchi},
  {Velasco}, {Vialle}, {Vitale}, {Vitillo}, {Wang}, {Wang}, {Wang}, {Wang},
  {Wang}, {Wang}, {Wei}, {Weng}, {Whitman}, {Wu}, {Wu}, {Xiong}, {Xu}, {Yan},
  {Yang}, {Yang}, {Yang}, {Yi}, {Yu}, {Yu}, {Zannoni}, {Zeissler}, {Zhang},
  {Zhang}, {Zhang}, {Zhang}, {Zhang}, {Zhang}, {Zheng}, {Zhuang}, {Zhukov},
  {Zichichi}, {Zimmermann}, {Zuccon}, \& {AMS
  Collaboration}}]{2017PhRvL.119y1101A}
{Aguilar}, M., {Ali Cavasonza}, L., {Alpat}, B., {et~al.} 2017, \prl, 119,
  251101, \dodoi{10.1103/PhysRevLett.119.251101}

\bibitem[{{Aguilar} {et~al.}(2018{\natexlab{a}}){Aguilar}, {Ali Cavasonza},
  {Ambrosi}, {Arruda}, {Attig}, {Aupetit}, {Azzarello}, {Bachlechner}, {Barao},
  {Barrau}, {Barrin}, {Bartoloni}, {Basara}, {Ba{\textcommabelow
  s}e{\v{g}}mez-du Pree}, {Battarbee}, {Battiston}, {Becker}, {Behlmann},
  {Beischer}, {Berdugo}, {Bertucci}, {Bindel}, {Bindi}, {de Boer}, {Bollweg},
  {Bonnivard}, {Borgia}, {Boschini}, {Bourquin}, {Bueno}, {Burger}, {Burger},
  {Cadoux}, {Cai}, {Capell}, {Caroff}, {Casaus}, {Castellini}, {Cervelli},
  {Chae}, {Chang}, {Chen}, {Chen}, {Chen}, {Cheng}, {Chou}, {Choumilov},
  {Choutko}, {Chung}, {Clark}, {Clavero}, {Coignet}, {Consolandi}, {Contin},
  {Corti}, {Creus}, {Crispoltoni}, {Cui}, {Dadzie}, {Dai}, {Datta}, {Delgado},
  {Della Torre}, {Demirk{\"o}z}, {Derome}, {Di Falco}, {Dimiccoli},
  {D{\'\i}az}, {von Doetinchem}, {Dong}, {Donnini}, {Duranti}, {D'Urso},
  {Egorov}, {Eline}, {Eronen}, {Feng}, {Fiandrini}, {Fisher}, {Formato},
  {Galaktionov}, {Gallucci}, {Garc{\'\i}a-L{\'o}pez}, {Gargiulo}, {Gast},
  {Gebauer}, {Gervasi}, {Ghelfi}, {Giovacchini}, {G{\'o}mez-Coral}, {Gong},
  {Goy}, {Grabski}, {Grandi}, {Graziani}, {Guo}, {Haino}, {Han}, {He}, {Heil},
  {Hsieh}, {Huang}, {Huang}, {Huh}, {Incagli}, {Ionica}, {Jang}, {Jia},
  {Jinchi}, {Kang}, {Kanishev}, {Khiali}, {Kim}, {Kim}, {Kirn}, {Konak},
  {Kounina}, {Kounine}, {Koutsenko}, {Kulemzin}, {La Vacca}, {Laudi},
  {Laurenti}, {Lazzizzera}, {Lebedev}, {Lee}, {Lee}, {Leluc}, {Li}, {Li}, {Li},
  {Li}, {Li}, {Li}, {Li}, {Lim}, {Lin}, {Lipari}, {Lippert}, {Liu}, {Liu},
  {Lordello}, {Lu}, {Lu}, {Luebelsmeyer}, {Luo}, {Luo}, {Lyu}, {Machate},
  {Ma{\~n}{\'a}}, {Mar{\'\i}n}, {Martin}, {Mart{\'\i}nez}, {Masi}, {Maurin},
  {Menchaca-Rocha}, {Meng}, {Mikuni}, {Mo}, {Mott}, {Nelson}, {Ni}, {Nikonov},
  {Nozzoli}, {Oliva}, {Orcinha}, {Palermo}, {Palmonari}, {Palomares},
  {Paniccia}, {Pauluzzi}, {Pensotti}, {Perrina}, {Phan}, {Picot-Clemente},
  {Pilo}, {Pizzolotto}, {Plyaskin}, {Pohl}, {Poireau}, {Quadrani}, {Qi}, {Qin},
  {Qu}, {R{\"a}ih{\"a}}, {Rancoita}, {Rapin}, {Ricol}, {Rosier-Lees},
  {Rozhkov}, {Rozza}, {Sagdeev}, {Schael}, {Schmidt}, {Schulz von Dratzig},
  {Schwering}, {Seo}, {Shan}, {Shi}, {Siedenburg}, {Son}, {Song}, {Tacconi},
  {Tang}, {Tang}, {Tescaro}, {Ting}, {Ting}, {Tomassetti}, {Torsti},
  {T{\"u}rko{\v{g}}lu}, {Urban}, {Vagelli}, {Valente}, {Valtonen}, {V{\'a}zquez
  Acosta}, {Vecchi}, {Velasco}, {Vialle}, {Vitale}, {Wang}, {Wang}, {Wang},
  {Wang}, {Wang}, {Wang}, {Wei}, {Weng}, {Whitman}, {Wu}, {Wu}, {Xiong}, {Xu},
  {Yan}, {Yang}, {Yang}, {Yang}, {Yi}, {Yu}, {Yu}, {Zannoni}, {Zeissler},
  {Zhang}, {Zhang}, {Zhang}, {Zhang}, {Zhang}, {Zhang}, {Zheng}, {Zhuang},
  {Zhukov}, {Zichichi}, {Zimmermann}, {Zuccon}, \& {AMS
  Collaboration}}]{2018PhRvL.120b1101A}
{Aguilar}, M., {Ali Cavasonza}, L., {Ambrosi}, G., {et~al.} 2018{\natexlab{a}},
  \prl, 120, 021101, \dodoi{10.1103/PhysRevLett.120.021101}

\bibitem[{{Aguilar} {et~al.}(2018{\natexlab{b}}){Aguilar}, {Ali Cavasonza},
  {Alpat}, {Ambrosi}, {Arruda}, {Attig}, {Aupetit}, {Azzarello}, {Bachlechner},
  {Barao}, {Barrau}, {Barrin}, {Bartoloni}, {Basara}, {Ba{\textcommabelow
  s}e{\v{g}}mez-du Pree}, {Battarbee}, {Battiston}, {Becker}, {Behlmann},
  {Beischer}, {Berdugo}, {Bertucci}, {Bindel}, {Bindi}, {de Boer}, {Bollweg},
  {Bonnivard}, {Borgia}, {Boschini}, {Bourquin}, {Bueno}, {Burger}, {Burger},
  {Cai}, {Capell}, {Caroff}, {Casaus}, {Castellini}, {Cervelli}, {Chang},
  {Chen}, {Chen}, {Chen}, {Chen}, {Cheng}, {Chou}, {Choumilov}, {Choutko},
  {Chung}, {Clark}, {Clavero}, {Coignet}, {Consolandi}, {Contin}, {Corti},
  {Creus}, {Crispoltoni}, {Cui}, {Dadzie}, {Dai}, {Datta}, {Delgado}, {Della
  Torre}, {Demirk{\"o}z}, {Derome}, {Di Falco}, {Dimiccoli}, {D{\'\i}az}, {von
  Doetinchem}, {Dong}, {Donnini}, {Duranti}, {Egorov}, {Eline}, {Eronen},
  {Feng}, {Fiandrini}, {Fisher}, {Formato}, {Galaktionov}, {Gallucci},
  {Garc{\'\i}a-L{\'o}pez}, {Gargiulo}, {Gast}, {Gebauer}, {Gervasi}, {Ghelfi},
  {Giovacchini}, {G{\'o}mez-Coral}, {Gong}, {Goy}, {Grabski}, {Grandi},
  {Graziani}, {Guo}, {Haino}, {Han}, {He}, {Heil}, {Hsieh}, {Huang}, {Huang},
  {Incagli}, {Jia}, {Jinchi}, {Kanishev}, {Khiali}, {Kirn}, {Konak}, {Kounina},
  {Kounine}, {Koutsenko}, {Kulemzin}, {La Vacca}, {Laudi}, {Laurenti},
  {Lazzizzera}, {Lebedev}, {Lee}, {Lee}, {Leluc}, {Li}, {Li}, {Li}, {Li}, {Li},
  {Li}, {Lin}, {Lipari}, {Lippert}, {Liu}, {Liu}, {Liu}, {Lordello}, {Lu},
  {Lu}, {Luebelsmeyer}, {Luo}, {Luo}, {Lyu}, {Machate}, {Ma{\~n}{\'a}},
  {Mar{\'\i}n}, {Martin}, {Mart{\'\i}nez}, {Masi}, {Maurin}, {Menchaca-Rocha},
  {Meng}, {Mikuni}, {Mo}, {Mott}, {Mussolin}, {Nelson}, {Ni}, {Nikonov},
  {Nozzoli}, {Oliva}, {Orcinha}, {Palermo}, {Palmonari}, {Palomares},
  {Paniccia}, {Pauluzzi}, {Pensotti}, {Perrina}, {Phan}, {Picot-Clemente},
  {Pilo}, {Plyaskin}, {Pohl}, {Poireau}, {Quadrani}, {Qi}, {Qin}, {Qu},
  {R{\"a}ih{\"a}}, {Rancoita}, {Rapin}, {Ricol}, {Rosier-Lees}, {Rozhkov},
  {Rozza}, {Sagdeev}, {Schael}, {Schmidt}, {Schulz von Dratzig}, {Schwering},
  {Seo}, {Shan}, {Shi}, {Siedenburg}, {Song}, {Tacconi}, {Tang}, {Tang},
  {Tescaro}, {Tian}, {Ting}, {Ting}, {Tomassetti}, {Torsti}, {Urban},
  {Vagelli}, {Valente}, {Valtonen}, {V{\'a}zquez Acosta}, {Vecchi}, {Velasco},
  {Vialle}, {Wang}, {Wang}, {Wang}, {Wang}, {Wang}, {Wang}, {Wei}, {Wei},
  {Weng}, {Whitman}, {Wu}, {Xiong}, {Xu}, {Yan}, {Yang}, {Yang}, {Yi}, {Yu},
  {Yu}, {Zannoni}, {Zeissler}, {Zhang}, {Zhang}, {Zhang}, {Zhang}, {Zhang},
  {Zhang}, {Zheng}, {Zhuang}, {Zhukov}, {Zichichi}, {Zimmermann}, {Zuccon}, \&
  {AMS Collaboration}}]{2018PhRvL.121e1103A}
{Aguilar}, M., {Ali Cavasonza}, L., {Alpat}, B., {et~al.} 2018{\natexlab{b}},
  \prl, 121, 051103, \dodoi{10.1103/PhysRevLett.121.051103}

\bibitem[{Aguilar {et~al.}(2020)Aguilar, Ali~Cavasonza, Ambrosi, Arruda, Attig,
  Barao, Barrin, Bartoloni, Ba\ifmmode \mbox{\c{s}}\else \c{s}\fi{}e\ifmmode
  \breve{g}\else \u{g}\fi{}mez-du Pree, Battiston, Becker, Behlmann, Beischer,
  Berdugo, Bertucci, Bindi, de~Boer, Bollweg, Borgia, Boschini, Bourquin,
  Bueno, Burger, Burger, Burmeister, Cai, Capell, Casaus, Castellini, Cervelli,
  Chang, Chen, Chen, Chen, Cheng, Chou, Chouridou, Choutko, Chung, Clark,
  Coignet, Consolandi, Contin, Corti, Cui, Dadzie, Dai, Delgado, Della~Torre,
  Demirk\"oz, Derome, Di~Falco, Di~Felice, D\'{\i}az, Dimiccoli, von
  Doetinchem, Dong, Donnini, Duranti, Egorov, Eline, Feng, Fiandrini, Fisher,
  Formato, Freeman, Galaktionov, G\'amez, Garc\'{\i}a-L\'opez, Gargiulo, Gast,
  Gebauer, Gervasi, Giovacchini, G\'omez-Coral, Gong, Goy, Grabski, Grandi,
  Graziani, Guo, Haino, Han, Hashmani, He, Heber, Hsieh, Hu, Huang, Incagli,
  Jang, Jia, Jinchi, Kanishev, Khiali, Kim, Kirn, Konyushikhin, Kounina,
  Kounine, Koutsenko, Kuhlman, Kulemzin, La~Vacca, Laudi, Laurenti, Lazzizzera,
  Lebedev, Lee, Lee, Li, Li, Li, Li, Li, Li, Light, Lin, Lippert, Liu, Lu, Lu,
  Luebelsmeyer, Luo, Lyu, Machate, Ma\~n\'a, Mar\'{\i}n, Marquardt, Martin,
  Mart\'{\i}nez, Masi, Maurin, Menchaca-Rocha, Meng, Mo, Molero, Mott,
  Mussolin, Ni, Nikonov, Nozzoli, Oliva, Orcinha, Palermo, Palmonari, Paniccia,
  Pashnin, Pauluzzi, Pensotti, Phan, Piandani, Plyaskin, Poluianov, Qi, Qin,
  Qu, Quadrani, Rancoita, Rapin, Reina~Conde, Rosier-Lees, Rozhkov, Rozza,
  Sagdeev, Schael, Schmidt, Schulz~von Dratzig, Schwering, Seo, Shan, Shi,
  Siedenburg, Solano, Sonnabend, Song, Sun, Sun, Tacconi, Tang, Tang, Tian,
  Ting, Ting, Tomassetti, Torsti, T\"uys\"uz, Urban, Usoskin, Vagelli, Vainio,
  Valente, Valtonen, V\'azquez~Acosta, Vecchi, Velasco, Vialle, Wallmann, Wang,
  Wang, Wang, Wang, Wang, Wang, Wei, Weng, Wu, Xiong, Xu, Yan, Yang, Yi, Yu,
  Yu, Zannoni, Zhang, Zhang, Zhang, Zhang, Zhang, Zhao, Zheng, Zhuang, Zhukov,
  Zichichi, Zimmermann, \& Zuccon}]{PhysRevLett.124.211102}
Aguilar, M., Ali~Cavasonza, L., Ambrosi, G., {et~al.} 2020, \prl, 124, 211102,
  \dodoi{10.1103/PhysRevLett.124.211102}

\bibitem[{Aguilar {et~al.}(2021)Aguilar, Cavasonza, Allen, Alpat, Ambrosi,
  Arruda, Attig, Barao, Barrin, Bartoloni, Ba\ifmmode \mbox{\c{s}}\else
  \c{s}\fi{}e\ifmmode \breve{g}\else \u{g}\fi{}mez-du Pree, Battiston,
  Behlmann, Beischer, Berdugo, Bertucci, Bindi, de~Boer, Bollweg, Borgia,
  Boschini, Bourquin, Bueno, Burger, Burger, Burmeister, Cai, Capell, Casaus,
  Castellini, Cervelli, Chang, Chen, Chen, Chen, Chen, Cheng, Chou, Chouridou,
  Choutko, Chung, Clark, Coignet, Consolandi, Contin, Corti, Cui, Dadzie,
  Delgado, Della~Torre, Demirk\"oz, Derome, Di~Falco, Di~Felice, D\'{\i}az,
  Dimiccoli, von Doetinchem, Dong, Donnini, Duranti, Egorov, Eline, Feng,
  Fiandrini, Fisher, Formato, Freeman, Galaktionov, G\'amez,
  Garc\'{\i}a-L\'opez, Gargiulo, Gast, Gervasi, Giovacchini, G\'omez-Coral,
  Gong, Goy, Grabski, Grandi, Graziani, Haino, Han, Hashmani, He, Heber, Hsieh,
  Hu, Incagli, Jang, Jia, Jinchi, Kanishev, Khiali, Kim, Kirn, Konyushikhin,
  Kounina, Kounine, Koutsenko, Kuhlman, Kulemzin, La~Vacca, Laudi, Laurenti,
  Lazzizzera, Lebedev, Lee, Lee, Li, Li, Li, Li, Li, Li, Liang, Light, Lin,
  Lippert, Liu, Liu, Lu, Lu, Luebelsmeyer, Luo, Luo, Lyu, Machate, Ma\~n\'a,
  Mar\'{\i}n, Marquardt, Martin, Mart\'{\i}nez, Masi, Maurin, Menchaca-Rocha,
  Meng, Mikhailov, Mo, Molero, Mott, Mussolin, Negrete, Nikonov, Nozzoli,
  Oliva, Orcinha, Palermo, Palmonari, Paniccia, Pashnin, Pauluzzi, Pensotti,
  Phan, Piandani, Plyaskin, Poluianov, Qin, Qu, Quadrani, Rancoita, Rapin,
  Conde, Robyn, Rosier-Lees, Rozhkov, Rozza, Sagdeev, Schael, von Dratzig,
  Schwering, Seo, Shakfa, Shan, Siedenburg, Solano, Song, Song, Sonnabend,
  Strigari, Su, Sun, Sun, Tacconi, Tang, Tang, Tian, Ting, Ting, Tomassetti,
  Torsti, T\"uys\"uz, Urban, Usoskin, Vagelli, Vainio, Valencia-Otero, Valente,
  Valtonen, V\'azquez~Acosta, Vecchi, Velasco, Vialle, Wang, Wang, Wang, Wang,
  Wang, Wang, Wang, Wang, Wang, Wei, Weng, Wu, Xiong, Xu, Yan, Yang, Yashin,
  Yi, Yu, Yu, Zannoni, Zhang, Zhang, Zhang, Zhang, Zhang, Zhao, Zheng, Zheng,
  Zhuang, Zhukov, Zichichi, Zimmermann, \& Zuccon}]{PhysRevLett.126.041104}
Aguilar, M., Cavasonza, L.~A., Allen, M.~S., {et~al.} 2021, Phys. Rev. Lett.,
  126, 041104, \dodoi{10.1103/PhysRevLett.126.041104}

\bibitem[{Ahlers \& Mertsch(2017)}]{Ahlers2017}
Ahlers, M., \& Mertsch, P. 2017, Progress in Particle and Nuclear Physics, 94,
  184

\bibitem[{{Ahn} {et~al.}(2006){Ahn}, {Seo}, {Adams}, {Bashindzhagyan},
  {Batkov}, {Chang}, {Christl}, {Fazely}, {Ganel}, {Gunasingha}, {Guzik},
  {Isbert}, {Kim}, {Kouznetsov}, {Panasyuk}, {Panov}, {Schmidt}, {Sina},
  {Sokolskaya}, {Wang}, {Wefel}, {Wu}, \& {Zatsepin}}]{ATIC06}
{Ahn}, H.~S., {Seo}, E.~S., {Adams}, J.~H., {et~al.} 2006, Advances in Space
  Research, 37, 1950, \dodoi{10.1016/j.asr.2005.09.031}

\bibitem[{{Ahn} {et~al.}(2008){Ahn}, {Allison}, {Bagliesi}, {Beatty},
  {Bigongiari}, {Boyle}, {Brand t}, {Childers}, {Conklin}, {Coutu},
  {Duvernois}, {Ganel}, {Han}, {Hyun}, {Jeon}, {Kim}, {Lee}, {Lee}, {Lutz},
  {Maestro}, {Malinin}, {Marrocchesi}, {Minnick}, {Mognet}, {Nam}, {Nutter},
  {Park}, {Park}, {Seo}, {Sina}, {Swordy}, {Wakely}, {Wu}, {Yang}, {Yoon},
  {Zei}, \& {Zinn}}]{2008APh....30..133A}
{Ahn}, H.~S., {Allison}, P.~S., {Bagliesi}, M.~G., {et~al.} 2008, Astroparticle
  Physics, 30, 133, \dodoi{10.1016/j.astropartphys.2008.07.010}

\bibitem[{{Ahn} {et~al.}(2010){Ahn}, {Allison}, {Bagliesi}, {Beatty},
  {Bigongiari}, {Childers}, {Conklin}, {Coutu}, {DuVernois}, {Ganel}, {Han},
  {Jeon}, {Kim}, {Lee}, {Lutz}, {Maestro}, {Malinin}, {Marrocchesi}, {Minnick},
  {Mognet}, {Nam}, {Nam}, {Nutter}, {Park}, {Park}, {Seo}, {Sina}, {Wu},
  {Yang}, {Yoon}, {Zei}, \& {Zinn}}]{2010ApJ...714L..89A}
{Ahn}, H.~S., {Allison}, P., {Bagliesi}, M.~G., {et~al.} 2010, \apjl, 714, L89,
  \dodoi{10.1088/2041-8205/714/1/L89}

\bibitem[{{Aloisio} {et~al.}(2015){Aloisio}, {Blasi}, \&
  {Serpico}}]{AloisioKink_2015}
{Aloisio}, R., {Blasi}, P., \& {Serpico}, P.~D. 2015, \aap, 583, A95,
  \dodoi{10.1051/0004-6361/201526877}

\bibitem[{{Amenomori} {et~al.}(2017){Amenomori}, {Bi}, {Chen}, {Chen}, {Chen},
  {Cui}, {Danzengluobu}, {Ding}, {Feng}, {Feng}, {Feng}, {Gou}, {Guo}, {He},
  {He}, {Hibino}, {Hotta}, {Hu}, {Hu}, {Huang}, {Jia}, {Jiang}, {Kajino},
  {Kasahara}, {Katayose}, {Kato}, {Kawata}, {Kozai}, {Labaciren}, {Le}, {Li},
  {Li}, {Li}, {Liu}, {Liu}, {Liu}, {Lu}, {Meng}, {Miyazaki}, {Mizutani},
  {Munakata}, {Nakajima}, {Nakamura}, {Nanjo}, {Nishizawa}, {Niwa}, {Ohnishi},
  {Ohta}, {Ozawa}, {Qian}, {Qu}, {Saito}, {Saito}, {Sakata}, {Sako}, {Shao},
  {Shibata}, {Shiomi}, {Shirai}, {Sugimoto}, {Takita}, {Tan}, {Tateyama},
  {Torii}, {Tsuchiya}, {Udo}, {Wang}, {Wu}, {Xue}, {Yamamoto}, {Yamauchi},
  {Yang}, {Yuan}, {Yuda}, {Zhai}, {Zhang}, {Zhang}, {Zhang}, {Zhang}, {Zhang},
  {Zhang}, {Zhaxisangzhu}, {Zhou}, \& {Tibet AS{\ensuremath{\gamma}}
  Collaboration}}]{Amenomori2017}
{Amenomori}, M., {Bi}, X.~J., {Chen}, D., {et~al.} 2017, \apj, 836, 153,
  \dodoi{10.3847/1538-4357/836/2/153}

\bibitem[{An {et~al.}(2019)An, Asfandiyarov, Azzarello, Bernardini, Bi, Cai,
  Chang, Chen, Chen, Chen, Chen, Cui, Cui, Dai, D{\textquoteright}Amone,
  De~Benedittis, De~Mitri, Di~Santo, Ding, Dong, Dong, Dong, Donvito, Droz,
  Duan, Duan, D{\textquoteright}Urso, Fan, Fan, Fang, Feng, Feng, Fusco, Gallo,
  Gan, Gao, Gargano, Gong, Gong, Guo, Guo, Guo, Han, Hu, Huang, Huang, Huang,
  Ionica, Jiang, Jin, Kong, Lei, Li, Li, Li, Li, Li, Liang, Liang, Liao, Liu,
  Liu, Liu, Liu, Liu, Liu, Loparco, Luo, Ma, Ma, Ma, Ma, Ma, Marsella,
  Mazziotta, Mo, Niu, Pan, Peng, Peng, Qiao, Rao, Salinas, Shang, Shen, Shen,
  Shen, Song, Su, Su, Sun, Surdo, Teng, Tykhonov, Vitillo, Wang, Wang, Wang,
  Wang, Wang, Wang, Wang, Wang, Wang, Wang, Wang, Wang, Wang, Wei, Wei, Wei,
  Wen, Wu, Wu, Wu, Wu, Wu, Xi, Xia, Xu, Xu, Xu, Xu, Xue, Yang, Yang, Yang,
  Yang, Yao, Yu, Yuan, Yue, Zang, Zhang, Zhang, Zhang, Zhang, Zhang, Zhang,
  Zhang, Zhang, Zhang, Zhang, Zhang, Zhang, Zhang, Zhao, Zhao, Zhao, Zhou,
  Zhou, Zhu, Zhu, \& Zimmer}]{eaax3793}
An, Q., Asfandiyarov, R., Azzarello, P., {et~al.} 2019, \sa, 5,
  \dodoi{10.1126/sciadv.aax3793}

\bibitem[{{Atkin} {et~al.}(2018){Atkin}, {Bulatov}, {Dorokhov}, {Gorbunov},
  {Filippov}, {Grebenyuk}, {Karmanov}, {Kovalev}, {Kudryashov}, {Kurganov},
  {Merkin}, {Panov}, {Podorozhny}, {Polkov}, {Porokhovoy}, {Shumikhin},
  {Tkachenko}, {Tkachev}, {Turundaevskiy}, {Vasiliev}, \&
  {Voronin}}]{2018JETPL.108....5A}
{Atkin}, E., {Bulatov}, V., {Dorokhov}, V., {et~al.} 2018, \jetpl, 108, 5,
  \dodoi{10.1134/S0021364018130015}

\bibitem[{{Atkin} {et~al.}(2019){Atkin}, {Bulatov}, {Vasiliev}, {Voronin},
  {Gorbunov}, {Grebenyuk}, {Dorokhov}, {Karmanov}, {Kovalev}, {Kudryashov},
  {Kurganov}, {Merkin}, {Panov}, {Podorozhny}, {Polkov}, {Porokhovoi},
  {Sveshnikova}, {Tkachev}, {Tkachenko}, {Turundaevskiy}, {Filippov}, \&
  {Shumikhin}}]{2019ARep...63...66A}
{Atkin}, E.~V., {Bulatov}, V.~L., {Vasiliev}, O.~A., {et~al.} 2019, \ar, 63,
  66, \dodoi{10.1134/S1063772919010013}

\bibitem[{{Axford}(1972)}]{1972NASSP.308..609A}
{Axford}, W.~I. 1972, {The Interaction of the Solar Wind With the Interstellar
  Medium. NASA Special Publication}, ed. C.~P. {Sonett}, P.~J. {Coleman}, \&
  J.~M. {Wilcox}, Vol. 308, 609

\bibitem[{{Baines} \& {Armstrong}(2012)}]{2012ApJ...744..138B}
{Baines}, E.~K., \& {Armstrong}, J.~T. 2012, \apj, 744, 138,
  \dodoi{10.1088/0004-637X/744/2/138}

\bibitem[{{Bell}(1978)}]{Bell78}
{Bell}, A.~R. 1978, \mnras, 182, 147

\bibitem[{{Bell}(2004)}]{Bell04}
---. 2004, \mnras, 353, 550, \dodoi{10.1111/j.1365-2966.2004.08097.x}

\bibitem[{{Bergh{\"o}fer} \& {Breitschwerdt}(2002)}]{2002A&A...390..299B}
{Bergh{\"o}fer}, T.~W., \& {Breitschwerdt}, D. 2002, \aap, 390, 299,
  \dodoi{10.1051/0004-6361:20020627}

\bibitem[{{Binns} {et~al.}(2016){Binns}, {Israel}, {Christian}, {Cummings}, {de
  Nolfo}, {Lave}, {Leske}, {Mewaldt}, {Stone}, {von Rosenvinge}, \&
  {Wiedenbeck}}]{2016Sci...352..677B}
{Binns}, W.~R., {Israel}, M.~H., {Christian}, E.~R., {et~al.} 2016, \sci, 352,
  677, \dodoi{10.1126/science.aad6004}

\bibitem[{{Blandford} \& {Ostriker}(1978)}]{BlandOst78}
{Blandford}, R.~D., \& {Ostriker}, J.~P. 1978, \apjl, 221, L29,
  \dodoi{10.1086/182658}

\bibitem[{{Blasi} {et~al.}(2012){Blasi}, {Amato}, \&
  {Serpico}}]{2012PhRvL.109f1101B}
{Blasi}, P., {Amato}, E., \& {Serpico}, P.~D. 2012, \prl, 109, 061101,
  \dodoi{10.1103/PhysRevLett.109.061101}

\bibitem[{{Boschini} {et~al.}(2020{\natexlab{a}}){Boschini}, {Della Torre},
  {Gervasi}, {Grand i}, {J{\o}hannesson}, {La Vacca}, {Masi}, {Moskalenko},
  {Pensotti}, {Porter}, {Quadrani}, {Rancoita}, {Rozza}, \&
  {Tacconi}}]{2020ApJ...889..167B}
{Boschini}, M.~J., {Della Torre}, S., {Gervasi}, M., {et~al.}
  2020{\natexlab{a}}, \apj, 889, 167, \dodoi{10.3847/1538-4357/ab64f1}

\bibitem[{{Boschini} {et~al.}(2020{\natexlab{b}}){Boschini}, {Della Torre},
  {Gervasi}, {Grand i}, {J{\o}hannesson}, {La Vacca}, {Masi}, {Moskalenko},
  {Pensotti}, {Porter}, {Quadrani}, {Rancoita}, {Rozza}, \&
  {Tacconi}}]{Boschini_2020}
---. 2020{\natexlab{b}}, \apjs, 250, 27, \dodoi{10.3847/1538-4365/aba901}

\bibitem[{{Boschini} {et~al.}(2021){Boschini}, {Della Torre}, {Gervasi},
  {Grandi}, {Johannesson}, {La Vacca}, {Masi}, {Moskalenko}, {Pensotti},
  {Porter}, {Quadrani}, {Rancoita}, {Rozza}, \&
  {Tacconi}}]{2021arXiv210112735B}
---. 2021, arXiv e-prints, arXiv:2101.12735.
\newblock \doarXiv{2101.12735}

\bibitem[{{Breitschwerdt} {et~al.}(2016){Breitschwerdt}, {Feige}, {Schulreich},
  {Avillez}, {Dettbarn}, \& {Fuchs}}]{2016Natur.532...73B}
{Breitschwerdt}, D., {Feige}, J., {Schulreich}, M.~M., {et~al.} 2016, \nat,
  532, 73, \dodoi{10.1038/nature17424}

\bibitem[{{Breitschwerdt} \& {Schmutzler}(1999)}]{Breitschwerdt1999}
{Breitschwerdt}, D., \& {Schmutzler}, T. 1999, \aap, 347, 650.
\newblock \doarXiv{astro-ph/9902268}

\bibitem[{Burgasser {et~al.}(2015)Burgasser, Gillon, Melis, Bowler, Michelsen,
  Gagliuffi, Gelino, Jehin, Delrez, Manfroid, \& Blake}]{Burgasser_2015}
Burgasser, A.~J., Gillon, M., Melis, C., {et~al.} 2015, The Astronomical
  Journal, 149, 104, \dodoi{10.1088/0004-6256/149/3/104}

\bibitem[{Bykov {et~al.}(2019)Bykov, Petrov, Krassilchtchikov, Levenfish,
  Osipov, \& Pavlov}]{Bykov2019}
Bykov, A., Petrov, A., Krassilchtchikov, A., {et~al.} 2019, The Astrophysical
  Journal Letters, 876, L8

\bibitem[{{Cook} {et~al.}(2009){Cook}, {Berger}, {Faestermann}, {Herzog},
  {Knie}, {Korschinek}, {Poutivtsev}, {Rugel}, \&
  {Serefiddin}}]{2009LPI....40.1129C}
{Cook}, D.~L., {Berger}, E., {Faestermann}, T., {et~al.} 2009, in \lps, 1129

\bibitem[{Cowsik \& Madziwa-Nussinov(2016)}]{cowsik2016spectral}
Cowsik, R., \& Madziwa-Nussinov, T. 2016, The Astrophysical Journal, 827, 119

\bibitem[{Cox \& Helenius(2003)}]{Cox_2003}
Cox, D.~P., \& Helenius, L. 2003, The Astrophysical Journal, 583, 205,
  \dodoi{10.1086/344926}

\bibitem[{Crawford {et~al.}(1998)Crawford, Lallement, \& Welsh}]{Crawford1998}
Crawford, I., Lallement, R., \& Welsh, B. 1998, Monthly Notices of the Royal
  Astronomical Society, 300, 1181

\bibitem[{{Cummings} {et~al.}(2016){Cummings}, {Stone}, {Heikkila}, {Lal},
  {Webber}, {J{\'o}hannesson}, {Moskalenko}, {Orlando}, \&
  {Porter}}]{2016ApJ...831...18C}
{Cummings}, A.~C., {Stone}, E.~C., {Heikkila}, B.~C., {et~al.} 2016, \apj, 831,
  18, \dodoi{10.3847/0004-637X/831/1/18}

\bibitem[{{D'Arrest}(1847)}]{1847MNRAS...8...16D}
{D'Arrest}, M. 1847, \mnras, 8, 16, \dodoi{10.1093/mnras/8.1.16}

\bibitem[{de~Wit {et~al.}(2020)de~Wit, Krasnoselskikh, Bale, Bonnell, Bowen,
  Chen, Froment, Goetz, Harvey, Jagarlamudi, {et~al.}}]{deWit2020}
de~Wit, T.~D., Krasnoselskikh, V.~V., Bale, S.~D., {et~al.} 2020, The
  Astrophysical Journal Supplement Series, 246, 39

\bibitem[{Dehnen(1999)}]{Dehnen_1999}
Dehnen, W. 1999, The Astrophysical Journal, 524, L35, \dodoi{10.1086/312299}

\bibitem[{{Dieterich} {et~al.}(2018){Dieterich}, {Weinberger}, {Boss}, {Henry},
  {Jao}, {Gagn{\'e}}, {Astraatmadja}, {Thompson}, \&
  {Anglada-Escud{\'e}}}]{2018ApJ...865...28D}
{Dieterich}, S.~B., {Weinberger}, A.~J., {Boss}, A.~P., {et~al.} 2018, \apj,
  865, 28, \dodoi{10.3847/1538-4357/aadadc}

\bibitem[{{Dorfi}(1984)}]{Dorfi84}
{Dorfi}, E. 1984, Advances in Space Research, 4, 205,
  \dodoi{10.1016/0273-1177(84)90313-2}

\bibitem[{{Drury}(1983)}]{Drury83}
{Drury}, L.~O. 1983, Reports on Progress in Physics, 46, 973

\bibitem[{{Drury} \& {Falle}(1986)}]{DruryFal86}
{Drury}, L.~O.~C., \& {Falle}, S.~A.~E.~G. 1986, \mnras, 223, 353

\bibitem[{{Drury} \& {Strong}(2017)}]{DruryStrong2017}
{Drury}, L. O.~C., \& {Strong}, A.~W. 2017, \aap, 597, A117,
  \dodoi{10.1051/0004-6361/201629526}

\bibitem[{{Dupuy} {et~al.}(2019){Dupuy}, {Liu}, {Best}, {Mann}, {Tucker},
  {Zhang}, {Baraffe}, {Chabrier}, {Forveille}, {Metchev}, {Tremblin}, {Do},
  {Payne}, {Shappee}, {Bond}, {Cetre}, {Chun}, {Delorme}, {Jovanovic},
  {Lilley}, {Mawet}, {Ragland }, {Wetherell}, \&
  {Wizinowich}}]{2019AJ....158..174D}
{Dupuy}, T.~J., {Liu}, M.~C., {Best}, W. M.~J., {et~al.} 2019, \aj, 158, 174,
  \dodoi{10.3847/1538-3881/ab3cd1}

\bibitem[{{Fang} {et~al.}(2020){Fang}, {Bi}, \& {Yin}}]{Fang2020}
{Fang}, K., {Bi}, X.-J., \& {Yin}, P.-F. 2020, arXiv e-prints,
  arXiv:2003.13635.
\newblock \doarXiv{2003.13635}

\bibitem[{Ferri{\`e}re(2015)}]{ferriere2015interstellar}
Ferri{\`e}re, K. 2015, in Journal of Physics: Conference Series, Vol. 577,
  012008

\bibitem[{{Fimiani} {et~al.}(2012){Fimiani}, {Cook}, {Faestermann}, {Gomez
  Guzman}, {Hain}, {Herzog}, {Korschinek}, {Ligon}, {Ludwig}, {Park}, {Reedy},
  \& {Rugel}}]{2012LPI....43.1279F}
{Fimiani}, L., {Cook}, D.~L., {Faestermann}, T., {et~al.} 2012, in \lps, 1279

\bibitem[{{Fimiani} {et~al.}(2014){Fimiani}, {Cook}, {Faestermann}, {Gomez
  Guzman}, {Hain}, {Herzog}, {Knie}, {Korschinek}, {Ligon}, {Ludwig}, {Park},
  {Reedy}, \& {Rugel}}]{2014LPI....45.1778F}
{Fimiani}, L., {Cook}, D.~L., {Faestermann}, T., {et~al.} 2014, in \lps, 1778

\bibitem[{{Fornieri} {et~al.}(2020){Fornieri}, {Gaggero}, {Guberman},
  {Brahimi}, \& {Marcowith}}]{Fornieri2020}
{Fornieri}, O., {Gaggero}, D., {Guberman}, D., {Brahimi}, L., \& {Marcowith},
  A. 2020, arXiv e-prints, arXiv:2007.15321.
\newblock \doarXiv{2007.15321}

\bibitem[{Frisch {et~al.}(2015)Frisch, Berdyugin, Piirola, Magalhaes,
  Seriacopi, Wiktorowicz, Andersson, Funsten, McComas, Schwadron,
  {et~al.}}]{Frisch2015}
Frisch, P., Berdyugin, A., Piirola, V., {et~al.} 2015, The Astrophysical
  Journal, 814, 112

\bibitem[{{Frisch} {et~al.}(2011){Frisch}, {Redfield}, \&
  {Slavin}}]{2011ARA&A..49..237F}
{Frisch}, P.~C., {Redfield}, S., \& {Slavin}, J.~D. 2011, \araa, 49, 237,
  \dodoi{10.1146/annurev-astro-081710-102613}

\bibitem[{{Fry} {et~al.}(2015){Fry}, {Fields}, \&
  {Ellis}}]{2015ApJ...800...71F}
{Fry}, B.~J., {Fields}, B.~D., \& {Ellis}, J.~R. 2015, \apj, 800, 71,
  \dodoi{10.1088/0004-637X/800/1/71}

\bibitem[{Furno {et~al.}(2006)Furno, Intrator, Ryutov, Abbate,
  Madziwa-Nussinov, Light, Dorf, \& Lapenta}]{furno2006current}
Furno, I., Intrator, T., Ryutov, D., {et~al.} 2006, Physical review letters,
  97, 015002

\bibitem[{{Giacinti} {et~al.}(2018){Giacinti}, {Kachelrieẞ}, \&
  {Semikoz}}]{Giacinti2018}
{Giacinti}, G., {Kachelrieẞ}, M., \& {Semikoz}, D.~V. 2018, \jcap, 2018, 051,
  \dodoi{10.1088/1475-7516/2018/07/051}

\bibitem[{{Goldreich} \& {Sridhar}(1997)}]{goldr97}
{Goldreich}, P., \& {Sridhar}, S. 1997, \apj, 485, 680, \dodoi{10.1086/304442}

\bibitem[{{Grebenyuk} {et~al.}(2019{\natexlab{a}}){Grebenyuk}, {Karmanov},
  {Kovalev}, {Kudryashov}, {Kurganov}, {Panov}, {Podorozhny}, {Tkachenko},
  {Tkachev}, {Turundaevskiy}, {Vasiliev}, \& {Voronin}}]{2019AdSpR..64.2546G}
{Grebenyuk}, V., {Karmanov}, D., {Kovalev}, I., {et~al.} 2019{\natexlab{a}},
  \asr, 64, 2546, \dodoi{10.1016/j.asr.2019.10.004}

\bibitem[{{Grebenyuk} {et~al.}(2019{\natexlab{b}}){Grebenyuk}, {Karmanov},
  {Kovalev}, {Kudryashov}, {Kurganov}, {Panov}, {Podorozhny}, {Tkachenko},
  {Tkachev}, {Turundaevskiy}, {Vasiliev}, \& {Voronin}}]{2019AdSpR..64.2559G}
---. 2019{\natexlab{b}}, \asr, 64, 2559, \dodoi{10.1016/j.asr.2019.06.030}

\bibitem[{{Gry} \& {Jenkins}(2014)}]{Gry2014}
{Gry}, C., \& {Jenkins}, E.~B. 2014, \aap, 567, A58,
  \dodoi{10.1051/0004-6361/201323342}

\bibitem[{{Gry} \& {Jenkins}(2017)}]{Gry2017}
---. 2017, \aap, 598, A31, \dodoi{10.1051/0004-6361/201628987}

\bibitem[{{Hanusch} {et~al.}(2019){Hanusch}, {Liseykina}, \&
  {Malkov}}]{Hanusch2019ApJ}
{Hanusch}, A., {Liseykina}, T.~V., \& {Malkov}, M. 2019, \apj, 872, 108,
  \dodoi{10.3847/1538-4357/aafdae}

\bibitem[{Haverkorn {et~al.}(2008)Haverkorn, Brown, Gaensler, \&
  McClure-Griffiths}]{Haverkorn2008}
Haverkorn, M., Brown, J., Gaensler, B., \& McClure-Griffiths, N. 2008, The
  Astrophysical Journal, 680, 362

\bibitem[{{Kang} {et~al.}(1992){Kang}, {Jones}, \& {Ryu}}]{KangJR92}
{Kang}, H., {Jones}, T.~W., \& {Ryu}, D. 1992, \apj, 385, 193,
  \dodoi{10.1086/170927}

\bibitem[{Kennel(1988)}]{Kennel1988}
Kennel, C.~F. 1988, Journal of Geophysical Research: Space Physics, 93, 8545,
  \dodoi{10.1029/JA093iA08p08545}

\bibitem[{{Knie} {et~al.}(2004){Knie}, {Korschinek}, {Faestermann}, {Dorfi},
  {Rugel}, \& {Wallner}}]{2004PhRvL..93q1103K}
{Knie}, K., {Korschinek}, G., {Faestermann}, T., {et~al.} 2004, \prl, 93,
  171103, \dodoi{10.1103/PhysRevLett.93.171103}

\bibitem[{{Knie} {et~al.}(1999){Knie}, {Korschinek}, {Faestermann}, {Wallner},
  {Scholten}, \& {Hillebrandt}}]{1999PhRvL..83...18K}
---. 1999, \prl, 83, 18, \dodoi{10.1103/PhysRevLett.83.18}

\bibitem[{{Koll} {et~al.}(2019){Koll}, {Korschinek}, {Faestermann},
  {G{\'o}mez-Guzm{\'a}n}, {Kipfstuhl}, {Merchel}, \&
  {Welch}}]{2019PhRvL.123g2701K}
{Koll}, D., {Korschinek}, G., {Faestermann}, T., {et~al.} 2019, \prl, 123,
  072701, \dodoi{10.1103/PhysRevLett.123.072701}

\bibitem[{{Kulsrud} \& {Pearce}(1969)}]{KulsrNeutr69}
{Kulsrud}, R., \& {Pearce}, W.~P. 1969, \apj, 156, 445, \dodoi{10.1086/149981}

\bibitem[{Kuznetsov(2001)}]{Kuznetsov2001}
Kuznetsov, E.~A. 2001, Journal of Experimental \& Theoretical Physics, 93,
  1052.
\newblock \url{https://link.springer.com/article/10.1134/1.1427116}

\bibitem[{Lallement {et~al.}(2005)Lallement, Qu{\'e}merais, Bertaux, Ferron,
  Koutroumpa, \& Pellinen}]{Lallement2005}
Lallement, R., Qu{\'e}merais, E., Bertaux, J.-L., {et~al.} 2005, Science, 307,
  1447

\bibitem[{{Landau} \& {Lifshitz}(1987)}]{Landau_Fluid}
{Landau}, L.~D., \& {Lifshitz}, E.~M. 1987, {Fluid Mechanics} (Pergamon Press)

\bibitem[{{Linsky} {et~al.}(2019){Linsky}, {Redfield}, \&
  {Tilipman}}]{Linsky2019}
{Linsky}, J.~L., {Redfield}, S., \& {Tilipman}, D. 2019, \apj, 886, 41,
  \dodoi{10.3847/1538-4357/ab498a}

\bibitem[{{Lipari} \& {Vernetto}(2020)}]{2020APh...12002441L}
{Lipari}, P., \& {Vernetto}, S. 2020, \app, 120, 102441,
  \dodoi{10.1016/j.astropartphys.2020.102441}

\bibitem[{Livshits \& Tsytovich(1970)}]{Livshits1970}
Livshits, M., \& Tsytovich, V. 1970, Nuclear Fusion, 10, 241,
  \dodoi{10.1088/0029-5515/10/3/003}

\bibitem[{{Ludwig} {et~al.}(2016){Ludwig}, {Bishop}, {Egli}, {Chernenko},
  {Deneva}, {Faestermann}, {Famulok}, {Fimiani}, {G{\'o}mez-Guzm{\'a}n},
  {Hain}, {Korschinek}, {Hanzlik}, {Merchel}, \& {Rugel}}]{2016PNAS..113.9232L}
{Ludwig}, P., {Bishop}, S., {Egli}, R., {et~al.} 2016, \nas, 113, 9232,
  \dodoi{10.1073/pnas.1601040113}

\bibitem[{Luthardt \& Roessler(2017)}]{Luthardt2017FossilFR}
Luthardt, L., \& Roessler, R. 2017, Geology, 45, 279

\bibitem[{{Lutomirski} \& {Sudan}(1966)}]{Lutom66}
{Lutomirski}, R.~F., \& {Sudan}, R.~N. 1966, Physical Review, 147, 156,
  \dodoi{10.1103/PhysRev.147.156}

\bibitem[{{Malkov}(1998)}]{m98}
{Malkov}, M.~A. 1998, \pre, 58, 4911

\bibitem[{{Malkov}(2015)}]{M_PoP2015}
---. 2015, Physics of Plasmas, 22, 091505, \dodoi{10.1063/1.4928941}

\bibitem[{{Malkov} \& {Aharonian}(2019)}]{MalkovAharonian2019}
{Malkov}, M.~A., \& {Aharonian}, F.~A. 2019, \apj, 881, 2,
  \dodoi{10.3847/1538-4357/ab2c01}

\bibitem[{{Malkov} \& {Diamond}(2009)}]{MD09}
{Malkov}, M.~A., \& {Diamond}, P.~H. 2009, \apj, 692, 1571,
  \dodoi{10.1088/0004-637X/692/2/1571}

\bibitem[{{Malkov} {et~al.}(2010{\natexlab{a}}){Malkov}, {Diamond},
  {O'C.~Drury}, \& {Sagdeev}}]{Milagro10}
{Malkov}, M.~A., {Diamond}, P.~H., {O'C.~Drury}, L., \& {Sagdeev}, R.~Z.
  2010{\natexlab{a}}, \apj, 721, 750, \dodoi{10.1088/0004-637X/721/1/750}

\bibitem[{{Malkov} {et~al.}(2010{\natexlab{b}}){Malkov}, {Diamond}, \&
  {Sagdeev}}]{MDS10PPCF}
{Malkov}, M.~A., {Diamond}, P.~H., \& {Sagdeev}, R.~Z. 2010{\natexlab{b}},
  Plasma Physics and Controlled Fusion, 52, 124006,
  \dodoi{10.1088/0741-3335/52/12/124006}

\bibitem[{{Mamajek} {et~al.}(2015){Mamajek}, {Barenfeld}, {Ivanov}, {Kniazev},
  {V{\"a}is{\"a}nen}, {Beletsky}, \& {Boffin}}]{2015ApJ...800L..17M}
{Mamajek}, E.~E., {Barenfeld}, S.~A., {Ivanov}, V.~D., {et~al.} 2015, \apjl,
  800, L17, \dodoi{10.1088/2041-8205/800/1/L17}

\bibitem[{{McKenzie} \& {Voelk}(1982)}]{McKVlk82}
{McKenzie}, J.~F., \& {Voelk}, H.~J. 1982, \aap, 116, 191

\bibitem[{{Niu}(2020)}]{Niu2020}
{Niu}, J.-S. 2020, arXiv e-prints, arXiv:2009.00884.
\newblock \doarXiv{2009.00884}

\bibitem[{{Panov} {et~al.}(2009){Panov}, {Adams}, {Ahn}, {Bashinzhagyan},
  {Watts}, {Wefel}, {Wu}, {Ganel}, {Guzik}, {Zatsepin}, {Isbert}, {Kim},
  {Christl}, {Kouznetsov}, {Panasyuk}, {Seo}, {Sokolskaya}, {Chang}, {Schmidt},
  \& {Fazely}}]{2009BRASP..73..564P}
{Panov}, A.~D., {Adams}, J.~H., {Ahn}, H.~S., {et~al.} 2009, \brasp, 73, 564,
  \dodoi{10.3103/S1062873809050098}

\bibitem[{{Pogorelov} {et~al.}(2015){Pogorelov}, {Borovikov}, {Heerikhuisen},
  \& {Zhang}}]{Pogorelov2015}
{Pogorelov}, N.~V., {Borovikov}, S.~N., {Heerikhuisen}, J., \& {Zhang}, M.
  2015, \apjl, 812, L6, \dodoi{10.1088/2041-8205/812/1/L6}

\bibitem[{{Pouquet} {et~al.}(1976){Pouquet}, {Frisch}, \& {Leorat}}]{Pouquet76}
{Pouquet}, A., {Frisch}, U., \& {Leorat}, J. 1976, Journal of Fluid Mechanics,
  77, 321, \dodoi{10.1017/S0022112076002140}

\bibitem[{Pouquet {et~al.}(2018)Pouquet, Rosenberg, Stawarz, \&
  Marino}]{pouquet2018helicity}
Pouquet, A., Rosenberg, D., Stawarz, J.~E., \& Marino, R. 2018, Helicity
  dynamics, inverse and bi-directional cascades in fluid and MHD turbulence: A
  brief review.
\newblock \doarXiv{1807.03239}

\bibitem[{{Redfield} \& {Linsky}(2008)}]{Redfield2008}
{Redfield}, S., \& {Linsky}, J.~L. 2008, \apj, 673, 283, \dodoi{10.1086/524002}

\bibitem[{{Rhines}(1975)}]{Rhines1975}
{Rhines}, P.~B. 1975, Journal of Fluid Mechanics, 69, 417,
  \dodoi{10.1017/S0022112075001504}

\bibitem[{Richardson(1918)}]{Richardson1918}
Richardson, R. G.~D. 1918, American Journal of Mathematics, 40, 283.
\newblock \url{http://www.jstor.org/stable/2370485}

\bibitem[{{Rugel} {et~al.}(2009){Rugel}, {Faestermann}, {Knie}, {Korschinek},
  {Poutivtsev}, {Schumann}, {Kivel}, {G{\"u}nther-Leopold}, {Weinreich}, \&
  {Wohlmuther}}]{2009PhRvL.103g2502R}
{Rugel}, G., {Faestermann}, T., {Knie}, K., {et~al.} 2009, \prl, 103, 072502,
  \dodoi{10.1103/PhysRevLett.103.072502}

\bibitem[{Ryutov {et~al.}(2006)Ryutov, Furno, Intrator, Abbate, \&
  Madziwa-Nussinov}]{Ryutov2006}
Ryutov, D., Furno, I., Intrator, T., Abbate, S., \& Madziwa-Nussinov, T. 2006,
  Physics of plasmas, 13, 032105

\bibitem[{{Sagdeev}(1966)}]{Sagdeev66}
{Sagdeev}, R.~Z. 1966, Reviews of Plasma Physics, 4, 23

\bibitem[{{Scholz}(2014)}]{2014A&A...561A.113S}
{Scholz}, R.~D. 2014, \aap, 561, A113, \dodoi{10.1051/0004-6361/201323015}

\bibitem[{{Sfeir} {et~al.}(1999){Sfeir}, {Lallement}, {Crifo}, \&
  {Welsh}}]{1999A&A...346..785S}
{Sfeir}, D.~M., {Lallement}, R., {Crifo}, F., \& {Welsh}, B.~Y. 1999, \aap,
  346, 785

\bibitem[{{Skilling}(1975)}]{Skill75a}
{Skilling}, J. 1975, \mnras, 172, 557

\bibitem[{{Snowden} {et~al.}(2014){Snowden}, {Chiao}, {Collier}, {Porter},
  {Thomas}, {Cravens}, {Robertson}, {Galeazzi}, {Uprety}, {Ursino},
  {Koutroumpa}, {Kuntz}, {Lallement}, {Puspitarini}, {Lepri}, {McCammon},
  {Morgan}, \& {Walsh}}]{Snowden2014}
{Snowden}, S.~L., {Chiao}, M., {Collier}, M.~R., {et~al.} 2014, \apjl, 791,
  L14, \dodoi{10.1088/2041-8205/791/1/L14}

\bibitem[{Spangler(2009)}]{Spangler2009}
Spangler, S.~R. 2009, Plasma Turbulence in the Local Bubble, ed. J.~L. Linsky,
  V.~V. Izmodenov, E.~M{\"o}bius, \& R.~von Steiger (New York, NY: Springer New
  York), 277--290, \dodoi{10.1007/978-1-4419-0247-4_22}

\bibitem[{Strong \& Moskalenko(1998)}]{Strong_1998}
Strong, A.~W., \& Moskalenko, I.~V. 1998, The Astrophysical Journal, 509, 212,
  \dodoi{10.1086/306470}

\bibitem[{{Vidal} {et~al.}(2015){Vidal}, {Dickinson}, {Davies}, \&
  {Leahy}}]{2015MNRAS.452..656V}
{Vidal}, M., {Dickinson}, C., {Davies}, R.~D., \& {Leahy}, J.~P. 2015, \mnras,
  452, 656, \dodoi{10.1093/mnras/stv1328}

\bibitem[{{Vladimirov} {et~al.}(2012){Vladimirov}, {J{\'o}hannesson},
  {Moskalenko}, \& {Porter}}]{2012ApJ...752...68V}
{Vladimirov}, A.~E., {J{\'o}hannesson}, G., {Moskalenko}, I.~V., \& {Porter},
  T.~A. 2012, \apj, 752, 68, \dodoi{10.1088/0004-637X/752/1/68}

\bibitem[{{V{\"o}lk} {et~al.}(1984){V{\"o}lk}, {Drury}, \&
  {McKenzie}}]{Voelk84}
{V{\"o}lk}, H.~J., {Drury}, L.~O., \& {McKenzie}, J.~F. 1984, \aap, 130, 19

\bibitem[{{Wallner} {et~al.}(2016){Wallner}, {Feige}, {Kinoshita}, {Paul},
  {Fifield}, {Golser}, {Honda}, {Linnemann}, {Matsuzaki}, {Merchel}, {Rugel},
  {Tims}, {Steier}, {Yamagata}, \& {Winkler}}]{2016Natur.532...69W}
{Wallner}, A., {Feige}, J., {Kinoshita}, N., {et~al.} 2016, \nat, 532, 69,
  \dodoi{10.1038/nature17196}

\bibitem[{Whitham(2011)}]{Whitham2011}
Whitham, G.~B. 2011, Linear and nonlinear waves, Vol.~42 (John Wiley \& Sons)

\bibitem[{{Wiedenbeck} {et~al.}(1999){Wiedenbeck}, {Binns}, {Christian},
  {Cummings}, {Dougherty}, {Hink}, {Klarmann}, {Leske}, {Lijowski}, {Mewaldt},
  {Stone}, {Thayer}, {von Rosenvinge}, \& {Yanasak}}]{1999ApJ...523L..61W}
{Wiedenbeck}, M.~E., {Binns}, W.~R., {Christian}, E.~R., {et~al.} 1999, \apjl,
  523, L61, \dodoi{10.1086/312242}

\bibitem[{Wood {et~al.}(2002)Wood, M{\"u}ller, Zank, \& Linsky}]{Wood2002}
Wood, B.~E., M{\"u}ller, H.-R., Zank, G.~P., \& Linsky, J.~L. 2002, The
  Astrophysical Journal, 574, 412

\bibitem[{{Yoon} {et~al.}(2011){Yoon}, {Ahn}, {Allison}, {Bagliesi}, {Beatty},
  {Bigongiari}, {Boyle}, {Childers}, {Conklin}, {Coutu}, {DuVernois}, {Ganel},
  {Han}, {Jeon}, {Kim}, {Lee}, {Lutz}, {Maestro}, {Malinine}, {Marrocchesi},
  {Minnick}, {Mognet}, {Nam}, {Nutter}, {Park}, {Park}, {Seo}, {Sina},
  {Swordy}, {Wakely}, {Wu}, {Yang}, {Zei}, \& {Zinn}}]{2011ApJ...728..122Y}
{Yoon}, Y.~S., {Ahn}, H.~S., {Allison}, P.~S., {et~al.} 2011, \apj, 728, 122,
  \dodoi{10.1088/0004-637X/728/2/122}

\bibitem[{{Yuan} {et~al.}(2020){Yuan}, {Qiao}, {Guo}, {Fan}, \&
  {Bi}}]{Yuan2020}
{Yuan}, Q., {Qiao}, B.-Q., {Guo}, Y.-Q., {Fan}, Y.-Z., \& {Bi}, X.-J. 2020,
  arXiv e-prints, arXiv:2007.01768.
\newblock \doarXiv{2007.01768}

\bibitem[{Zirnstein {et~al.}(2016)Zirnstein, Heerikhuisen, Funsten, Livadiotis,
  McComas, \& Pogorelov}]{Zirnstein2016}
Zirnstein, E., Heerikhuisen, J., Funsten, H., {et~al.} 2016, The Astrophysical
  Journal Letters, 818, L18

\end{thebibliography}



\end{document}